\documentclass{statsoc}
\renewcommand{\normalsize}{\fontsize{10}{10}\selectfont}
\usepackage[margin=0.8in]{geometry}

\usepackage{amsgen,amsmath,amstext,amsbsy,amsopn,amssymb,subfigure,stmaryrd}
\usepackage[dvips]{graphicx}
\usepackage{comment}
\usepackage{enumitem}
\usepackage{algorithm}
\usepackage{algorithmic}
\usepackage{url}
\usepackage{color}
\usepackage{multirow}
\usepackage{float}

\newtheorem{Theorem}{Theorem}
\newtheorem{Assumption}{Assumption}
\newtheorem{Proposition}{Proposition}
\newtheorem{Conjecture}{Conjecture}

\newtheorem{Lemma}{Lemma}
\newtheorem{Remark}{Remark}

\newcommand{\subscript}[2]{$#1 _ #2$}
\newcommand{\cC}{\mathcal{C}}

\newcommand{\cS}{\mathcal{S}}

\newcommand{\cO}{\mathcal{O}}

\newcommand{\cI}{{\mathcal{I}}}

\newcommand{\cA}{{\mathcal{A}}}

\newcommand{\cE}{{\mathcal{E}}}

\newcommand{\cX}{{\mathcal{X}}}
\newcommand{\cY}{{\mathcal{Y}}}
\newcommand{\cZ}{{\mathcal{Z}}}
\newcommand{\tF}{\textrm{F}}

\newcommand{\X}{{\mathbf{X}}}
\newcommand{\B}{{\mathbf{B}}}

\newcommand{\I}{{\mathbf{I}}}
\newcommand{\M}{{\mathbf{M}}}
\newcommand{\N}{{\mathbf{N}}}
\newcommand{\E}{{\mathbf{E}}}

\renewcommand{\S}{{\mathbf{S}}}
\newcommand{\U}{{\mathbf{U}}}
\newcommand{\V}{{\mathbf{V}}}
\newcommand{\Var}{{\rm Var}}

\newcommand{\W}{{\mathbf{W}}}
\newcommand{\A}{{\mathbf{A}}}
\newcommand{\Y}{{\mathbf{Y}}}
\newcommand{\Z}{{\mathbf{Z}}}

\newcommand{\rank}{{\rm rank}}

\newcommand{\SSigma}{\boldsymbol{\Sigma}}

\newcommand{\diag}{{\rm diag}}

\newcommand{\SVD}{{\rm SVD}}
\newcommand{\SNR}{{\rm SNR}}

\newcommand{\argmin}{\mathop{\rm arg\min}}
\newcommand{\argmax}{\mathop{\rm arg\max}}
\newcommand{\bbR}{\mathbb{R}}
\newcommand{\bbI}{\mathbb{I}}

\newcommand{\bbO}{\mathbb{O}}
\newcommand{\bbP}{\mathbb{P}}
\newcommand{\bbE}{\mathbb{E}}
\newcommand{\cM}{\mathcal{M}}

\newcommand{\stat}{{\rm stat}}
\newcommand{\comp}{{\rm comp}}

\newcommand{\normSize}[2]{#1\lVert#2#1\rVert}

\title[High-order Tensor Clustering]{Exact Clustering in Tensor Block Model: Statistical Optimality and Computational Limit\footnote{R. Han, Y. Luo, and A. R. Zhang were partly supported by NSF Grants CAREER-2203741, DMS-2023239, NIH Grant R01 GM131399, and funding from Wisconsin Alumni Research Foundation. M. Wang was partly supported by NSF CAREER DMS-2141865, DMS-1915978, DMS-2023239, EF-2133740, and funding from Wisconsin Alumni Research Foundation.}}
\author[R. Han, Y. Luo, M. Wang and A. R. Zhang]{Rungang Han$^{1,2}$, Yuetian Luo$^1$, Miaoyan Wang$^1$, Anru R. Zhang$^{1,2, *}$}
\address{$^1$University of Wisconsin-Madison, USA}
\address{$^2$Duke University, USA}
\address{$^*$Address of correspondence: \texttt{anru.zhang@duke.edu}}

\begin{document}
\begin{abstract}
High-order clustering aims to identify heterogeneous substructures in multiway datasets that arise commonly in neuroimaging, genomics, social network studies, etc. The non-convex and discontinuous nature of this problem pose significant challenges in both statistics and computation. In this paper, we propose a tensor block model and the computationally efficient methods, \emph{high-order Lloyd algorithm} (HLloyd), and \emph{high-order spectral clustering} (HSC), for high-order clustering. The convergence guarantees and statistical optimality are established for the proposed procedure under a mild sub-Gaussian noise assumption. Under the Gaussian tensor block model, we completely characterize the statistical-computational trade-off for achieving high-order exact clustering based on three different signal-to-noise ratio regimes. The analysis relies on new techniques of high-order spectral perturbation analysis and a ``singular-value-gap-free'' error bound in tensor estimation, which are substantially different from the matrix spectral analyses in the literature. Finally, we show the merits of the proposed procedures via extensive experiments on both synthetic and real datasets.
\end{abstract}

\begin{sloppypar}
\section{Introduction}\label{sec:intro}

High-order tensors have received increasing recent attention in many fields including social networks~\citep{kolda2006tophits}, computer vision~\citep{koniusz2016sparse}, neuroscience~\citep{zhang2019tensor}, and genomics~\citep{hore2016tensor}. Tensors provide an effective representation of the hidden structures in multiway data. One of the popular structure utilized in tensor data analysis is the so-called low-rankness, which decomposes the signal tensor into a low-dimensional core tensor and multiple matrix factors, one on each modes~\citep{kolda2001orthogonal}. Despite many celebrated results in tensor data analysis under the low-rank formulation, such as tensor regression \citep{zhou2013tensor}, tensor completion \citep{xia2020statistically}, tensor PCA \citep{zhang2018tensor}, and generalized tensor learning \citep{han2020optimal}, another important model, \emph{multiway tensor block model}, has not been well studied yet. 

\begin{figure}[htbp]
\centering
\includegraphics[width=.8\textwidth]{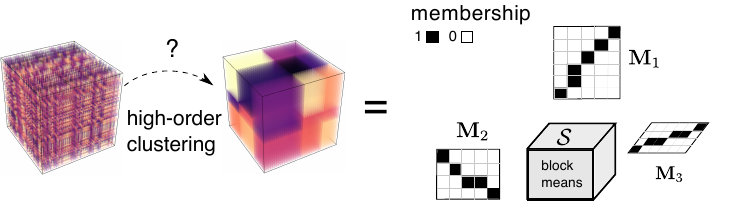}
\caption{High-order clustering aims to identify the block structure represented by membership matrices on each of the modes.}\label{fig:model-illu}
\end{figure}

Figure~\ref{fig:model-illu} shows an example of  multiway tensor block model for an order-3 tensor in which each of the modes is partitioned into several clusters. The goal is to identify the block structure (clustering), as well as to recover the whole tensor data (estimation), from observed data. In comparison with the low-rankness, the discrete block structure is more interpretable, because of the membership information encoded by loading matrices. The tensor block model and high-order clustering arise commonly in practical applications. For example,
\begin{itemize}
    \item \emph{Multi-tissue Gene Expression Analysis.} Gene expression profiles such as scRNA-seq and microarrays are collected from multiple individuals across numbers of tissues~\citep{mele2015human,wang2019three}. Genes involved in the same biological function typically exhibit similar expressions for some group of tissues and individuals, while these expression values vary from group to group. Similarly, tissues/individuals exhibit clustering patterns due to the similarity therein. Investigating the complex interactions among these three entities is of great scientific interest. 
    
    \item \emph{Multilayer Network Analysis.} Multilayer networks arise commonly in longitudinal studies of network~\citep{lei2019consistent} and multi-relational data~\citep{nickel2011three}. A multilayer network consists of multiple directed/undirected graphs (or adjacency matrices), where each graph represents the connection among the same set of vertices. The multilayer network data can be organized as an order-3 tensor with the first two modes being vertices and the third mode being the contexts under which the graph is observed. Depending on how the connection edges are encoded, the tensor entries can be either binary (indicating presence/absence of connection) or continuous (representing weighted connection strength). 
    
    \item \emph{Online Click-through Prediction.} In e-commerce, predicting click-through for user-item pairs in a time-specific way plays an important role in online recommendation system  \citep{sun2015provable,shan2016predicting}. The click-through data in a specific day can be organized as an order-3 tensor, where each entry indexed by (users, items, time) represents the total number of user-item interactions in a time period (e.g., 24 different hours in that day). The users/items often exhibit clustering structures due to similar preferences/attributes. In addition, the shopping behaviour also varies in time, and this heterogeneity depends on the specific group of users and items. 
\end{itemize}
Additional applications of tensor block model include hypergraph clustering~\citep{ke2019community,chien2019minimax}, collaborative filtering~\citep{zhang2020dynamic} and signal detection in 3D/4D imaging~\citep{zhang2020denoising}, among others.

The multi-way structure of a data tensor imposes unique challenges to clustering analysis. In the vector case, clustering methods find subgroups of the observations based on entrywise similarity; while in the matrix case, bi-clustering algorithms seek to simultaneously identify the block structure of observations (rows) and features (columns)~\citep{busygin2008biclustering}. Recent developments of tensor-based high-order clustering methods roughly fall into two types. The first approach utilizes the maximum likelihood estimation (MLE)~\citep{wang2019multiway} to search for the tensor block structure. The MLE, however, is a mixed-integer programming and therefore NP-hard to compute in general~\citep{aloise2009np}. The computational intractability renders the statistical inference less useful in practice. The second approach adopts polynomial-time algorithms for surrogate objectives. Efforts in this vein include convex relaxation~\citep{chi2020provable} and spectral relaxation~\citep{zha2002spectral,wu2016general}. Despite the popularity, these methods often sacrifice the statistical accuracy for computational feasibility. To our best knowledge, a provable scheme that achieves both statistical and computational efficiency has yet to be developed.

In this paper, we develop a computationally efficient procedure for the task of high-order clustering in tensor block model. The procedure operates in steps: \emph{High-order Spectral Clustering} (HSC) and \emph{High-order Lloyd} (HLloyd). The proposed HSC algorithm involves a power iteration procedure. While the statistical property of the power iteration has been recently established under a strong singular value gap condition \citep{zhang2018tensor,luo2021sharp}, the previous result is not applicable to our analysis, because of the possible lack of singular value gap in block tensors even under the model identifiablility conditions. This difference originates from the unique ``discrete'' low-rank structure in tensor block model as opposed to the ``continuous'' low-rank structure considered in \cite{zhang2018tensor}. The discrete setting requires new  theoretical analysis for high-order clustering methods (see Section \ref{sec:TBM} and Remark \ref{rmk:singular-gap}). We then establish the clustering error rate for HSC algorithm under modest conditions (see Section \ref{sec:HSC-theory}). The second component of the proposed procedure, HLloyd algorithm, can be seen as a high-order extension of Lloyd algorithm for 1-dimensional \emph{$k$-means} to order-$d$ clustering. Compared to the analysis of Lloyd algorithm for vector clustering with a single discrete structure~\citep{lu2016statistical}, the multiple discrete structures in high-order clustering make the analysis more challenging (see details in Remark \ref{rmk:compare to single structure}). We prove that, under warm initialization, our HLloyd algorithm solves the high-order clustering problem with optimality guarantees in tensor block model.

\begin{figure}[htbp]
\vspace{.4cm}
\centering
\includegraphics[width=.8\textwidth]{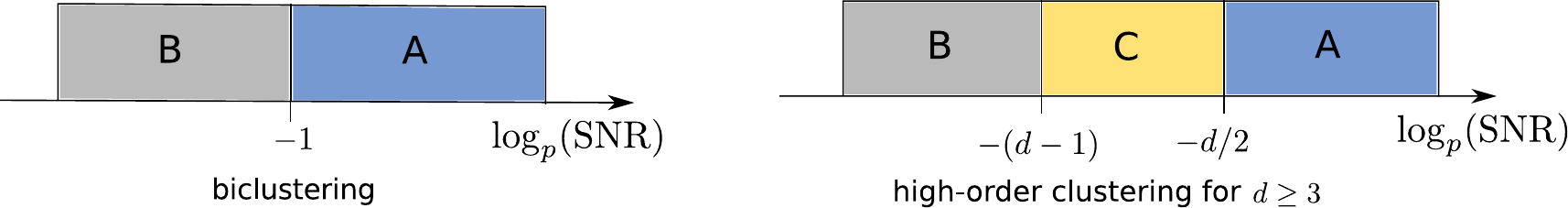}
\caption{Phase transition diagram for high-order clustering in order-$d$ tensor block model with dimension $(p,\ldots,p)$. A statistical-computational gap (Region C) arises only for tensors of order 3 or greater.}\label{fig:phase-transition}
\end{figure}

Apart from the newly proposed algorithm, we discover an intriguing interplay between statistical optimality and computational efficiency of high-order clustering in tensor block model. Specifically, we introduce a notion of signal-to-noise ratio (SNR) for tensor block model that quantifies the minimum gaps between block means (see formal definition in Section \ref{sec:theory-assumption}) over the noise level. This notion completely characterizes the hardness of the high-order clustering in tensor block model. Our main phase transition results can be informally summarized as follows.
\begin{Theorem}[Informal results]\label{thm:informal}
    Consider the high-order clustering on an order-$d$ dimension-$p$ tensor under the Gaussian tensor block model (see \eqref{eq:model-tensor} in Section \ref{sec:TBM}). Suppose $\SNR = p^\gamma$ and $p \to \infty$. 
    \begin{itemize}
        \item When $\gamma > -d/2$, the proposed HLloyd + HSC algorithm performs exact clustering (Theorem \ref{coro:exact-recovery});
        \item When $\gamma < -(d-1)$, no algorithm can achieve exact clustering (Theorem \ref{thm:mcr-lower-bound});
        \item When $-(d-1) < \gamma < -d/2$, MLE achieves exact clustering at the cost of being computationally intractable; and no polynomial-time algorithm can achieve exact clustering under a computational hardness assumption for hypergraphic planted clique detection (Theorem \ref{thm:computational-lower-bound}).
    \end{itemize}
    Here, the exact clustering means that the clustering labels are precisely recovered with high probability.
\end{Theorem}
Figure~\ref{fig:phase-transition} summarizes the phase transitions in high-order clustering under the tensor block model. In the strong SNR region A ({$\SNR = p^\gamma$, $\gamma > -d/2$}), we prove that the combination of HSC and HLloyd achieves exact clustering in polynomial time. We find that the estimation error bound of the target tensor is free of the tensor dimension (Theorem \ref{thm:tensor-est-hp}), which is distinct from the tensor estimation error bounds in the literature under the continuous low-rank structure (see Remark \ref{rmk: dimension free tensor est}). In the weak SNR region B ($\gamma < -(d-1)$), we develop a minimax lower bound to show that no algorithm succeeds in high-order clustering for tensor block model. In the modest SNR region C ($-(d-1) < \gamma < -d/2$), we show that the problem is statistically possible while computationally infeasible. That is, computing any estimate that achieves exact clustering is as hard as solving a version hypergraphic planted clique detection problem which is conjectured to be polynomial-time unsolvable (see details in Section \ref{sec:comp-limits}). Note that the former two SNR regions apply to matrix biclustering $(d=2)$ while the latter statistical-computational gap region exists only for high-order tensors with $d\geq 3$. To the best of our knowledge, we are among the first to establish both of the statistical and computational limits for high-order clustering in tensor block models.

\subsection{Related Literature}\label{sec:literature}

Our work is related to but also clearly distinctive from several lines of existing work. Classic clustering algorithms such as $k$-means~\citep{jain2010data} and spectral clustering~\citep{von2007tutorial} have been widely used in statistics and machine learning. In the order-1 (vector) case, the clustering problem is usually formulated under Gaussian mixture model. The optimal statistical guarantee has been developed for the state-of-art clustering algorithms, including spectral clustering \citep{loffler2019optimality}, EM algorithm \citep{wu2019randomly} and Lloyd algorithm \citep{lu2016statistical}. In the order-2 (matrix) case, clustering methods have been studied under the stochastic block model \citep{abbe2017community}, biclustering \citep{gao2016optimal}, and bipartite community detection~\citep{zhou2019analysis}. 
These existing methods and theories are inapplicable to high-order clustering due to the distinct algebraic properties of general order-$d$ tensors. The high-order clustering also exhibits many distinct features compared to the clustering for vectors or biclustering for matrices, e.g., the statistical-computational gap arises only from tensors of order 3 or greater as shown in Figure~\ref{fig:phase-transition}.

In addition, our work is related to the recent development on low-rank tensor decomposition, in which the main goal is to find the best low-rank approximation of a tensor. Numerous algorithms have been proposed, such as truncated power iteration~\citep{anandkumar2014tensor}, high-order SVD~\citep{de2000best}, high-order orthogonal iteration~\citep{de2000multilinear}, sparse tensor SVD~\citep{zhang2019optimal}. Although the block structure implies low-rankness (see Section \ref{sec:TBM}), the classic low-rank tensor spectral methods fail to fully utilize the structural information of tensor block model. Moreover, the singular value gap condition, which was commonly imposed in the literature on tensor decomposition, does not generally hold in the tensor block model, and thus new technical tools are required for our high-order clustering problem. 

Another related topic is on the statistical and computational trade-off in high-dimensional statistics. This topic focuses on the gap between signal-to-noise ratio requirements under which the problem is information-theoretically solvable vs. polynomial-time solvable. Some common structures yielding these trade-offs include sparsity \citep{ma2015computational,chen2016statistical}, robustness \citep{diakonikolas2017statistical}, tensors \citep{richard2014statistical,barak2016noisy}, etc. In the last decade, a number of schemes have been proposed to provide rigorous evidence for the gap between computational and statistical limits, such as average-case reduction \citep{wang2016statistical,brennan2018reducibility}, sum of squares \citep{barak2019nearly}, statistical query \citep{feldman2017statistical}, low-degree polynomials \citep{hopkins2017bayesian}, etc. We refer readers to the recent survey \citep{wu2018statistical} for a review. In this work, we reveal the statistical and computational trade-off for high-order clustering in the tensor block model and give rigorous evidence for the computational limit via average-case reduction. 

\subsection{Organizations}\label{sec:organization}

The rest of the paper is organized as follows. After a brief review of basic tensor algebra in Section \ref{sec:notation}, we introduce the tensor block model and the high-order tensor clustering problem in Section \ref{sec:TBM}. Then, the HSC and HLloyd algorithms are proposed to solve this problem in Section \ref{sec:alg}. We present the statistical proprieties of the algorithms in Section \ref{sec:theory}. We study the fundamental statistical and computational limits of high-order clustering in tensor block model in Section \ref{sec:comp-limits}. Extensive numerical analyses on simulated data, flight route network data, and online click-through data are presented in Sections \ref{sec: simulation} and \ref{sec:real-data}, respectively. Conclusion and discussion are given in Section \ref{sec:discussions}. Proof sketches for the main results are provided in Section \ref{sec:proof-sketch}, and all detailed proofs are collected in the Supplementary Materials. Codes for running the simulations and real data analyses of this paper are available at \url{https://github.com/Rungang/HLloyd}.

\section{Multiway Tensor Block Model}\label{sec:model}

\subsection{Notations and Preliminaries}\label{sec:notation}

We use lowercase letters ($a,b,x\ldots$) to denote scalars/vectors. For any $a, b \in \bbR$, let $a\wedge b$ and $a\vee b$ be the minimum and maximum of $a$ and $b$, respectively. Suppose $\{a_n\}, \{b_n\}$ are two sequences of positive numbers. We denote $a_n\lesssim b_n$ or $a_n=\cO(b_n)$ (respectively, $a_n\gtrsim b_n$) if there exists a constant $C>0$ such that $a_n\leq Cb_n$ (respectively, $a_n\geq Cb_n$) for all $n$; and we denote $a_n \asymp b_n $ if there exists $c, C$ such that $c a_n\leq b_n \leq C a_n$ for all $n$. We denote matrices by bold uppercase letters ($\A,\B,\U,\ldots$).  Let $\bbO_{p,r}$ be the collection of all $p$-by-$r$ matrices with orthonormal columns: $\bbO_{p,r} = \{\U \in \bbR^{p\times r}: \U^\top \U = \I\}$, where $\I$ is the identity matrix. We use $\A_{ij}, \A_{i:}$, and $\A_{: j}$ to denote the $(i,j)$th entry, the $i$th row, and the $j$th column of $\A$, respectively. Let $\lambda_1(\A) \geq \lambda_2(\A) \geq \cdots \geq 0$ be the singular values of $\A$ in descending order and we use $\SVD_r(\A)$ to denote the matrix comprised of the top $r$ left singular vectors of $\A$. We use $\|\A\| = \lambda_1(\A)$ to denote the spectral norm of $\A$ and use $\|\A\|_\tF = \sqrt{\sum_{i=1}^{p_1}\sum_{j=1}^{p_2}\A_{ij}^2} = \sqrt{\sum_{i=1}^{p_1\wedge p_2}\lambda_i^2(\A)}$ to denote its Frobenius norm. For any matrix $\A=[a_1,\ldots, a_J]\in \mathbb{R}^{I\times J}$ and $\B\in \mathbb{R}^{K\times L}$, the \emph{Kronecker product} is defined as the $(IK)$-by-$(JL)$ matrix $\A\otimes\B = [a_1\otimes \B \cdots a_J\otimes \B]$. 

Recall that tensors are multi-way arrays. We call the number of modes of a tensor as its order. We use calligraphy letters ($\cA,\cX,\cY\ldots$) to denote tensors of order three or higher. For example, if a tensor $\cX$ represents a $d$-way array of size $p_1\times p_2 \times \cdots \times p_d$, we say $\cX$ is an order-$d$ tensor and write $\cX \in \bbR^{p_1\times \cdots \times p_d}$. The $(i_1,\ldots,i_d)$th element of a tensor $\cX$ is written as $\cX_{i_1,\ldots,i_d}$. Let $\cX$ and $\cY$ be two tensors of the same dimension, the inner product of them is defined as $\langle \cX, \cY \rangle = \sum_{i_1,\ldots,i_d}\cX_{i_1,\ldots, i_d}\cY_{i_1,\ldots, i_d}$. The Frobenius norm of the tensor $\cX$ is defined as $\|\cX\|_\tF = \langle \cX,\cX\rangle^{1/2}$. The multilinear multiplication of a tensor $\cS\in \bbR^{r_1\times \cdots \times r_d}$ by matrices $\U_k \in\mathbb{R}^{p_k\times r_k}$ is defined as
\begin{equation*}
\left(\cS \times_1 \U_1 \times \cdots \times_d \U_d\right)_{i_1,\ldots, i_d} =\textstyle\sum_{j_1=1}^{r_1} \cdots\sum_{j_d=1}^{r_d}  \cS_{j_1,\ldots, j_d}(\U_{1})_{i_1j_1}\cdots(\U_{d})_{i_dj_d}, 
\end{equation*}
which results in an order-$d$ $(p_1,\ldots,p_d)$-dimensional tensor.
 We also introduce the matricization operator that transforms tensors to matrices. Particularly the mode-$1$ matricization for $\cX \in \bbR^{p_1 \times \cdots \times p_d}$ is defined as
\begin{equation*}
    \begin{split}
	    \cM_1 (\cX) \in \bbR^{p_1 \times (p_2\cdots p_d)}, \text{ where}~[\cM_1(\cX)]_{j_1,j_2 + p_2(j_3-1)+\cdots+p_2\cdots p_{d-1}(j_d-1)} = \cX_{j_1,\ldots,j_d}. 
	\end{split}
\end{equation*}
Each row of $\cM_k(\cX)$ is the vectorization of a mode-$k$ slice. The following identity that relates the matrix-tensor product and matricization will be used extensively in our analysis:
\begin{equation*}
    \cM_k(\cS \times_1 \U_1 \times \cdots \times_d \U_d) = \U_k \cM_k(\cS) \left(\U_{k+1} \otimes \cdots \otimes \U_d \otimes \U_{1} \otimes \cdots \otimes \U_{k-1}\right)^\top.
\end{equation*}
We say a tensor $\cX \in \bbP^{p_1\times \cdots \times p_d}$ has Tucker-rank $(r_1,\ldots,r_d)$ if $r_k = \rank(\cM_k(\cX))$. In this case $\cX$ admits a \emph{Tucker decomposition}:
\begin{equation}\label{eq:Tucker-def}
    \cX = \cS \times_1 \U_1 \times \cdots \times_d \U_d
\end{equation}
for some $\cS \in \bbR^{r_1\times \cdots \times r_d}$ and $\U_k \in \bbR^{p_k \times r_k}$.
The readers are referred to \cite{kolda2009tensor} for a more comprehensive tutorial on tensor decomposition.

A clustering that partitions $p$ entities into $r$ clusters is represented by a vector $z \in [r]^p$ such that the $i$th entry of $z$ equals to $s$ if and only if the $i$th entity belongs to the $s$th cluster. Here, we use shorthand $[r]:=\left\{1,2,\ldots,r\right\}$ to denote the $r$-set. For two clusters $a=(a_1,\ldots,a_p)^\top, b=(b_1,\ldots,b_p)^\top\in [r]^{p}$ on the same set, we denote the \emph{misclassification rate} as: 
\[
h(a,b)=\min_{\pi \in \Pi_{r}} \frac{1}{p}\sum_{i=1}^p \bbI\{a_i \neq \pi(b_i)\},
\]
where $\Pi_r$ is the collection of all permutations $\pi$ on $[r]$ and $\bbI(\cdot)$ is the indicator function. Let $\hat z$ be an estimator of $z \in [r]^p$. As sample size goes to infinity, we say $\hat z$ is consistent if
\begin{equation*}
    \bbP\left(h(\hat z, z) > \varepsilon\right) \rightarrow 0,\qquad \forall \varepsilon > 0;
\end{equation*}
and we say $\hat z$ exactly recovers $z$ if
\begin{equation*}
    \bbP\left(h(\hat z, z) = 0\right) \rightarrow 1.
\end{equation*}

Finally, we use $C, C_0, C_1, \ldots$ and $c, c_0, c_1,\ldots$ to represent generic large and small positive constants, respectively. The actual values of these generic symbols may differ from line to line. We introduce the following notions for $p_1\times \cdots \times p_d$ tensors with rank $(r_1,\cdots,r_d)$:
\begin{equation*}
    \begin{split}
        & \overline p = \max_{k\in [d]} p_k,~ \underline p = \min_{k\in [d]} p_k, ~ p_* = \prod_{k\in [d]} p_k,~ p_{-k} = p_*/p_k,\\
        & \overline r = \max_{k\in [d]} r_k~, r_* = \prod_{k\in [d]} r_k,~ r_{-k} = r_*/r_k.
    \end{split}
\end{equation*}

\subsection{Tensor block model}\label{sec:TBM}

Let $\cY\in \bbR^{p_1 \times \cdots \times p_d}$ be an order-$d$ $(p_1,\ldots,p_d)$-dimensional data tensor of interest. The tensor block model assumes an underlying checkerbox structure in the signal tensor (see Figure~\ref{fig:model-illu}). Specifically, suppose there are $r_k$ clusters in the $k$th mode of the signal tensor for all $k \in [d]$, and we represent the clustering along $k$th mode by a vector $z_k \in [r_k]^{p_k}$. Then, the entries $\cY_{j_1,\ldots,j_d}$ are  realizations from the following block model:
\begin{equation}\label{eq:model-tensor-entry}
\cY_{j_1,\ldots,j_d}=\cS_{(z_1)_{j_1},\ldots,(z_d)_{j_d}}+\cE_{j_1,\ldots,j_d},\quad \forall (j_1,\ldots,j_d)\in[p_1]\times \cdots \times[p_d],
\end{equation}
where $\cS\in \bbR^{r_1 \times \cdots \times r_d}$ is the core tensor with collected block means, and $\cE_{j_1,\ldots,j_d}$s are some mean-zero observational noises. Model~\eqref{eq:model-tensor-entry} can be equivalently written in a form of tensor-matrix product: 
\begin{equation}\label{eq:model-tensor}
	\cY = \cS \times_1 \M_1 \times \cdots \times_d \M_d + \cE,
\end{equation}
where $\cE\in \mathbb{R}^{p_1\times \cdots \times p_d}$ is the noise tensor,  $\M_k \in \{0,1\}^{p_k\times r_k}$ is the membership matrix associated with $z_k$ such that $(\M_k)_{ij} = 1$ if and only if $(z_k)_{i} = j$. That is, $\M_k$ has one copy of 1 and $(r_k-1)$ copies of 0s in each row. We will use forms \eqref{eq:model-tensor-entry} and \eqref{eq:model-tensor} interchangeably throughout the paper. Note that $\bbE\cY$ admits a Tucker low-rank structure \eqref{eq:Tucker-def} with Tucker-rank bounded by, but may not equal to, $(r_1,\ldots,r_d)$. The discrete structure in $\M_k$ makes the model more informative and brings new challenges as we mentioned in the introduction. 

We impose the following distributional assumption on the noise tensor $\cE$:
\begin{Assumption}[Sub-Gaussian noise]\label{asmp:sub-Gaussian-distribution}
Suppose each entry of $\cE$ follows an independent zero-mean sub-Gaussian distribution with sub-Gaussian norm bounded by $\sigma$:
\begin{equation*}
    \bbE \exp\left(\lambda\cE_{j_1,\ldots,j_d}\right) \leq e^{\lambda^2\sigma^2/2},\qquad \forall \lambda \in \bbR.
\end{equation*}
\end{Assumption}
Assumption \ref{asmp:sub-Gaussian-distribution} holds in some specific problems arising in applications:
\begin{itemize}
    \item Gaussian Tensor Block Models (GTBM): each entry $\cY_{j_1,\ldots,j_d}$ is Gaussian-distributed with mean $\cS_{(z_1)_{j_1},\ldots,(z_d)_{j_d}}$ and variance $\sigma^2$. This setting is suitable to model a tensor $\cY$ with continuous entries on $\bbR$;
    \item Stochastic Tensor Block Models (STBM): the core tensor $\cS \in [0,1]^{r_1\times \cdots \times r_d }$ and each entry $\cY_{(z_1)_{j_1},\ldots,(z_d)_{j_d}} \sim \text{Bernoulli}(\cS_{(z_1)_{j_1},\ldots,(z_d)_{j_d}})$. Let $\cE = \cY - \bbE \cY$. Then Assumption \ref{asmp:sub-Gaussian-distribution} holds for $\sigma \leq 1/4$. This setting is suitable to model a tensor $\cY$ with binary entries on $\{0,1\}$.
\end{itemize}

In this paper, we mainly focus on two tasks on the inference for tensor block model:
\begin{itemize}[noitemsep]
\item Clustering. Recover the membership matrix $\M_k$, or equivalently the label vector $z_k$, for each mode. 
\item Estimation. Estimate the underlying signal tensor $\cX:=\mathbb{E}(\cY)$.
\end{itemize}
\section{Algorithms for High-order Clustering}\label{sec:alg}

We introduce the procedure for high-order clustering in this section. The procedure includes two algorithms, High-order Spectral Clustering (HSC) and High-order Lloyd (HLloyd), which will be elaborated in the next two subsections. 

\subsection{High-order Lloyd Algorithm (HLloyd)}\label{sec:HLloyd}

As a starting point, it is natural to consider the following least squares estimator of $\cS$ and $z_k$:
\begin{equation}\label{eq:MLE}
	\left(\hat\cS, \hat z_1,\ldots \hat z_d\right) = \argmin_{\cS,z_k \in [r_k]^{p_k},k=1,\ldots,d} \sum_{j_1,\ldots,j_d}  \left(\cY_{j_1,\ldots,j_d} -  \cS_{(z_1)_{j_1},\ldots,(z_d)_{j_d}}\right)^2.
\end{equation}
This scheme is a mixed-integer programming with one continuous ($\cS$) and $d$ discrete ($z_1,\ldots,z_d$) arguments. In general, \eqref{eq:MLE} is non-convex and computationally intractable.

Therefore, we propose a new iterative method to solve this problem. Suppose at step $t$, we have estimators $\cS^{(t)}, z_1^{(t)}, \cdots, z_d^{(t)}$ and want to update them at step $t+1$. On one hand, given the block membership vectors $(z_1^{(t)}, \cdots, z_d^{(t)})$, the optimization objective function \eqref{eq:MLE} becomes quadratic of $\cS$ and the optimal solution can be computed via a block-wise average:
\begin{equation}\label{eq:S-update}
    \begin{split}
        \cS^{(t+1)} = \argmin_{\cS} \sum_{j_1,\ldots,j_d}  \left(\cY_{j_1,\ldots,j_d} -  \cS_{(z_1^{(t)})_{j_1},\ldots,(z_d^{(t)})_{j_d}}\right)^2 \\
        \implies\quad  \cS^{(t+1)}_{i_1,\ldots,i_d} = \text{Average}\left(\left\{\cY_{j_1,\ldots,j_d}\colon (z^{(t)}_{k})_{j_k} = i_k, \forall k \in [d]\right\}\right),
    \end{split}
\end{equation} 
where $\text{Average}(\cdot)$ computes the sample mean given a set of values. 

On the other hand, given $\cS^{(t)}$ and $(d-1)$ block memberships ($z_1^{(t)},\ldots, z_{k-1}^{(t)}, z_{k+1}^{(t)},\ldots, z_{d}^{(t)}$), the update of membership vector $z_k^{(t+1)}$ can be obtained by performing a nearest neighbor search in a dimension-reduced space. Specifically, we first aggregate all mode-$k$ slices of $\cY$ to a dimension-reduced space using $\{z_{k'}^{(t)}\}_{k'\neq k}$, the block information from the other $d-1$ clusters. That is, we calculate $\cY_k^{(t)} \in \bbR^{r_1\times \cdots \times r_{k-1} \times p_k \times r_{k+1} \times \cdots \times r_d}$ as
\begin{equation}\label{eq:z-update-aggregation}
    (\cY_k^{(t)})_{i_1,\cdots,i_{k-1},j,i_{k+1},\ldots,i_d} = \text{Average}\left(\left\{\cY_{j_1,\cdots,j_{k-1},j,j_{k+1},\ldots,j_d}\colon (z^{(t)}_{l})_{j_l} = i_l, \forall l \in [d]/k\right\}\right).
\end{equation}
Intuitively speaking, when the other $d-1$ memberships ($z_1^{(t)},\ldots, z_{k-1}^{(t)}, z_{k+1}^{(t)},\ldots, z_{d}^{(t)}$) are close to the truth, such an aggregation can significantly reduce the noise level within $\cY_k^{(t)}$. Then, we perform the nearest neighbor search to update the estimate for $(z_k)_j$:
\begin{equation}\label{eq:z-update}
    (z_k^{(t+1)})_j = \argmin_{a \in [r_k]} \left\|\left(\cM_{k}(\cY_k^{(t)})\right)_{j:} - \left(\cM_{k}(\cS^{(t)})\right)_{a:}\right\|_2^2.
\end{equation}
The full procedure is presented in Algorithm~\ref{alg:HO-Lloyd} and is referred to as the \emph{high-order Lloyd} (HLloyd) algorithm.
\begin{Remark}
Our updating scheme of $z_k^{(j+1)}$ is slightly different from the updating scheme in the classic Lloyd algorithm. In particular, \eqref{eq:z-update} is not an exact local greedy solution for~\eqref{eq:S-update} for fixed $\cS^{(t)}$ and ($z_1^{(t)},\ldots, z_{k-1}^{(t)}, z_{k+1}^{(t)},\ldots, z_{d}^{(t)}$). In comparison, the computational complexity for the proposed updating scheme, i.e., \eqref{eq:z-update-aggregation} and \eqref{eq:z-update}, is $O(p_* + p_kr_*)$, while the complexity of local greedy method is $O(p_*r_k)$, since the proposed updating scheme performs nearest neighbor search only on an $r_{-k}$-dimensional space rather than the original $p_{-k}$-dimensional space. 
\end{Remark}

\begin{algorithm}[!ht]
\caption{High-order Lloyd Algorithm (HLloyd)}
\label{alg:HO-Lloyd}
\begin{algorithmic}
\REQUIRE Data tensor $\cY\in \bbR^{p_1\times\cdots\times p_d}$, initialization labels $\{z^{(0)}_k\in[r_k]^{p_k}\}$, iteration number $T$
\FORALL {$t=0$ to $T-1$} 
	\STATE Update the block means $\cS^{(t)}$ via
	$$\cS^{(t)}_{i_1,\ldots,i_d} = \text{Average}\left(\left\{\cY_{j_1,\ldots,j_d}\colon (z^{(t)}_{k})_{j_k} = i_k, \forall k \in [d]\right\}\right).$$
	\FORALL {$k=1,\ldots,d$}
		\FORALL {$j =1,\ldots,p_k$}
		\STATE Calculate $ \cY_k^{(t)} \in \bbR^{r_1\times \cdots \times r_{k-1} \times p_k \times r_{k+1} \times \cdots \times r_d}$ such that
		\begin{equation*}
			\begin{split}
		    & (\cY_k^{(t)})_{i_1,\cdots,i_{k-1},j,i_{k+1},\ldots,i_d} \\
		    = &  \text{Average}\left(\left\{\cY_{j_1,\cdots,j_{k-1},j,j_{k+1},\ldots,j_d}\colon (z^{(t)}_{l})_{j_l} = i_l, \forall l \in [d]/k\right\}\right).
		    \end{split}
		\end{equation*}
		\STATE 
		Update the mode-$k$ membership for the $j$th entity $(z_{k}^{(t+1)})_j$ via 
		\begin{equation*}
		(z_k^{(t+1)})_j = \argmin_{a \in [r_k]} \left\|\left(\cM_{k}(\cY_k^{(t)})\right)_{j:} - \left(\cM_{k}(\cS^{(t)})\right)_{a:}\right\|_2^2.
		\end{equation*}
		\ENDFOR
	\ENDFOR
\ENDFOR
\RETURN {Estimated block memberships $\{z^{(T)}_k, k=1,\ldots,d\}$}
\end{algorithmic}
\end{algorithm}

\subsection{High-order Spectral Clustering (HSC)}\label{sec:HSC}

While HLloyd provides an iterative optimization strategy for tensor clustering, initialization of labels $z_k^{(0)}$ remains unaddressed and turns out to be crucial to the clustering performance. In this section, we propose the \emph{High-order Spectral Clustering} (HSC) algorithm for initialization. This algorithm is a generalization of the classic spectral clustering method but with a number of fundamental novelties as we describe below. Note that the main step for spectral clustering is to estimate the singular subspace of the membership matrix $\M_k$. By tensor algebra, we have
\begin{equation*}
    \bbE \cM_k(\cY) = \cM_k(\cX) = \M_k \cM_k(\cS) \left(\M_{k+1} \otimes \cdots \otimes \M_d \otimes \M_1 \otimes \cdots \otimes \M_{k-1}\right).
\end{equation*}
It is tempting to estimate the singular subspace of $\M_k$ by 
\begin{equation}\label{eq:HOOI-1}
	\tilde\U_k = \SVD_{r_k}(\cM_k(\cY)), \quad k=1,\ldots,d.
\end{equation}
This method is referred to as the high-order singular value decomposition (HOSVD) \citep{de2000multilinear}. However, we find that $\tilde\U_k$ here does not fully utilize the low-rank structure in all modes of the data. We propose to further improve the estimate by making the projection of $\cY$ onto the pre-estimated subspaces of the other $(d-1)$ modes, i.e.,
\begin{equation}\label{eq:HOOI-2}
    \hat \U_k = \SVD_{\min\{r_k, r_{-k}\}}\left(\cM_k(\cY \times_1 \tilde\U_1^\top \times \cdots \times_{k-1}\tilde\U_{k-1}^\top \times_{k+1} \tilde\U_{k+1}^\top \times\cdots \times_d \tilde\U_d^\top)\right).
\end{equation}

After obtaining $\{\hat{\U}_k\}_{k=1}^d$, we get the initialization labels $z_k^{(0)}$ by performing $k$-means on $p_k$ rows of the following projected matrix:
\begin{equation}\label{eq:low-dim-loadings}
    \hat \Y_k = \hat\U_k\hat\U_k^\top\cM_k\left(\cY \times_1 \hat\U_1^\top \times \cdots \times_{k-1}\hat\U_{k-1}^\top \times_{k+1} \hat\U_{k+1}^\top \times\cdots \times_d \hat\U_d^\top\right).
\end{equation}
The pseudocode of HSC is given in Algorithm \ref{alg:HO-SC}. In particular, since the exact $k$-means may be computationally difficult, we use a relaxed $k$-means in Step~\eqref{ineq:K-means}, which can be efficiently solved by approximation algorithms, such as $k$-means++ with relaxation factor $M = \cO(\log \bar r)$~\citep{arthur2006k}.  
\begin{algorithm}
\caption{High-order spectral clustering (HSC)}
\label{alg:HO-SC}
\begin{algorithmic}
\REQUIRE $\cY\in \bbR^{p_1\times \cdots \times p_d}$, $r_1,\ldots,r_d$, relaxation factor in $k$-means: $M>1$
    \STATE Compute $\tilde\U_k = \SVD_{r_k}(\cM_k(\cY))$ for  $k=1,\ldots,d$
\FORALL {$k=1$ to $d$}
	\STATE Compute the singular space estimator $\hat\U_k$ via
	\begin{equation*}
	\begin{split}
    \hat \U_k = \SVD_{\min\{r_k, r_{-k}\}}(\cM_k(\cY \times_1 \tilde\U_1^\top \times \cdots \times_{k-1}\tilde\U_{k-1}^\top \times_{k+1} \tilde\U_{k+1}^\top \times\cdots \times_d \tilde\U_d^\top)).
    \end{split}
	\end{equation*}
\ENDFOR
\FORALL {$k=1$ to $d$}
	\STATE {Calculate $\hat \Y_k = \hat\U_k\hat\U_k^\top\cM_k\left(\cY \times_1 \hat\U_1^\top \times \cdots \times_{k-1}\hat\U_{k-1}^\top \times_{k+1} \hat\U_{k+1}^\top \times\cdots \times_d \hat\U_d^\top\right)$} 	
	\STATE {Find $z_k^{(0)} \in [r_k]^{p_k}$ and centroids $\hat x_1, \ldots, \hat x_{r_k} \in \bbR^{r_{-k}}$ such that}
	\begin{equation}\label{ineq:K-means}
		\sum_{j=1}^{p_k}\normSize{}{(\hat\Y_k)_{j:}^\top - \hat x_{(z_k^{(0)})_j}}_2^2 \leq M\min_{\substack{x_{1}, \ldots, x_{r_k} \in \bbR^{r_{-k}}\\ z_k\in [r_k]^{p_k}}}\sum_{j=1}^{p_k}\normSize{}{(\hat \Y_k)_{j:}^\top - x_{(z_k)_j}}_2^2
	\end{equation} 
\ENDFOR
\RETURN {$\{z_k^{(0)}\in[r_k]^{p_k}, k=1,\ldots, d\}$}
\end{algorithmic}
\end{algorithm}

Our algorithm takes $r_k$'s as inputs. In our theory, $r_k$'s are allowed to grow with the tensor dimension. In our simulation studies, we assume the true $r_k$'s are known for simplicity; while in practice, we recommend rank selection using data-driven criteria (see \eqref{eq:BIC} in Section \ref{sec:real-data}) such as the Bayesian information criterion (BIC) and/or prior knowledge.

\begin{Remark}
In the matrix bi-clustering setting ($d=2$), the step of refined SVD~\eqref{eq:low-dim-loadings} can be removed from the algorithm, since $\hat{\U}_k$ and $\tilde{\U}_k$ are provably the same. Moreover, $\hat{\U}_k$ has guaranteed accuracy due to the Eckart-Young-Mirsky theorem \citep{eckart1936approximation,mirsky1960symmetric}. In contrast, computing the best low-rank approximation of $\cY$ when $d \geq 3$ is NP-hard in general \citep{hillar2013most}, and the estimator $\tilde{\U}_k$ does not yield the desired accuracy. We thus introduce an indispensable refinement~\eqref{eq:low-dim-loadings}, because the additional projection of $\cY$ on the pre-estimated multilinear subspaces substantially reduces the noise in $\hat{\U}_k$. This step makes the proposed HSC method distinct from a simple extension of the classic spectral methods in the matrix setting.
\end{Remark}

\section{Statistical Theory}\label{sec:theory}

In this section, we study the statistical properties of the proposed algorithms.

\subsection{Assumptions} \label{sec:theory-assumption}

We first assume a non-degenerate condition on the separation among block means (i.e., core tensor) to ensure the identifiability for clustering:
\begin{equation}\label{eq:delta-define}
	\Delta_k^2 = \Delta_k^2(\cS) := \min_{i_1 \neq i_2} \left\|(\cM_k(\cS))_{i_1:} - (\cM_k(\cS))_{i_2:}\right\|_2^2 > 0.
\end{equation} 
In particular, we set $\Delta_k^2 = \infty$ if $r_k=1$. Roughly speaking, this condition means all mode-$k$ slices of the core tensor $\cS$ are distinct; otherwise the number of blocks, $(r_1,\ldots,r_d)$, should be reduced to smaller. Generally speaking, clustering is easier to achieve when the separation is larger or the noise level is smaller. Recall $\sigma$ is the sub-Gaussian norm of the noise distribution in Assumption \ref{asmp:sub-Gaussian-distribution}. Therefore, we define the signal-to-noise ratio (SNR) as 
\begin{equation}\label{eq:SNR}
    \text{SNR}:= \Delta_{\min}^2/\sigma^2,\quad \text{~where } \Delta_{\min}^2 := \min_{k\in [d]} \Delta_k^2.
\end{equation}

\begin{Remark}\label{rmk:singular-gap}
It is worth mentioning that an identifiable core in tensor block model may have degenerate ranks, i.e., $\rank(\cM_k(\cS)) < r_k$. This is significantly different from most literatures on low-rank tensor decomposition, where the singular value gap $\lambda_{r_k}(\cM_k(\cS))$ was assumed to be sufficiently large~\citep{richard2014statistical,zhang2018tensor,xia2020statistically}. For example, consider the following core tensor $\cS$ representing $2$-by-$2$-by-$2$ clusters:
\begin{equation*}
    \cS_{1::} = \begin{bmatrix}
         1 & -1 \\
         -1 & 1
    \end{bmatrix}, \qquad \cS_{2::} = \begin{bmatrix}
         -1 & 1 \\
         1 & -1
    \end{bmatrix}.
\end{equation*}
Note that the rank of $\cM_k(\cS)$ is 1, which is smaller than the number of clusters at mode-$k$. The above example has non-zero separation $\Delta_{\min}^2(\cS) = 16$, so the two clusters on each mode are still identifiable. 
\end{Remark}

Since permuting the cluster labels does not alter the clustering result (e.g., naming $\{1, 3\}$, $\{2, 4\}$ as Cluster I/II is equivalent to naming them as Cluster II/I), the cluster label vector $z_k$ on mode-$k$ is estimable only up to a permutation of cluster labels. Given the initialization label $z_k^{(0)}$, let $\pi_k^{(0)}\colon [r_k]\to [r_k]$ be the optimal permutation that minimizes the mismatches between $z_k^{(0)}$ and $z_k$, i.e.,
\begin{equation*}
    \pi_k^{(0)} := \argmin_{\pi \in \Pi_{r_k}}\frac{1}{p_k}\sum_{j=1}^{p_k} \bbI\left\{(z_{k}^{(0)})_j \neq (\pi \circ z_{k})_j\right\}, \quad \text{where}\ (\pi\circ z_k)_j := \pi((z_k)_j).
\end{equation*}
Let $t$ be the iteration index in HLloyd algorithm. We define $h_k^{(t)}$ as the mode-$k$ {\it misclassification rate} at the $t$th iteration of HLloyd algorithm:
\begin{equation}\label{eq:def-class-error}
	\begin{split}
		& h_k^{(t)} := \frac{1}{p_k}\sum_{j=1}^{p_k} \bbI\left\{(z_{k}^{(t)})_j \neq (\pi_k^{(0)} \circ z_{k})_j\right\}.	\end{split}
\end{equation}
We also impose the following ``balanced cluster size'' assumption for technical convenience. Such an assumption is widely used in the literature of mixture model clustering~\citep{loffler2019optimality,gao2019iterative,wu2020optimal}.
\begin{Assumption}\label{asmp:balance-size}
There exists universal positive constants $0<\alpha<1<\beta$ such that
\begin{equation}\label{ineq:asmp-balance}
	\alpha p_k/r_k \leq \left|\{j \in [p_k]: (z_k)_j = a\}\right| \leq \beta p_k/r_k,\qquad \forall a \in [r_k],~ k\in [d],
\end{equation}
where $|\cdot|$ is the cardinality of a given set. 
\end{Assumption}

\subsection{Algorithmic Theoretical Guarantees}\label{sec:HSC-theory}
Now we are in a position to establish the theoretical guarantees for High-order Lloyd and High-order spectral clustering. In order to prove our main theoretical result in Theorem \ref{coro:exact-recovery}, we introduce the following more convenient measure of misclassification loss in addition to the classification error rate $h_k^{(t)}$:
\begin{equation}\label{eq:loss}
	\begin{split}
		l_k^{(t)}:= \frac{1}{p_k}\sum_{j=1}^{p_k} \left\|\left(\cM_k(\cS)\right)_{(z_k^{(t)})_j:} - \left(\cM_k(\cS)\right)_{(\pi_k^{(0)} \circ z_k)_j :} \right\|_2^2.
	\end{split}
\end{equation}
The following lemma establishes a relationship between $h_k^{(t)}$ and $l_k^{(t)}$, which implies that it suffices to bound $l_k^{(t)}$ in order to develop the target upper bound for $h_k^{(t)}$.
\begin{Lemma}\label{lm:loss-relation}
    Define $h_k^{(t)}$ and $l_k^{(t)}$ as in \eqref{eq:def-class-error} and~\eqref{eq:loss}. Then,
    $h_k^{(t)} \leq l_k^{(t)}/\Delta_k^2$.
\end{Lemma}
Our first result is the local convergence of High-order Lloyd algorithm.
\begin{Theorem}[Local convergence of HLloyd]\label{thm:HLloyd}
    Suppose Assumptions \ref{asmp:sub-Gaussian-distribution} and \ref{asmp:balance-size} hold. Let $\{z_k^{(0)}\}_{k=1}^d$ be the initialization of HLloyd algorithm and $\{z_k^{(t)}\}_{k=1}^d$ be the estimations at step $t$. Assume the initialization satisfies
    \begin{equation}\label{ineq:HLloyd-initia-cond}
        l_k^{(0)} \leq c\Delta_{\min}^2/r_k, 
    \end{equation}
   and the SNR satisfies
	\begin{equation}\label{ineq:SNR-HLloyd}
	    \Delta_{\min}^2/\sigma^2 \geq \frac{C\bar p   r_*^2 \bar r \log \bar p}{p_*}.
	\end{equation}
	Then with probability at least $1-\exp(-c\underline p)-\exp\left(-\frac{c p_*}{r_{*}\bar p}\frac{\Delta_{\min}^2}{\sigma^2}\right)$, for all $t \geq 0$,
	\begin{equation}\label{ineq:step-t-lk-bound}
		l_k^{(t)} \leq C\sigma^2\exp\left(-\frac{cp_*\Delta_{\min}^2}{r_* \bar p\sigma^2}\right) + \frac{\Delta_{\min}^2}{2^t},\qquad \forall k\in [d].
	\end{equation}
\end{Theorem}
\begin{Remark}
	\label{rmk:compare to single structure}
Recently, \cite{gao2019iterative} provided the convergence analysis of iterative algorithms for the inference problems with a single discrete structure, including the Lloyd algorithm for Gaussian mixture model as a special instance. Their techniques do not apply to our problem as the tensor block model admits multiple discrete structures (i.e., clusters in each mode). One specific technical challenge for proving Theorem \ref{thm:HLloyd} is to characterize the impact to misclassification rate in mode-$k$ by those in the other $d-1$ modes. See more discussions in the proof sketch in Section \ref{sec:proof-sketch}. 
\end{Remark}

The inequality \eqref{ineq:step-t-lk-bound} and Lemma \ref{lm:loss-relation} together imply that when $T \geq 2\lceil \log \bar p \rceil$,
\begin{equation} \label{eq:classification-error-decreasing}
    h_k^{(T)} \leq C\cdot\SNR^{-1}\exp\left(-\frac{cp_*\SNR}{r_* \bar{p} }\right) + 2^{-T} < \frac{1}{p_k}.
\end{equation}
Therefore, HLloyd algorithm guarantees exact recovery of clustering under the SNR condition \eqref{ineq:SNR-HLloyd}.

We also need a good initialization satisfying \eqref{ineq:HLloyd-initia-cond} in order to apply Theorem \ref{thm:HLloyd}. Our initialization accuracy is guaranteed by the proposed high-order spectral clustering (HSC, Algorithm \ref{alg:HO-SC}).
\begin{Theorem}[Upper bound on misclassification rate of HSC]\label{thm:HO-SC}
	Suppose Assumptions \ref{asmp:sub-Gaussian-distribution} and \ref{asmp:balance-size} hold and each entry of $\cE$ has equal variance. If the SNR satisfies
	\begin{equation}\label{ineq:SNR-HOSC}
		\Delta_{\min}^2/\sigma^2 \geq C M\left(\bar pr_*^2\bar r\log \bar p/p_* + r_*\bar r/p_*^{1/2}\right),
	\end{equation}
	then, with probability at least $1-C\exp(-c\underline p)$, $\forall k \in [d]$,
    \begin{equation*}
        \begin{split}
            l_k^{(0)} & \leq C M\cdot \sigma^2 \cdot (r_{-k}/p_*)\left(r_* + \bar p \bar r^2 + p_*^{1/2}\bar r\right) \leq c\Delta_{\min}^2/r_k,\\
            h_k^{(0)} & \leq C M \cdot (r_{-k}/p_*)\left(r_* + \bar p \bar r^2 + p_*^{1/2}\bar r\right) \sigma^2/\Delta_{\min}^2.
        \end{split}
    \end{equation*}
\end{Theorem}

The combination of Theorems \ref{thm:HLloyd} and \ref{thm:HO-SC} yields the following main result of our paper.
\begin{Theorem}\label{coro:exact-recovery}
	Denote $\{z_k^{(t)}\}_{k=1}^d$ as the membership vectors in the iteration $t$ of HLloyd algorithm with $\{z_k^{(0)}\}_{k=1}^d$ being the output of HSC. Under the same conditions of Theorem \ref{thm:HO-SC}, with probability at least $1-\exp(-c\underline p)-\exp\left(-\frac{c p_*}{4r_{*}\bar p}\frac{\Delta_{\min}^2}{\sigma^2}\right)$, we have exact recovery of $\{z_k\}_{k=1}^d$ when $T \geq 2\lceil \log \bar p \rceil$; i.e., there exist a set of permutations $\{\pi_k\}_{k=1}^d$ such that
	\begin{equation*}
		z_k^{(T)} = \pi_k \circ  z_k,\qquad \forall k=1,\ldots,d.
	\end{equation*}
	In particular, if the numbers of clusters $r_k$ are fixed, $T \geq \lceil 2\log p\rceil$, $p_1 \asymp \cdots \asymp p_d \asymp p$, and the SNR satisfies
\begin{equation}\label{ineq:SNR-equal-p}
    \Delta_{\min}^2/\sigma^2 \geq C\left( p^{-d/2} \vee p^{-(d-1)}\log p\right),
\end{equation}
then $z_k^{(T)}$ achieves exact clustering with high probability.
\end{Theorem}
In comparison, if one only applies the proposed HSC algorithm without HLloyd refinement, the same condition \eqref{ineq:SNR-equal-p} would only guarantee consistent clustering according to Theorem \ref{thm:HO-SC}.

\begin{Remark}[Comparison with theory for matrix spectral method]\label{eq:rm5}
A key intermediate step for the proof of Theorem \ref{thm:HO-SC} is to evaluate the estimation error for the projected tensor observation: 
\begin{equation}\label{eq:Y-tilde}
    \|\tilde \cY - \cX\|_\tF^2,\qquad \text{where~} \quad \tilde\cY := \cY \times_1 \hat\U_1\hat\U_1^\top \times \cdots \times_d \hat\U_d\hat \U_d^\top \quad \text{ for } \hat\U_k \text{ in } \eqref{eq:HOOI-2}.
\end{equation}
In matrix case ($d=2$), the estimation error bound can be simply derived by the algebraic property of SVD. However, for high-order tensors ($d \geq 3$), such an error bound was usually established under strong singular value gap condition \citep{zhang2018tensor} that does not hold in tensor block model. To overcome this issue, we develop the following new \emph{singular-value-gap-free} bound on tensor estimation for $\tilde \cY$. 
\begin{Proposition}\label{prop:HOOI-singular-free}
    \emph{(A singular-value-gap-free tensor estimation error bound).} Suppose $\cY = \cX + \cE \in \bbR^{p_1\times\cdots \times p_d}$, $\cX$ has Tucker-rank $(r_1,\ldots,r_d)$, and $\cE$ satisfies Assumption \ref{asmp:sub-Gaussian-distribution} with equal variance $\Var(\cE_{j_1,\ldots,j_d}) = \sigma^2$, $\forall (j_1,\ldots,j_d)\in[p_1]\times \cdots \times[p_d]$. Let $\tilde \cY$ be defined according to \eqref{eq:Y-tilde}.  
    Then, with probability at least $1-C\exp(-c\underline p)$,
    \begin{equation*}
        \left\|\tilde \cY - \cX\right\|_{\tF}^2 \leq  C\sigma^2\left(p_*^{1/2}\bar r+\bar p\bar r^2 + r_*\right).
    \end{equation*} 
\end{Proposition}
The proof sketch of Theorem \ref{thm:HO-SC} and Proposition \ref{prop:HOOI-singular-free} are presented in Section \ref{sec:proof-sketch}.  
\end{Remark}

\begin{Remark}\emph{(Heteroskedastic errors).}
  The guarantee for local convergence in Theorem \ref{thm:HLloyd} applies to the general sub-Gaussian noise under Assumption \ref{asmp:sub-Gaussian-distribution}, which covers both GTBMs and STBMs. Our theoretical guarantee for initialization (Theorem \ref{thm:HO-SC}) is established under an additional homoscedasticity condition, which excludes some interesting STBMs with Bernoulli distributions. Nevertheless, our simulation study in Section \ref{sec: simulation} illustrates that our algorithm still performs well in practice for a wide range of STBMs. The theoretical error bound for heteroskedastic high-order spectral clustering will be left as future work.
\end{Remark}

In addition to recovering cluster memberships in the tensor block model, another important task is to recover the block means $\cS$, or equivalently, to denoise the observed tensor $\cY$ and obtain an estimate of $\cX$. Given the estimates of labels $ z_1^{(T)}, \ldots, z_d^{(T)}$, a natural estimator for $\cX$ is the aggregated mean in each estimated block, i.e.,
\begin{equation*}
    \hat \cX_{j_1,\ldots,j_d} = \text{Average}\left(\left\{\cY_{j_1',\ldots,j_d'}: ( z_k^{(T)})_{j_k'} = ( z_k^{(T)})_{j_k}, j_k' \in [p_k], \forall k\in[d]\right\} \right).
\end{equation*}
We have the following guarantee for $\hat\cX$. 
\begin{Theorem}[Upper bound of estimation error for HLloyd + HSC]\label{thm:tensor-est-hp}
	Under the same conditions of Theorem \ref{thm:HO-SC}, we have with probability at least $1-\exp(-cr_*)-\exp(-c\underline p) - \exp\left(-\frac{c p_*}{4r_{*}\bar p}\frac{\Delta_{\min}^2}{\sigma^2}\right)$ that
	\begin{equation}\label{ineq:loss-X-hp}
		\left\|\hat\cX - \cX\right\|_\tF^2 \leq C \sigma^2r_*.
	\end{equation}
\end{Theorem}

\begin{Remark}[Comparison with HOOI] \label{rmk: dimension free tensor est}
Recall the tensor block model \eqref{eq:model-tensor} naturally admits a Tucker low-rank structure. A tempting strategy is to apply high-order orthogonal iteration (HOOI) on $\cY$ to estimate $\cX$. By Theorem 1 of \cite{zhang2018tensor}, the HOOI estimator $\hat \cX_{HOOI}$ achieves the following statistical rate under a strong singular gap condition: 
\begin{equation}\label{ineq:HOOI-rate}
    \bbE \left\|\hat \cX_{HOOI} - \cX\right\|_\tF^2 \leq C\sigma^2\left(r_* +\sum_{k=1}^{d} p_k r_k\right).
\end{equation}
Compared to Theorem \ref{thm:tensor-est-hp}, \eqref{ineq:HOOI-rate} has an additional dimension-dependent term. The intuition behind this phenomenon is that HOOI only fits a low-rank model while fails to capture the discrete structure in the tensor block model. This theoretical result is also supported by our numeric experiments in Section \ref{sec:num-properties}.
\end{Remark}

\section{Statistical and Computational Trade-offs}\label{sec:comp-limits}

In this section, we study the statistical and computational limits of high-order clustering in the tensor block model \eqref{eq:model-tensor}. We assume the distribution of $\cE_{j_1,\ldots,j_d}$ is i.i.d. Gaussian.

We specifically focus on the following parameter space,
\begin{equation}\label{eq:parameter_class}
    \begin{split}
   	& \Theta\left(\{\Delta_k\}_{k=1}^d, \alpha, \beta\right) = \\
   	& \left\{\Big(\cS, z_1,\ldots,z_d\Big)\colon \begin{array}{l}
    \cS \in \bbR^{r_1\times \cdots \times r_d}, \ \ \Delta_k(\cS) \geq \Delta_k,\ \  z_k \in [r_k]^{p_k} \\ 
    \alpha \frac{p_k}{r_k} \leq  \left|\{j \in [p_k]\colon (z_k)_j = a\}\right|\ \leq \beta \frac{p_k}{r_k},\ \forall a \in [r_k],k \in [d]
    \end{array}\right\}.
	\end{split}
\end{equation}
Here, the constraints in the parameter space \eqref{eq:parameter_class} correspond to assumptions in Section \ref{sec:theory-assumption}. We further introduce the following parameter regime
\begin{equation} \tag{A1} \label{eq:comp limit region}
	\Delta^2_{\min}/\sigma^2 = p^\gamma \quad \text{ and } \quad p_1 \asymp \cdots \asymp p_d \asymp p.
\end{equation}

\subsection{Statistical Limit}\label{sec:statistical-limit}

The following theorem establishes the SNR lower bound for exact label recovery, which reveals the statistical limit of high-order clustering in the tensor block model. 
\begin{Theorem}[Statistical lower bound]\label{thm:mcr-lower-bound}
Consider the tensor block model~\eqref{eq:model-tensor}. Suppose $r_k = o(p_k^{1/3})$, and $\frac{\Delta_k^2}{\sigma^2}\frac{p_{-k}}{r_{-k}} < c_0$ for some constant $c_0>0$.
Then, for any estimator $\hat z_k$,
\begin{equation} \label{ineq:stat limits}
		\sup_{(\cS, z_1,\ldots, z_d)\in \Theta} \bbE \min_{\pi_k \in \Pi_{r_k}}\sum_{j=1}^{p_k} \bbI\{ (\hat z_k)_j \neq (\pi_k \circ z_k)_j\} \geq 1.
\end{equation}
\end{Theorem}
Theorem \ref{thm:mcr-lower-bound} suggests the impossibility of exact clustering under the parameter regime \eqref{eq:comp limit region}, when $r_k$'s are constants and $\gamma < -(d-1)$. Furthermore, it was shown that the MLE \eqref{eq:MLE}, while being computationally intractable, achieves consistent clustering given $\gamma >-(d-1)$~\citep{wang2019multiway}. Therefore, $\gamma_{\stat} := -(d-1)$ serves as the statistical limit for the clustering in tensor block model: when $\gamma > \gamma_{\stat}$, there exists an algorithm that can successfully recover the clustering labels; when $\gamma < \gamma_{\stat}$, all algorithms, regardless the computational complexity, fail to do so.

In comparison, according to the discussion in Section \ref{sec:HSC-theory}, the combination of HSC and HLloyd algorithms achieves exact clustering when $\gamma > -d/2 =: \gamma_{\comp}$, which seems more stringent than the statistical limit $\gamma_{\stat}$ for $d\geq 3$. However, we should point out that, unlike MLE, our algorithm (HSC and HLloyd) is polynomial-time implementable. In the next section, we show that $\gamma > \gamma_{\comp}$ is indeed necessary for any polynomial-time algorithm to succeed. 

\subsection{Computational Limit}\label{sec:computational-limit}

In this section, we establish the computational limit for high-order clustering under model \eqref{eq:model-tensor}. We first introduce the hypergraphic planted clique (HPC) detection problem and its hardness conjecture, which are building blocks for the main results on computational limit. 

A $d$-hypergraph can be seen as an order-$d$ extension of regular graph. In a $d$-hypergraph $G = (V(G), E(G))$, each hyperedge $e \in E$ includes a set of $d$ different vertices in $V$. Define $\mathcal{G}_d(N, 1/2)$ as the Erd\H{o}s-R{\'e}nyi $d$-hypergraph with $N$ vertices, where each hyperedge $(i_1, \ldots, i_d)$ is independently included in $E$ with probability $1/2$. Also we define $\mathcal{G}_d(N, 1/2, \kappa)$ as the hypergraphic planted clique (HPC) model with the clique size $\kappa$. To generate $G \sim \mathcal{G}_d(N, 1/2, \kappa)$, we sample a random hypergraph from $\mathcal{G}_d(N, 1/2)$, pick $\kappa$ vertices uniformly at random from $[N]$, denote them as $K$, and connect all hyperedges $e$ if all vertices of $e$ are in $K$. The HPC detection can be formulated as the following hypothesis testing problem:
\begin{equation}\label{eq: HPC detection problem}
H_0: G \sim \mathcal{G}_d \left(N, 1/2\right)\quad \text{v.s.} \quad H_1: G \sim \mathcal{G}_d \left(N, 1/2, \kappa \right).
\end{equation}
We consider the following version of HPC detection conjecture, which was introduced and studied in the literature \citep{zhang2018tensor,brennan2020reducibility}. 
\begin{Conjecture}[HPC detection conjecture]\label{conj: hardness of tensor clique detection}
Suppose $d\geq 2$ is a fixed integer. Suppose
$$\limsup_{N \to \infty} \log \kappa / \log \sqrt{N} \leq 1 - \epsilon \quad \text{for any }\epsilon > 0.$$
Then, for any sequence of polynomial-time tests $\{\phi \}_N: G \to \{0,1\}$, $\liminf_{N \to \infty} \bbP_{H_0} \left(\phi(G) = 1\right) + \bbP_{H_1} \left(\phi(G) = 0\right) > 1/2$. 
\end{Conjecture}
\cite{zhang2018tensor} observed that spectral method solves HPC detection efficiently if $\kappa = \Omega(\sqrt{N})$ but fails when $\kappa = N^{1/2 - \epsilon}$ for any $\epsilon>0$. Recently, it has also been shown that many classes of powerful algorithms, including metropolis algorithms and low-degree polynomial algorithms, fail to solve the HPC detection problem in polynomial time under the conjectured hard regime \citep{luo2020tensor,brennan2020reducibility}. Several open questions on HPC detection---in particular, whether HPC detection is equivalently hard as PC detection---are discussed in \cite{luo2020open}. 

With the HPC detection hardness conjecture, we have the following computational lower bound for high-order clustering in the tensor block model.
\begin{Theorem}[Computational Lower Bound]\label{thm:computational-lower-bound}
	Consider the tensor block model \eqref{eq:model-tensor} under the parameter regime \eqref{eq:comp limit region} and Conjecture \ref{conj: hardness of tensor clique detection}. If $\gamma < -d/2 =:\gamma_{\comp}$, then for any polynomial-time estimator $(\hat{z}_1, \ldots, \hat{z}_d)$, we have
	\begin{equation*}
	\liminf_{p \to \infty} \sup_{(\cS, z_1,\ldots, z_d)\in \Theta}  \bbP\left(\exists k \in [d] \text{ s.t. } \min_{\pi \in \Pi_{r_k}} \sum_{j=1}^{p_k} \bbI\left\{(\hat{z}_{k})_j \neq (\pi \circ z_{k})_j \right\} \geq 1  \right) \geq 1/2.
	\end{equation*}
\end{Theorem}
Combining Theorems \ref{thm:HLloyd}, \ref{thm:mcr-lower-bound}, and \ref{thm:computational-lower-bound}, we have finished the proof for the informal statement of Theorem \ref{thm:informal} and established the phase transition diagram of Figure \ref{fig:phase-transition} in the introduction section. These results render the whole picture of the statistical and computational limits of the high-order clustering in the tensor block model.

\section{Numerical Studies} \label{sec: simulation}

In this section, we first study the performance of the proposed high-order Lloyd (HLloyd) and high-order spectral clustering (HSC) algorithms. Then we compare the proposed algorithms with other state-of-the-art methods. Unless otherwise noted, we consider order-$d$ tensor block models with $p_1 = \cdots = p_d = p$, $r_1 = \cdots = r_d = r$ and balanced cluster sizes across all modes throughout the simulations. In each experiment, we report the averaged statistics and the standard error across 100 replications.

\subsection{Properties of High-order Lloyd and High-order Spectral Clustering} \label{sec:num-properties}

We first study the clustering performance of the proposed HLloyd and HSC algorithms using clustering error rate (CER). The CER is calculated using the disagreement between estimated and true partitions, i.e.,\ one minus adjusted random index~\citep{milligan1986study}. A lower CER implies a better clustering and CER = 0 means exact clustering. In the first four experiments, we consider the Gaussian tensor block models with variance $\sigma^2 = 1$, and in the fifth experiment, we study the performance of the proposed algorithms on stochastic tensor block models.

\vskip.2cm

\noindent{\bf Statistical and Computational Phase Transition.} The first experiment investigates the phase transition of high-order clustering with respect to SNR. We perform the proposed polynomial-time HLloyd algorithm with HSC initialization on both matrix and order-3 tensor block models. We set $r=5$, $p\in\{200,400\}$ for the matrix case, and $p\in\{80,100\}$ for the order-3 tensor case. We assess the SNR phases by plotting CER as a function of the signal strength $\gamma$ for which $\Delta^2_{\min}=\cO(p^{\gamma})$. The performance of our estimator is compared to MLE \eqref{eq:MLE}, i.e., the global optimum. Since MLE is NP-hard to compute, we approximate the global optimum using oracle initialization refined by high-order Lloyd and call the estimate the ``oracle estimate''. The oracle initialization is specified as the perturbed ground truth contaminated by 20\% random labeling. 

\begin{figure}[htbp]
	\centering
	\subfigure[Matrix Clustering]{\includegraphics[height = 1.7in]{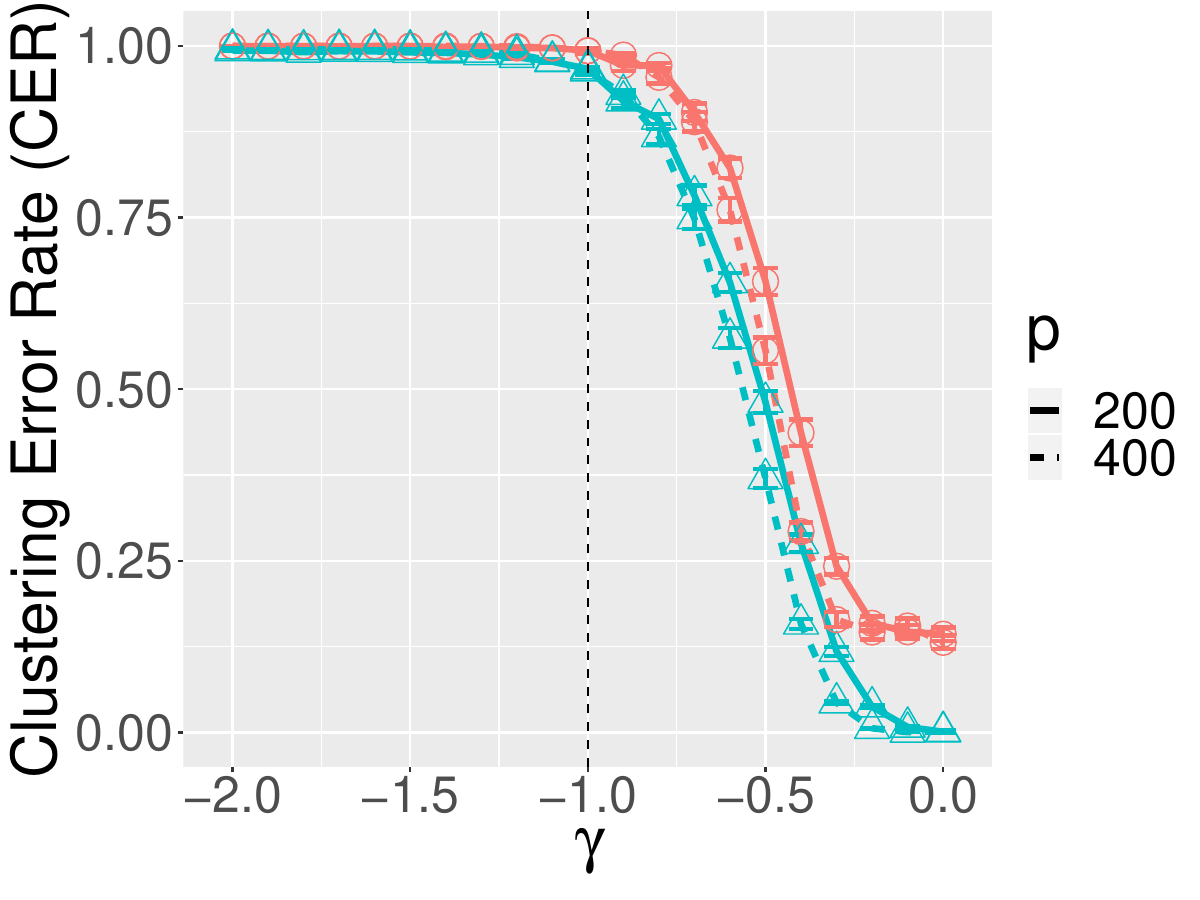}}
	\subfigure[Tensor Clustering]{\includegraphics[height = 1.7in]{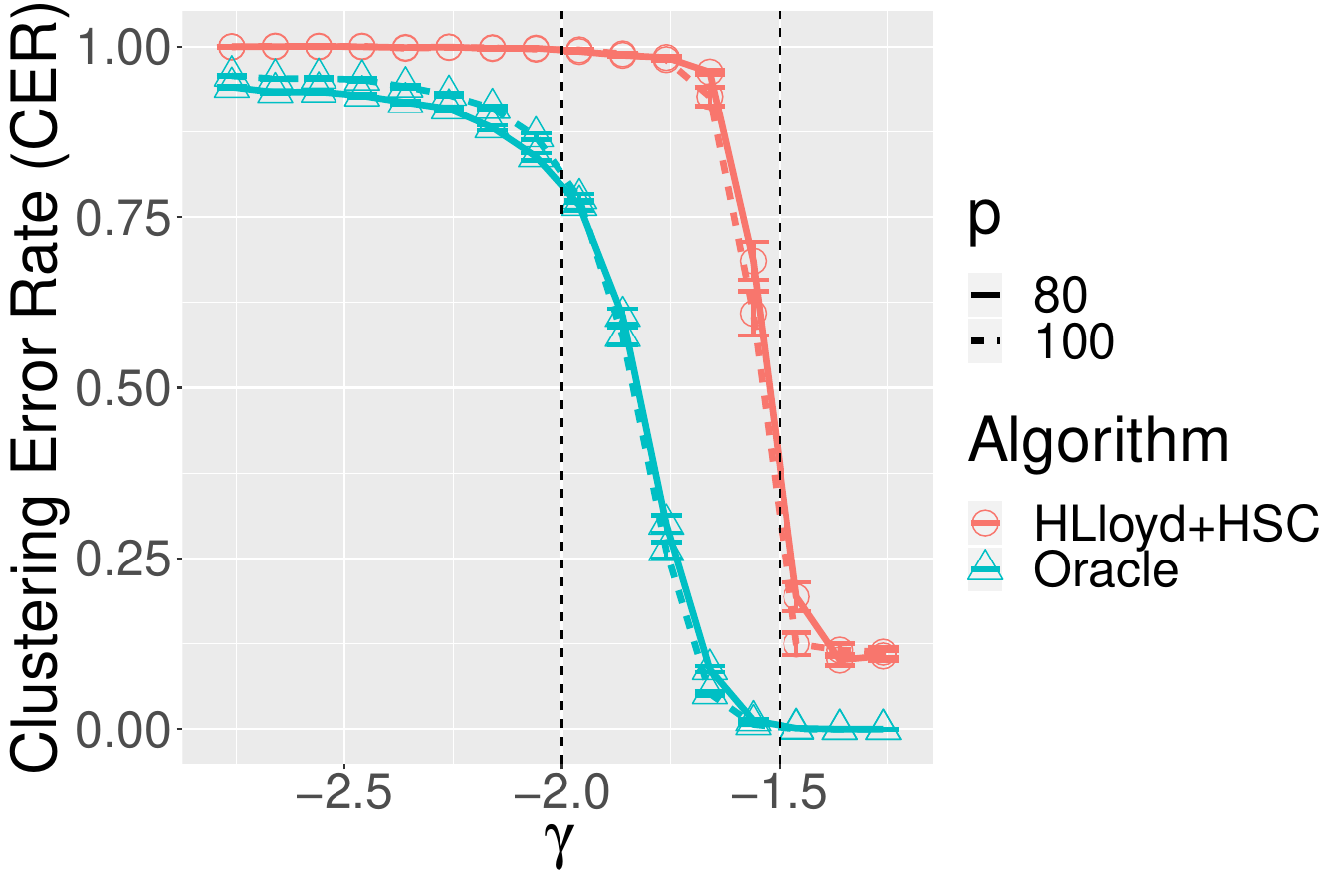}}
	\caption{SNR phase transition for clustering in tensor block model under the setting $r = 5, \Delta_{\min}^2 = \cO(p^{\gamma})$. The error bar on each point represents the standard error. (a) Matrix clustering $p = 200,400$, $\gamma \in (-2,0)$; (b) Tensor Clustering $d=3, p =80,100$, $\gamma \in (-2.8, -1.2)$.} \label{fig: phase transition}
\end{figure}

Figure~\ref{fig: phase transition} shows the SNR phases in the clustering problem for matrices and for order-3 tensors. We find that, in the matrix case, our method and oracle estimates undergo similar SNR phase transitions. A flat, high clustering error is observed for both estimates when $\gamma < -1$, and the error immediately decreases as $\gamma > -1$. Note that our theory in Section~\ref{sec:HSC-theory} has implied an optimal SNR $\cO(p^{-1})$ for order-2 cases, and this critical ratio is indeed achieved by the proposed algorithm. In contrast to the matrix case, the order-3 tensor clustering reveals a striking gap between HLloyd and oracle estimates. In particular, the phase transition occurs around $\gamma = -2$ for oracle estimates, whereas $\gamma = -1.5$ for HLloyd estimates. This gap reflects the statistical-computational gap $p^{-(d-1)}\ll \text{SNR}\ll p^{-d/2}$ for clustering in tensor block model when $d\geq 3$, which corroborates our theoretical results in Section \ref{sec:comp-limits}. 

\vskip.2cm

\begin{figure}[h!]
	\centering
	\includegraphics[height = 2in]{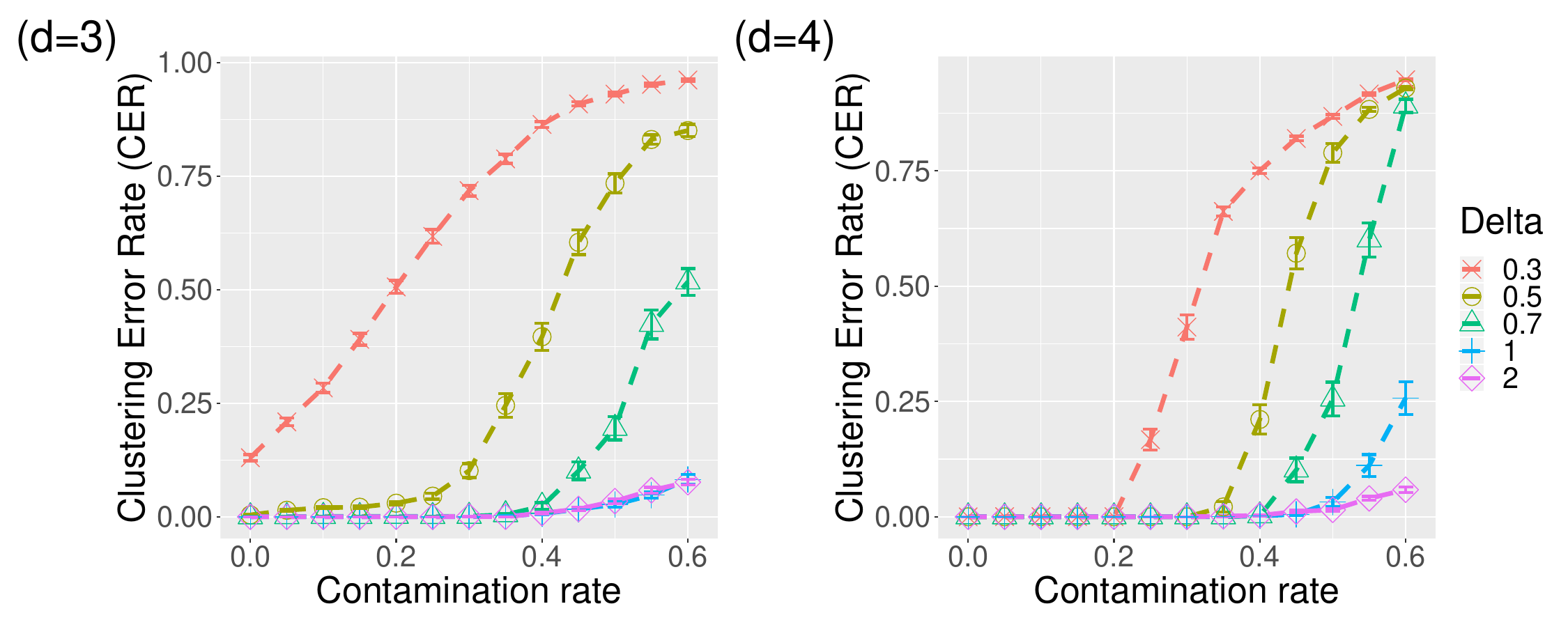}
		\caption{Impact of initialization to the performance of HLloyd algorithm. The CER and the standard error are plotted against the contamination rate from $0$ to $0.6$ under the settings $\Delta_{\min} \in \{ 0.3,0.5,0.7, 1,2\}$, $p = 50, r = 5$.
	}\label{fig: init impact lloyd} 
	\vspace{-.3cm}
\end{figure}

\noindent{\bf Impact of Initialization to HLloyd}. The second simulation examines the impact of initialization to the performance of the high-order Lloyd algorithm. We consider CER for $p = 50$, $r = 5$, $d \in \{3,4\}$, and $\Delta_{\min} \in \{0.3,0.5,0.7,1,2\}$. Given a contamination rate $\varepsilon$, we generate initialization labels by randomly shuffling $100\varepsilon$\% labels of the ground truth and then run HLloyd. Figure \ref{fig: init impact lloyd} shows that the clustering error decreases as the signal strength $\Delta_{\min}$ increases or the contamination rate decreases in both $d = 3$ and $4$ settings. This shows that stronger signal and better initialization enhance the clustering performance of the HLloyd algorithm.  The experiment also indicates that a proper initialization is crucial for the success of HLloyd. 
We will show next that the proposed HSC algorithm can achieve a proper initialization.

\vskip.2cm

\begin{figure}[h!]
	\centering
	\includegraphics[width=\linewidth]{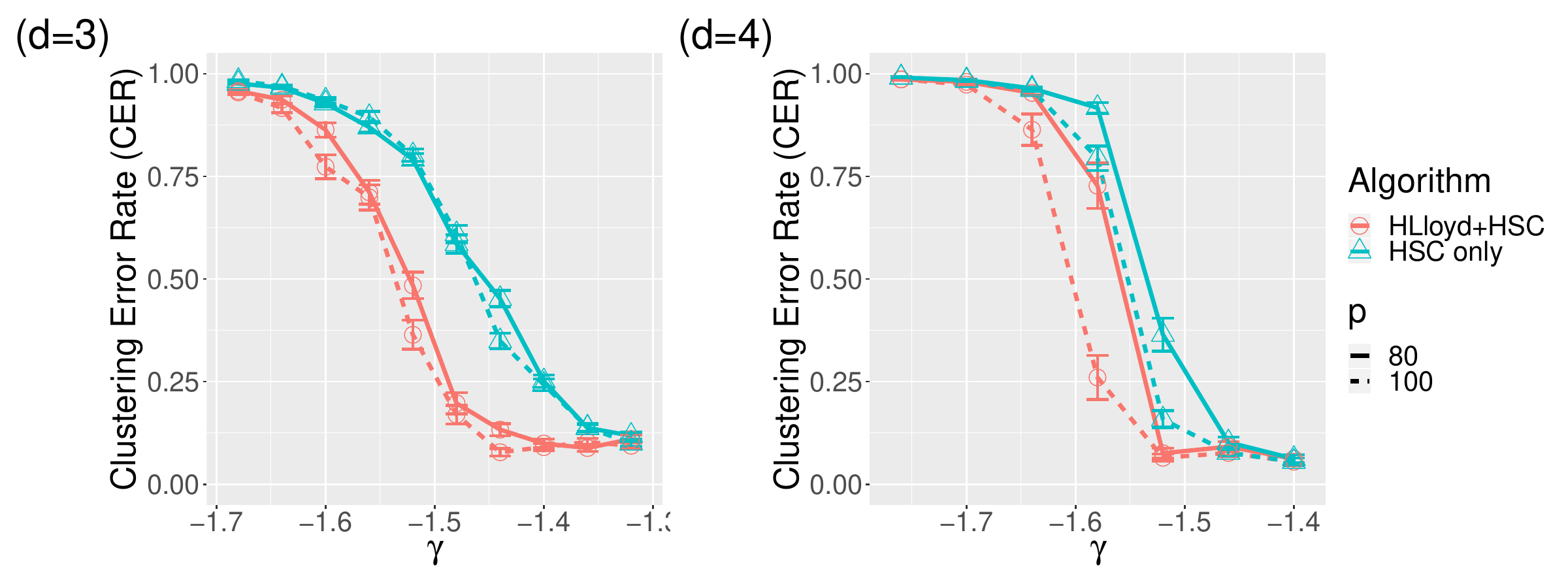}
	\caption{Clustering of high-order spectral clustering and spectral initialization with high-order Lloyd refinement, for $r = 5$, $p \in \{80,100\}$, and varying $\gamma$. 
	}\label{fig: lloyd} 
	\vspace{-.3cm}
\end{figure}

\noindent{\bf Clustering via HLloyd + HSC algorithms}. In the third simulation, we assess the clustering accuracy for the proposed approaches: HSC-only algorithm and the combined algorithm (HLloyd + HSC). We consider the settings $r = 5, p \in \{80,100\}$, and $\Delta_{\min}^2 = \cO(p^{-\gamma})$ for a range of $\gamma$. Figure \ref{fig: lloyd} shows that both approaches achieve nearly exact recovery as the signal level increases in the tensor block model. For a wide range of settings, given the initialization by HSC, HLloyd can greatly improves the HSC-only algorithm and achieves more accurate clustering. This confirms our theoretical results that HLloyd algorithm effectively boosts the clustering performance of HSC. 

\vskip.2cm

\begin{figure}[htb]
	\centering
	\subfigure[Tensor Estimation Error]{\includegraphics[height = 1.7in]{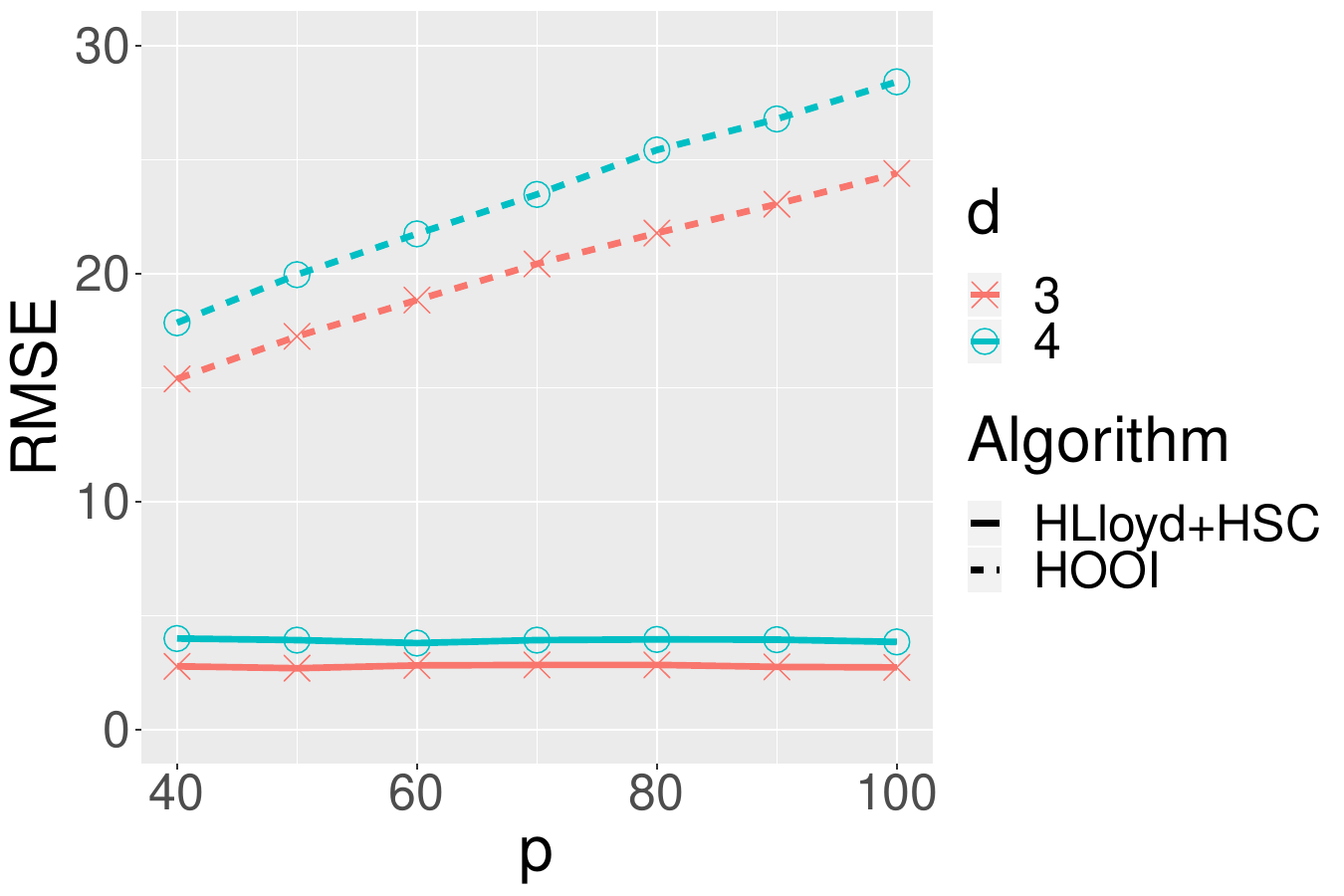}} 
	\subfigure[Clustering in STBM]{\includegraphics[height = 1.7in]{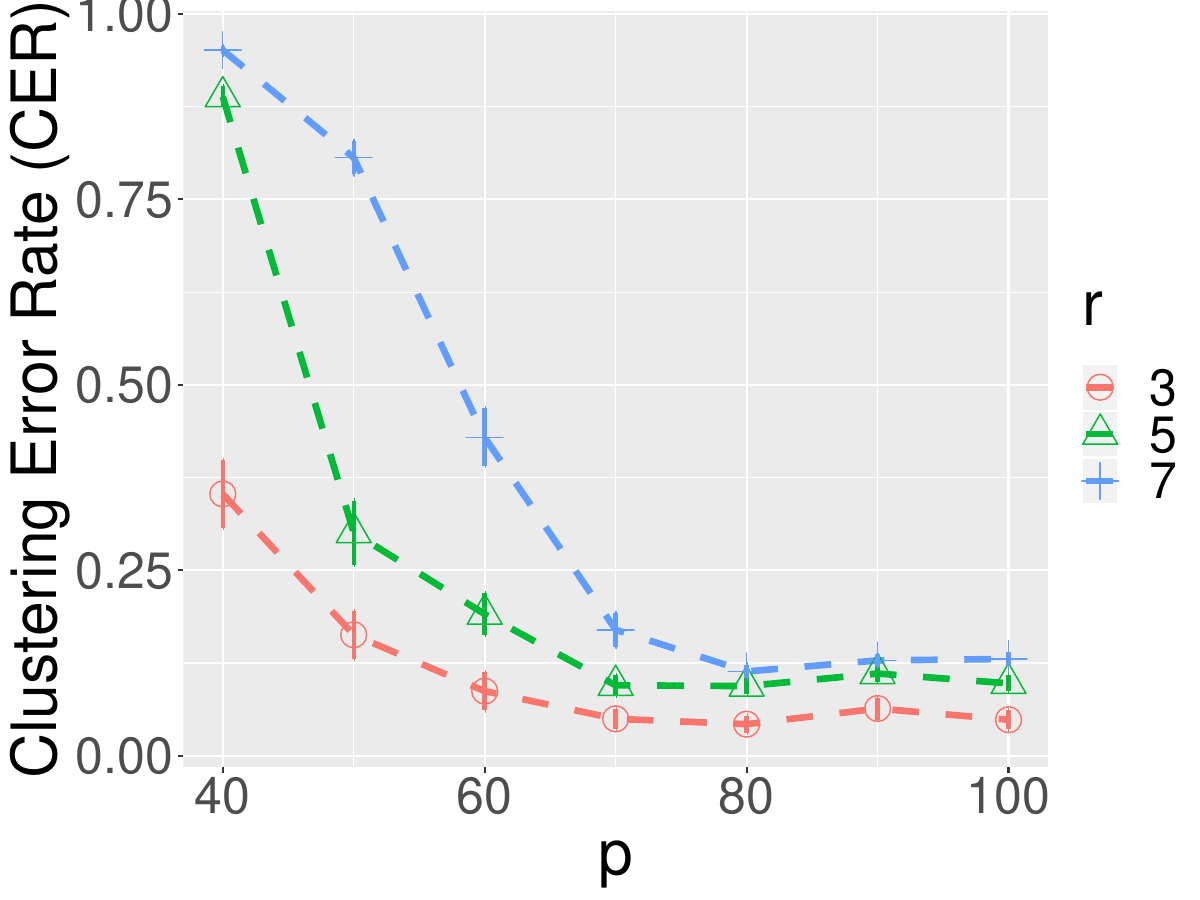}}
	\caption{(a) Comparison of tensor estimation error of HLloyd + HSC and HOOI for $r =2$, $p \in \{40,50,\ldots,100\}$; (b) Clustering of high-order Lloyd with spectral initialization for stochastic tensor block model with $d = 3$, $r \in\{3,5,7\}$, and $p \in \{40,50,\ldots,100\}$.}\label{fig: tensor estimation} 
\end{figure}

\noindent{\bf Tensor estimation via HLloyd + HSC algorithms}. We use the following experiment to compare the tensor estimation errors of HLloyd + HSC and high-order orthogonal iteration (HOOI). Figure \ref{fig: tensor estimation}(a) shows the root mean squared error (RMSE) $\|\hat{\cX} - \cX\|_\tF$ for $p \in \{40,50,\ldots,100\}$, $r =2$, $d \in \{3,4\}$, and $\Delta_{\min} = 2$. As $p$ increases, the tensor estimation error of HLloyd + HSC is almost flat over the range of $p$. This matches our theoretical results in Theorem \ref{thm:tensor-est-hp} that the tensor estimation bound of the proposed algorithm is free of dimension $p$. In contrast, the estimation error of HOOI grows almost linearly with respect to $p$. This demonstrates the benefit of HLLoyd + HSC over previously tensor algorithm HOOI on tensor estimation in tensor block model. 

\vskip.2cm

\noindent{\bf Simulations on Stochastic Tensor Block Models.} We also assess our algorithm on stochastic tensor block models. 
Let $r \in \{3,5,7\}$, $p \in \{40,50,\ldots, 100 \}$, $d=3$. The memberships $z_1,\ldots, z_d$ are generated in the same way as before. Each entry in the core tensor $\cS$, which encodes the connection probability in each cluster, is generated uniformly at random from $[0,0.1]$. Then we generate data tensor $\cY$ with independent Bernoulli entries $\cY_{(z_1)_{j_1},\ldots,(z_d)_{j_d}} \sim \text{Bernoulli}(\cS_{(z_1)_{j_1},\ldots,(z_d)_{j_d}})$. We consider low-probability because the binary tensors in practice often have low average connection probability (see estimated connection probability in real data in the forthcoming Figures \ref{fig: airline heatmap} and \ref{fig: click-through}). The simulation results are provided in Figure \ref{fig: tensor estimation}(b). We can see as $p$ increases, the clustering error of the proposed algorithm decreases as expected. The results validate our main results in Section \ref{sec:theory}. In particular, our algorithm performs well under a general class of sub-Gaussian noises in tensor block model.

\subsection{Comparison with Other Algorithms} \label{sec:num-comparison}
In this section, we compare HLloyd + HSC with two other classic tensor clustering algorithms: \begin{enumerate}[label=(\alph*)]
\item HOSVD + Kmeans: apply high-order SVD (HOSVD) on $\cY$ \citep{de2000multilinear,ghoshdastidar2015spectral}, then perform $k$-means on the outcome factors of HOSVD; 
\item CP + Kmeans: apply CANDECOMP/PARAFAC (CP) decomposition on $\cY$ \citep{carroll1970analysis}, then perform $k$-means on the outcome factors of CP decomposition.
\end{enumerate}
In the first three simulations, we compare these algorithms under the Gaussian tensor block model; in the fourth simulation, we compare these algorithms under the stochastic tensor block model.

\begin{figure}[htb!]
	\centering
	\includegraphics[width=\linewidth]{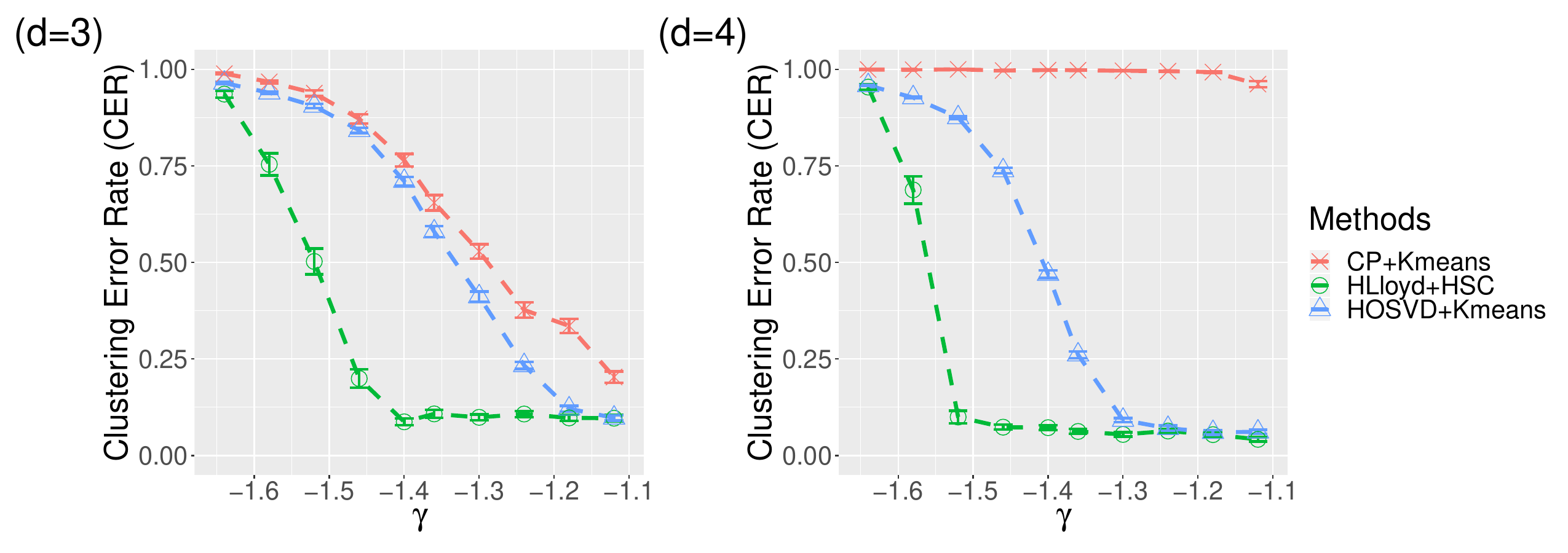}
	\caption{Comparison of HLloyd + HSC with HOSVD-/CP-decomposition-based clustering methods for $p = 80, r= 5$ and varying $\gamma$.}\label{fig: TC compare} 
	\vspace{-.3cm}
\end{figure}

In the first simulation, we let $r = 5, p = 80$ and $\Delta^2_{\min}\asymp p^{\gamma}$ with varying $\gamma$. The comparison results in Figure \ref{fig: TC compare} show that HLloyd greatly improves the other two methods for most of $\gamma$ considered. When $d = 4$ and $\gamma$ is large, the HOSVD-based clustering method has a similar performance to HLloyd, and both of them achieve better accuracy than the CP-decomposition-based algorithm.  However, the performance of CP based algorithm becomes even worse as $d$ increases, which is because the CP decomposition can hardly capture dense core tensor structure in the tensor block model.

\begin{figure}[htb!]
	\centering
	\includegraphics[width=\linewidth]{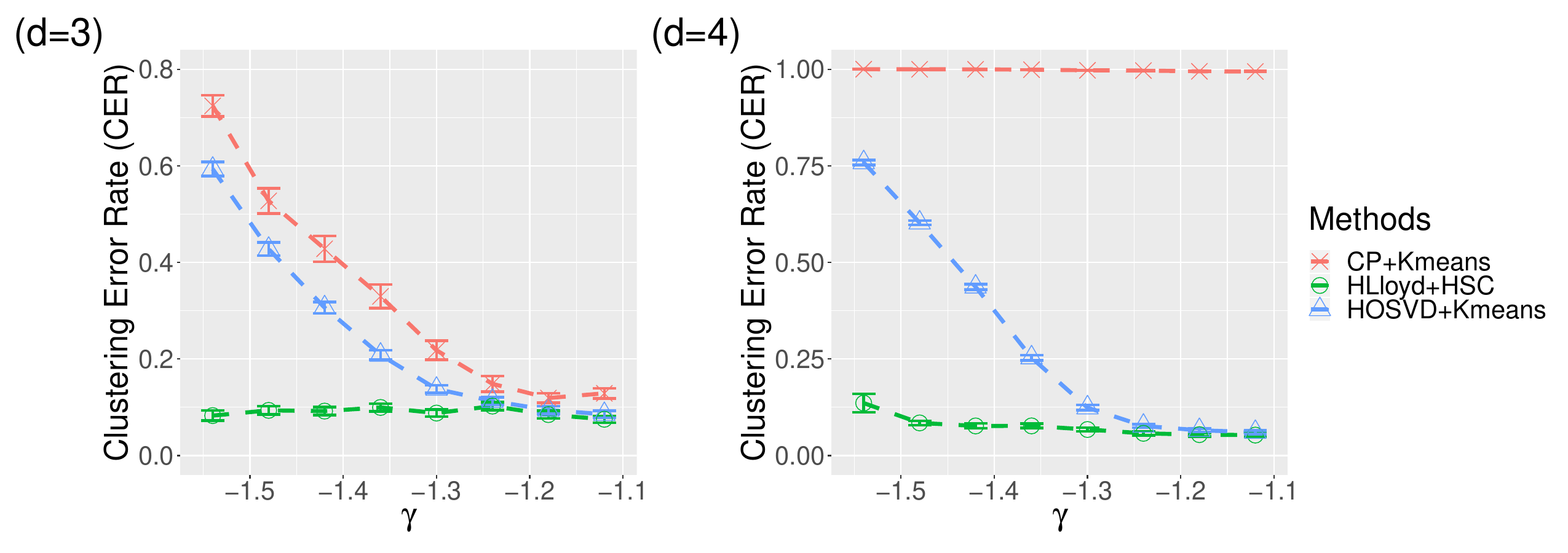}
	\caption{Comparison of HLloyd + HSC with HOSVD-/CP-decomposition-based clustering methods with different numbers of clusters along each mode.}\label{fig: TC diff r compare}
\end{figure}

Next, we compare the performance of these algorithms when the number of clusters along each mode differs. We set $p$ and $\Delta_{\min}$ to be the same as the previous experiment and $r_1 = 3, r_2 = 5, r_3 = 7$ when $d = 3$; $r_1 = 3, r_2 = 5, r_3 = 7, r_4 = 9$ when $d=4$. Figure \ref{fig: TC diff r compare} shows the averaged CER for these methods. We find that HLloyd + HSC still outperforms the other two algorithms and its CER standard error is low and stable for all the ranges of $\gamma$ considered.

\begin{figure}[htp!]
	\centering
	\includegraphics[width=\linewidth]{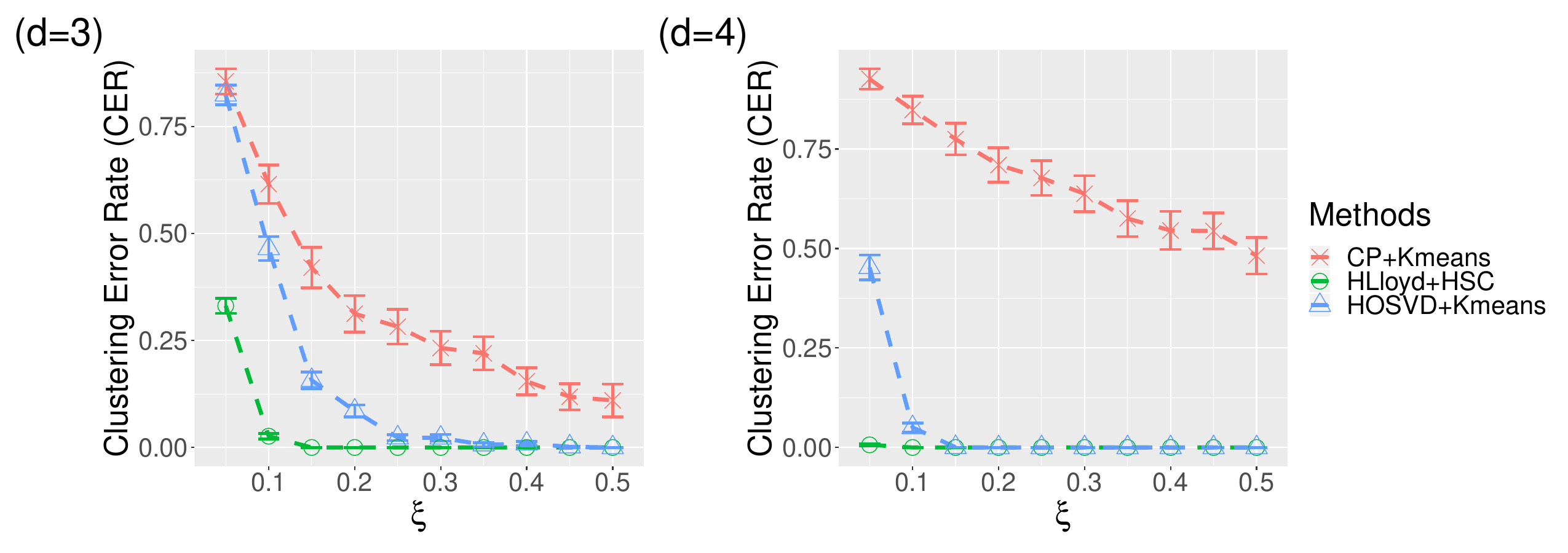}
	\caption{Comparison of HLloyd + HSC with HOSVD-/CP-decomposition-based clustering methods when cluster sizes are imbalanced.
	}\label{fig: TC imbalance compare} 
\end{figure}

Third, we examine the effect of imbalanced cluster size on the performance of these algorithms. In this setting, we let each mode have two clusters, i.e., $r = 2$, and the proportion of cluster 1 in each mode is $\xi$. We let $p, \Delta_{\min}$ be the same as before with fixed $\gamma = -1.4$. Figure \ref{fig: TC imbalance compare} illustrates the clustering performance of these algorithms as one gradually increases $\xi$ from $0.05$ to $0.5$ and it can be observed that the proposed algorithm is more robust against the imbalanced cluster size than the other two baseline methods. 

\begin{figure}[htp!]
	\centering
	\includegraphics[width=\linewidth]{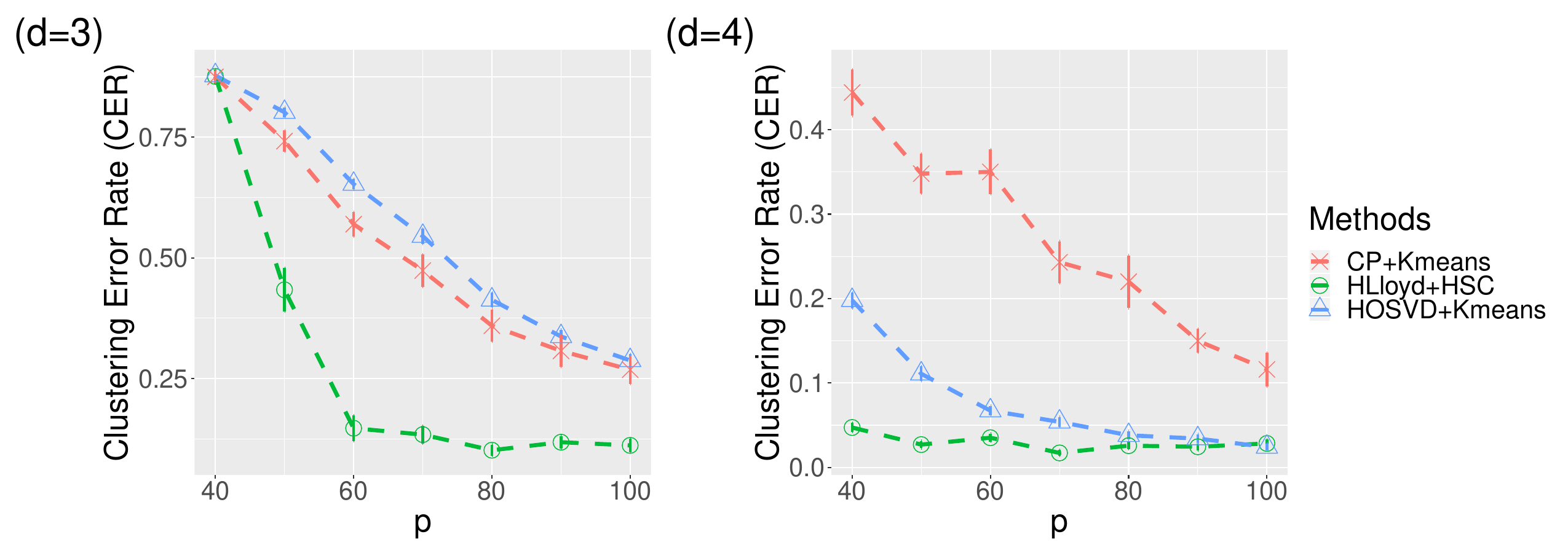}
	\caption{Comparison of HLloyd+HSC with HOSVD-/CP-decomposition-based clustering methods in stochastic tensor block model.}\label{fig: TC STBM compare} 
\end{figure}

Finally, we compare the performance of these algorithms in the stochastic tensor block models. Consider the setting $r = 5,p \in \{40,50,\ldots,100\}$ and every entry in the core tensor $\cS$ is generated uniformly at random from $[0,0.1]$. The comparison result is given in Figure \ref{fig: TC STBM compare}. We can see that in both settings $d = 3$ and $d = 4$, the proposed algorithm performs much better than the other two algorithms. 

\section{Real Data Analysis}\label{sec:real-data}

\subsection{Flight Route Network}\label{sec:flight-route}

In the first application, we study a subset of the worldwide air routes networks\footnote{Available at:  \url{https://openflights.org/data.html\#route}} based on 66,765 global flight routes from 568 airlines and 3,409 airports. We focus on the top $50$ airports with the highest numbers of flight routes and obtain an order-$3$ tensor $\cY$ of size $39 \times 50 \times 50$, where the entry $\cY_{ijk}$ equals to $1$ if there exists a flight route from airport $j$ to airport $k$ in airline $i$ and equals to $0$ otherwise. We perform tensor clustering using HLloyd initialized by HSC with rank $(r_1, r_2, r_2)=(5,5,5)$. Here, $r_1$ and $r_2$ are chosen from $\{3,4,5,6\}$ that they (i) do not result in a singleton in any cluster and (ii) minimize the following Bayesian information criterion (BIC) for block models~\citep{wang2019multiway},
\begin{equation}\label{eq:BIC}
	\text{BIC}(r_1,\ldots,r_d) = p_*\log( \|\hat\cX -\cY \|_\tF^2) + \left(r_* + \sum_{k=1}^d p_k \log r_k \right)\log p_*.
\end{equation}

Tables \ref{tab: airline clustering} and \ref{tab: airport clustering} show the clustering results for airlines and airports, respectively. The clusters well capture the underlying geographic and traffic information. Meaningful regions, such as the USA, Europe, China, Southeast Asia, were identified in the airline clusters and/or airport clusters. We also find several mixture clusters: Airline Cluster 3 is a mixture cluster consisting relatively small airlines around the world; Airline Cluster 5 consists of three airlines from Europe and Delta Airlines from USA.

\begin{table}
	\caption{\label{tab: airline clustering}Clustering of airlines based on the global flight routes network.}
	\centering
	\begin{tabular}{c | c  }
	\hline
	 & Airlines  \\
	\hline
	Cluster 1  & AA, UA, US (USA)  \\
	Cluster 2 &  BA, AY, IB (Europe)  \\
	Cluster 3 &  SU, AB, AI, AM, NH, AC, AS, FL, DE, etc (Mixture)  \\
	Cluster 4 &  CA, MU, CZ, HU, 3U, ZH (China)  \\
	Cluster 5 & AF, AZ, KL (Europe), DL (USA) \\
	\end{tabular}
\end{table}

\begin{table}
	\caption{\label{tab: airport clustering}Clustering of airports based on global flight route network.} 
	\centering
	\begin{tabular}{c | c  }
	\hline
	 & Airports  \\
	\hline
	\multirow{1}{4em}{Cluster 1}  & PHX, SFO, LAX, EWR, IAH, ATL, DEN, LAS, YYZ,\\
	\multirow{1}{4em}{}  & MEX (North America)  \\
	\multirow{1}{4em}{Cluster 2} & TPE, HKG, DEL, KUL, SIN, BKK, ICN,   \\
	\multirow{1}{4em}{} & DME, etc (Southeast Asian)  \\
	Cluster 3 & BRU, FRA, DUS, MUC, MAN, AMS, BCN, MAD, FCO, \\
	Cluster 3 &  ZRH (Europe) \\
	Cluster 4 & PEK, CAN, XIY, KMG, HGH, CKG, CTU, PVG(China)\\
	Cluster 5 & MIA, DEW, PHL, JFK, ORD, CLT (USA), LHR (UK)\\
	\end{tabular}
\end{table}

\begin{figure}[htbp]
	\centering
	\subfigure[US airlines]{\includegraphics[height = 1.8in]{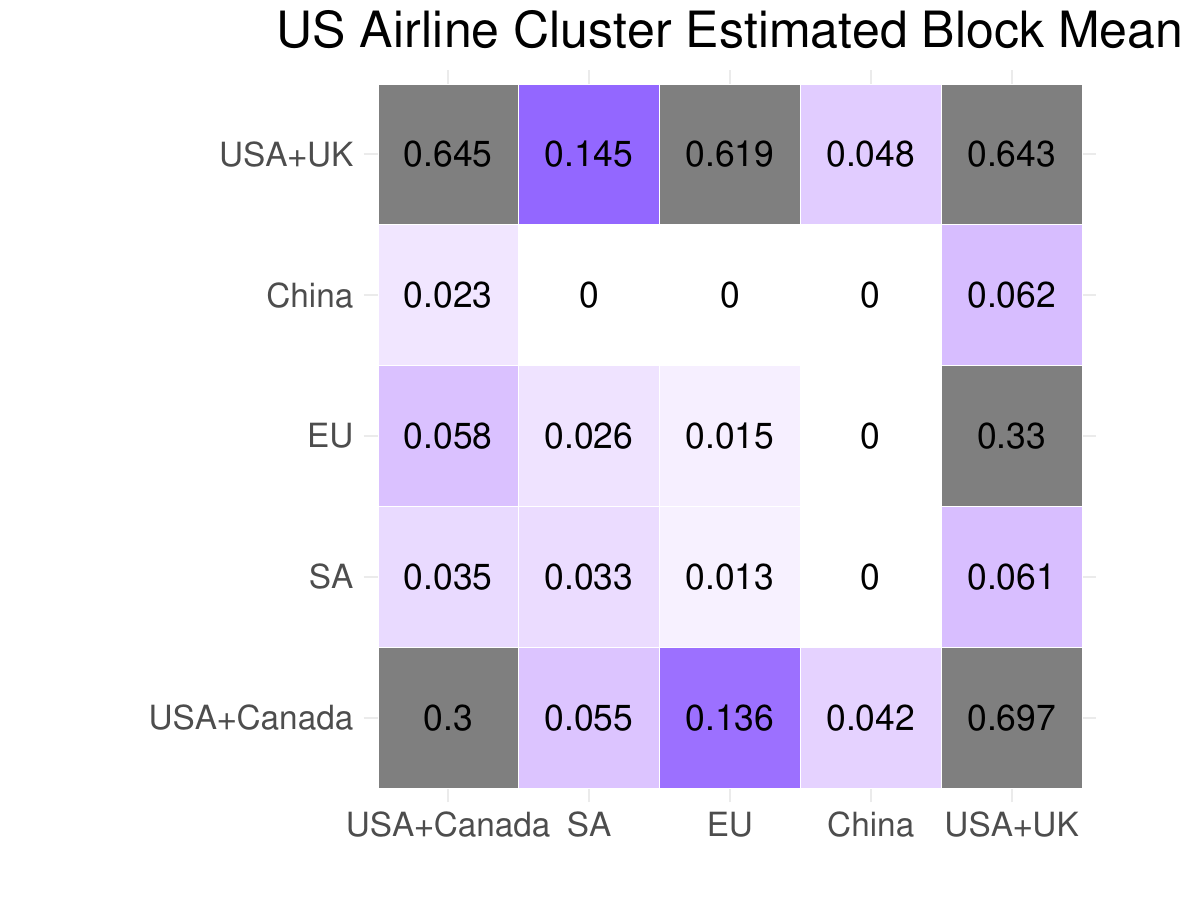}}
	\hskip.2cm
	\subfigure[Mixture airlines]{\includegraphics[height = 1.8in]{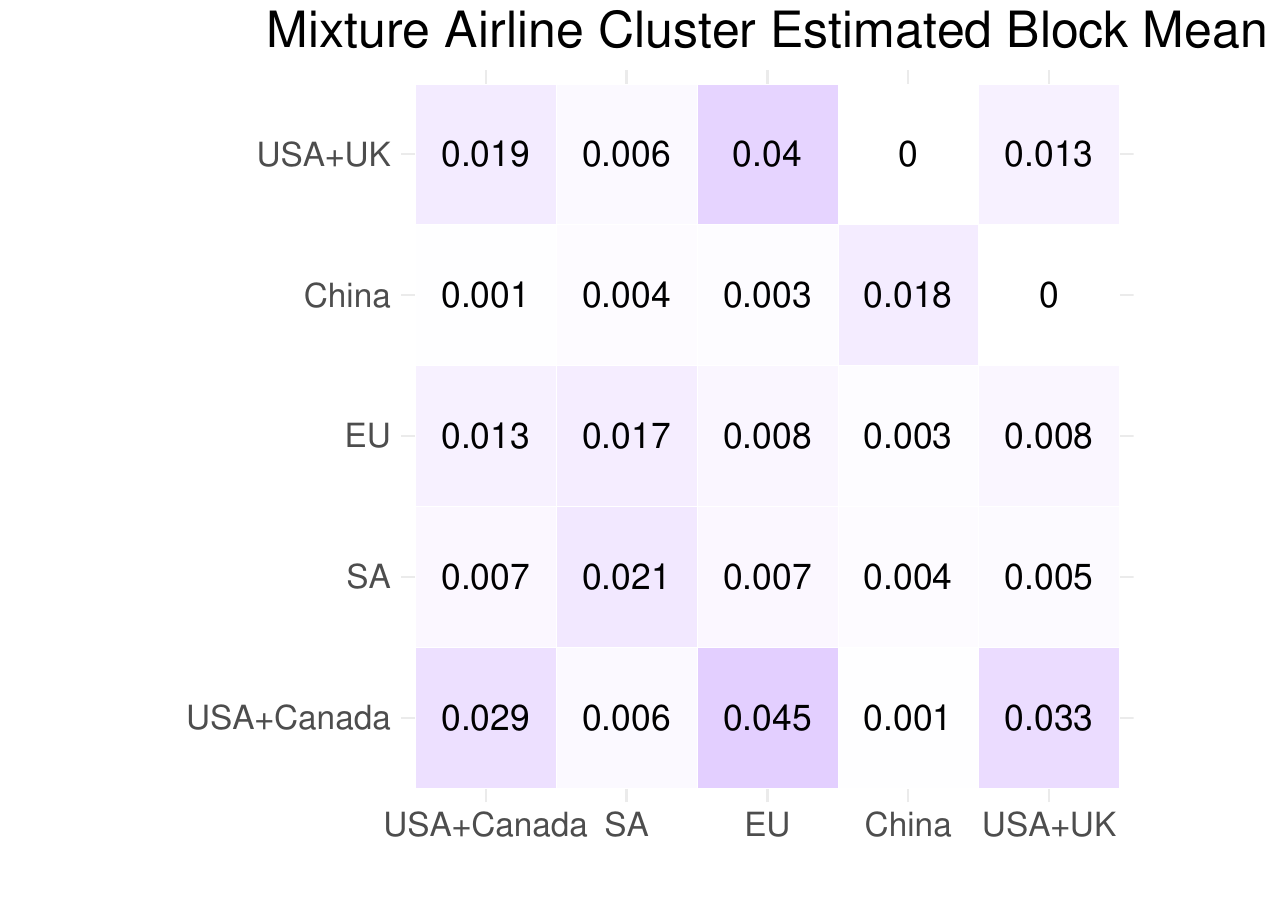}}
	\caption{Heatmaps of two matrix slices in the estimated block mean corresponding to the clusters of US airlines and mixture airlines in clustering results of global flight routes network. Here ``EU'' is short for Europe and ``SA'' is short for Southeast Asia.}
	\label{fig: airline heatmap}
\end{figure}

Figure \ref{fig: airline heatmap} shows the estimated block mean $\hat\cS$ corresponding to the US airline cluster (Airline Cluster 1) and the mixture airline cluster (Airline Cluster 3), respectively. The rows and columns of each matrix represent Airport Clusters 1-5 listed in Table \ref{tab: airport clustering}. The value in each matrix represents the connectivity of airports from each pair of clusters. We find that the US airline block mean matrix shows multiple zeros/small values among South Asian, European and Chinese airports. The matrix also shows high values among airports connected to USA. This result reveals that US airlines operate few flights between airports in Europe or Asia. In contrast, for the mixture airline cluster, the mean matrix has many non-zero but small entries. This reflects that this cluster consists of many small-scale airports scattered around the world.

We also compare the proposed HLloyd+HSC method with the HOSVD- and CP-decomposition-based clustering algorithms (see Section \ref{sec:num-comparison}). We find the proposed method yields more meaningful results than the other two. For example, Table \ref{tab: airline clustering-HOSVD} collects the airline clustering result of the HOSVD based clustering algorithm. Comparing Tables \ref{tab: airline clustering} and \ref{tab: airline clustering-HOSVD}, the proposed algorithm has a better performance on clustering USA and China airlines. More clustering results of the HOSVD-/CP-decomposition-based algorithms can be found in Appendix \ref{sec:additional-numerics}.

\begin{table}
	\caption{\label{tab: airline clustering-HOSVD}Clustering of airlines of the HOSVD based algorithm.}
	\centering
	\begin{tabular}{c| c  }
	\hline
	 & Airlines  \\
	\hline
	Cluster 1  & AA, US (USA)  \\
	Cluster 2 &  UA (USA)  \\
	Cluster 3 &  3U (China), SU, AB, AI, AM, NH, AC, AS, FL, DE, etc (Mixture)  \\
	Cluster 4 &  CA, MU, CZ, HU, ZH (China)  \\
	Cluster 5 & AF, AZ, KL (Europe), DL (USA) \\
	\end{tabular}
\end{table}

Finally, we perform a similar analysis for the USA flight routes network. Specifically, we pick the top $50$ airports and $9$ airlines with the most traffic and present the clustering results based on proposed algorithm in Tables \ref{tab: USA airline clustering} and \ref{tab: USA airport clustering}. We find the airports are grouped mainly based on their traffic sizes rather than their geographic information. We also find the three main airlines -- United Airlines, Delta Airlines, American Airlines -- are in different clusters, although they share similar numbers of flight routes. One possible reason is that these major airline companies are competitors, and the geometric distributions of their flight routes complement each other.

\begin{table}\caption{\label{tab: USA airline clustering}Clustering of US airlines based on the USA flight route network.} 
	\centering
	\begin{tabular}{c | c  }
	\hline
	 & Airlines  \\
	\hline
	Cluster 1  & UA(547), DL(633), WN(664), FL(420)  \\
	Cluster 2 & AA(640), US(590)  \\
	Cluster 3 & G4(6), B6(136), NK(134)  \\\hline
	\end{tabular}
\end{table}

\begin{table}
 	\caption{\label{tab: USA airport clustering}Clustering of US airports based on USA flight route network. }
	\centering
	\begin{tabular}{c | c  }
	\hline
	 & Airports  \\
	\hline
	Cluster 1  & PHX(137),LAX(148),DFW(130),ORD(151),PHL(110),JFK(108),  \\
	& CLT(159)  \\
	Cluster 2 & BOS(91),MCI(68),SFO(76),CLE(52),CVG(50),EWR(73), etc \\
	Cluster 3 & OAK(28),MEM(46),HOU(47),SAT(48),IAD(57),HNL(22),\\
	& SJC(33), etc\\\hline
	\end{tabular}
\end{table}

\subsection{Online Click-through Data}\label{sec:online-click-through}

In this section, we illustrate the application of proposed algorithms to time-dependent user-item collaborative filtering on an e-commerce dataset. The goal is to identify user clusters and item clusters in a longitudinal study. Specifically, we use the users' online click-through behavior data on \emph{Taobao.com}, one of the most popular online shopping website in China. The data\footnote{Available at:  \url{https://tianchi.aliyun.com/dataset/dataDetail?dataId=649}} of user-item interaction records are collected over eight consecutive days from Nov 25 to Dec 02, 2017. Due to the high dimensionality of the original dataset ($\approx 10^6$ users and $\approx 10^4$ item categories), we only select the most active 100 users and the most popular 50 items in our analysis. For the $m$th day, we construct a binary tensor $\cY_m \in \{0,1\}^{100\times 50 \times 24}$, where the $(i,j,k)$th entry of $\cY_m$ equals to one if and only if the $i$th user has an interaction with the $j$th item (i.e., make a click) in the $k$th hour in that day. Let $\cY = \frac{1}{8}\sum_{m=1}^8 \cY_i$ be the averaged observation, and we apply the proposed method to $\cY$. For the hour-mode, we set the number of clusters $r_3 = 4$, as we expect the behaviours might be separated into four time periods including dawn, morning, afternoon, and evening; for the other two modes, we set the cluster numbers $r_1$, $r_2$ to be the largest possible values that do not result in a singleton in any cluster. This leads to $r_1 = 4$ and $r_2 = 4$. 

\begin{figure}[htbp]
	\centering
	\includegraphics[width = 1\textwidth]{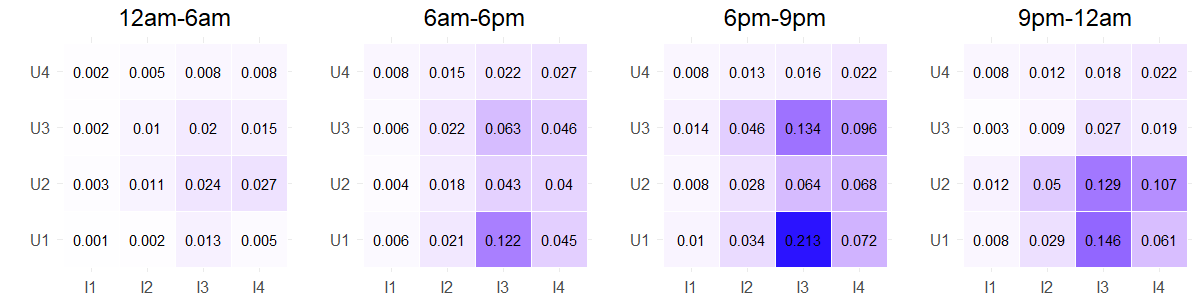}
	\caption{Clustering outcomes of online click-through data. Each matrix represents a mode-3 slice of $\hat \cS$ for different daily time cluster. The users clusters $U1, U2, U3, U4$ have $13,25,22,40$ members; the item clusters $I1,I2,I3,I4$ have $31, 11, 3, 5$ members, respectively.}\label{fig: click-through} 
\end{figure}

Figure \ref{fig: click-through} shows the estimation of block means $\hat \cS \in [0,1]^{4\times 4 \times 4}$, where $\hat\cS_{ijk}$ is the estimated probability that a user from the $i$th group click some item from the $j$th group in the daily time period of the $k$th group. The clustering results show that the time mode is well separated into four consecutive periods: 12am-6am (before dawn), 6am-6pm (daytime), 6pm-9pm (evening), and 9pm-12am (late night). We find the average activity is extremely low in Period I (before dawn) and high in Period III (evening), as most people are sleeping in Period I and could spend more time on online shopping after their daily work in Period III. For fixed user/item groups, we find some specific time-dependent behaviours. For example, by comparing the last heatplot with the previous two in Figure \ref{fig: click-through}, we identify a particular group of users U3, whose activities almost vanish after 9pm; this might corresponds to middle-aged people or seniors who sleep and rise early. On the other hand, U1 could be the group of young people since they have the most clicks in the late-night while being the least active in the early morning before 6am. Due to the lack of users/items features, we can not exactly verify our analysis. Nevertheless, the identified similarities among entries without external annotations illustrate the applicability of our method to clustering analysis.

\section{Discussion}\label{sec:discussions}

This paper develops a polynomial-time high-order clustering algorithm consisting of HLloyd iterations and HSC initialization. Critical thresholds for signal-to-noise ratio are established, revealing the intrinsic distinctions between (vector) clustering, (matrix) biclustering and (tensor) triclustering/high-order clustering. In particular, we provide both statistical and computational limits for high-order clustering in the tensor block model. 

It is worth mentioning that while our focus is on clustering for tensor block models, the developed results are useful in a broader variety of applications. For example, the block-wise constant structure in our tensor model is closely related to nonparametric graphon modeling~\citep{amini2018semidefinite,klopp2017oracle}. In the high-order case, our framework lends itself well to triclustering~\citep{hore2016tensor} and multilayer pattern recognition~\citep{lei2019consistent,lee2020tensor}. The tensor block model is a building block for more complex structures including latent space models~\citep{wang2018learning}, low-rank models~\citep{young2018universality}, and isotonic models~\citep{pananjady2020isotonic}. In principle, complex tensor data can still be fitted by block models, at a cost of a large $r$. In this regard, we review our block models as the first step towards theoretical understanding of more complicated models. Building flexible models for general tensor data is an important future question.

While this paper mainly considers the tensor block model under Gaussian noise, our algorithm and proof techniques can also be generalized to other statistical settings, such as missing observations, non-continuous observations (e.g., dichotomous or count data), and heavy-tailed noises. For example, when the tensor entries are observed under Huber's $\varepsilon$-contamination model, one can change the aggregated mean procedure in HLloyd algorithm to aggregated median for robustness. On the other hand, it is also interesting to combine the discrete block assumption with other popular low-dimensional structures such as sparsity, monotonicity, and smoothness. In addition, our current lower bounds focus on settings with i.i.d. Gaussian noise; the statistical and computational lower bounds for stochastic tensor block models with Bernoulli distribution remain a challenging and open problem.

\section{Proof Sketches and Technical Overview of Main Results}\label{sec:proof-sketch}

In this section, we briefly discuss the high-level ideas and sketches of proofs for Theorem \ref{thm:HO-SC}, Proposition \ref{prop:HOOI-singular-free}, and Theorem \ref{thm:HLloyd}. The complete proofs are deferred to Section \ref{sec:proofs-main} in the Supplementary Materials. 

\subsection{Proof Sketch of Theorem \ref{thm:HO-SC}}

Our analysis is conducted on the misclassification loss $l_k^{(t)}$ defined in \eqref{eq:loss}. We specifically discuss how to bound $l_1^{(0)}$ while the same argument applies to the other $l_k^{(0)}$s. Recall $z_1^{(0)}$ is the clustering outcome of applying $k$-means on rows of $\hat\Y_1$ ($\hat\Y_1$ is defined in \eqref{eq:low-dim-loadings}). One can check applying $k$-means on the rows of $\hat\Y_1$ is equivalent to applying $k$-means on the rows of 
\begin{equation*}
    \tilde\Y_1 := \bbP_{\hat \U_1}\cM_1\left(\cY \times_2 \bbP_{\hat\U_2} \times  \cdots  \times_d \bbP_{\hat\U_d}\right) \in \bbR^{p_1 \times p_{-1}},
\end{equation*}
where $\bbP_{\hat\U}:=\hat\U\hat\U^\top$ is the projection operator.
Under proper regularity conditions, we can prove that
\begin{equation}\label{ineq:lk0-bound}
    l_1^{(0)} \lesssim \frac{Mr_{-1}}{p_*} \left\|\tilde\Y_1 - \cM_1(\cX)\right\|_\tF^2. 
\end{equation}
Thus, to bound $l_1^{(0)}$, we only need to bound 
$\|\tilde\Y_1 - \mathcal{M}_1(\cX)\|_\tF^2$, 
i.e., the difference of $\tilde\Y_1$ to its population counterpart $\bbE \cM_1(\cY) = \cM_1(\cX)$, which is exactly the goal of Proposition \ref{prop:HOOI-singular-free}. Then Theorem \ref{thm:HO-SC} can be concluded by combining \eqref{ineq:lk0-bound}, Proposition \ref{prop:HOOI-singular-free}, and Lemma \ref{lm:loss-relation}. 

\subsection{Proof Ideas of Proposition \ref{prop:HOOI-singular-free}}

We assume $r_k \leq r_{-k}$ for the convenience of illustration. One can show by tensor algebra that
\begin{equation}\label{ineq:HOOI-sketch-1}
    \left\|\tilde{\cY} - \cX\right\|_\tF^2 \lesssim \sum_{k=1}^d \left\|\hat\U_{k\perp}\cM_k(\cX)\right\|_\tF^2 + \left\|\cZ \times_1 \hat \U_1^\top \times \cdots \times_d \hat\U_d^\top\right\|_\tF,
\end{equation}
where $\hat\U_{k\perp}$ is the perpendicular subspace of $\hat\U_k$. The second term in \eqref{ineq:HOOI-sketch-1} can be bounded by sub-Gaussian concentration inequalities while the analysis on the first term ($\|\hat\U_{k\perp}\cM_k(\cX)\|_\tF^2$) is more involved. Recall that
\begin{equation*}
    \begin{split}
    	& \hat\U_k = \SVD_{r_k}\left(\cM_k(\cY)\left(\tilde\U_{k+1} \otimes \cdots \otimes \tilde\U_d \otimes \tilde\U_1 \otimes \cdots \otimes \tilde\U_{k-1}\right)^\top\right), \\
    	& \tilde\U_k = \SVD_{r_k}(\mathcal{M}_k(\cY)).
    \end{split}
\end{equation*}
A classic scheme to analyze $\|\hat\U_{k\perp}\cM_k(\cX)\|_\tF$ is by establishing an upper bound on the principle angles between the subspaces spanned by the preliminary estimated singular vectors ($\tilde \U_k$) and the true singular vectors ($\U_k := \SVD_{r_k}(\cM_k(\cX))$) in the sin-theta distance: $\|\sin\Theta(\tilde \U_k, \U_k)\| = \|\tilde \U_{k\perp}^\top \U_{k}\|.$ 
To obtain an upper bound on $\|\sin\Theta(\tilde\U_k, \U_k)\|$, a singular value gap condition, i.e., some lower bound on $\lambda_{r_k}(\cM_k(\cX)) -  \lambda_{r_k+1}(\cM_k(\cX))$, is crucial as indicated by the classic literature on matrix perturbation theory \citep{davis1970rotation,wedin1972perturbation}. Since there is no singular value gap condition in the context of Proposition \ref{prop:HOOI-singular-free}, it is difficult to prove the desired bounds via bounding $\|\sin\Theta(\tilde{\U}_k, \U_k)\|$.

Let $\U_k'=\SVD_{r_k'}\left(\cM_k(\cX)\right)$ for some $r_k' \leq r_k$. Our main idea for proving Proposition \ref{prop:HOOI-singular-free} is to decompose $\cX$ into two parts:
\begin{equation*}
    \cX = \cX' + (\cX - \cX'),\quad \text{where } \cX':= \cX \times_1 \bbP_{\U_1'} \times \cdots \times_d \bbP_{\U_d'}. 
\end{equation*}
By appropriately choosing $r_k'$, $\cX'$ can be taken as the ``strong signal'' tensor with large singular values; while $\cX-\cX'$ is the ``weak signal'' tensor with small singular values. With such a decomposition, we only focus on estimating $\cX'$ while leaving $\cX-\cX'$ as a bias term. To this end, we first introduce a deterministic matrix perturbation bound for subspaces of different dimensions.
\begin{Lemma}\label{lm:new-perturbation-determin}
Suppose the first $r$ and the rest $p_1-r$ singular vectors of $\Y\in \bbR^{p_1 \times p_2}$ are $\hat \U \in \bbO_{p_1,r}$ and $\hat \U_{\perp} \in \bbO_{p_1,p_1-r}$, respectively. For some $1 < r' \leq r$, let $\W \in \bbO_{p_1,r'}$ be any $p_1$-by-$r'$ column orthonormal matrix and $\W_{\perp} \in \bbO_{p_1,p_1-r'}$ be the orthogonal complement of $\W$. Given that $\sigma_{r'}(\W^\top\Y) > \sigma_{r+1}(\Y)$, we have
\begin{equation*}
    \left\|\hat\U_{\perp}^\top\W\right\|\leq \frac{\sigma_{r'}(\W^\top \Y)\|\W^\top_\perp \Y \bbP_{\Y^\top \W}\|}{\sigma_{r'}^2(\W^\top\Y) - \sigma_{r+1}^2(\Y)}.
\end{equation*}
Here $\bbP_\A = \A(\A^\top \A)^{\dagger}\A^\top$ is the projection operator.
\end{Lemma}
To apply it to our problem, we further obtain the following probabilistic bound under the additive heterogeneous sub-Gaussian noises.
\begin{Lemma}\label{lm:new-perturbation-subGaussian}
    Let $\Y = \X + \Z \in \bbR^{p_1 \times p_2}$ where $\X$ is a rank-$r$ matrix. Let $\hat\U = \SVD_{r}(\Y)$, $\U_{r'} = \SVD_{r'}(\X)$ for some $1\leq r' \leq r$. Suppose $\Z_{ij}$ are independent mean-zero sub-Gaussian entries with $\bbE \Z_{ij}^2=1$ and $\|\Z_{ij}\|_{\psi_2} \leq C$. Then, with probability at least $1-\exp(-cp_1\wedge p_2)$,
    \begin{equation*}
        \left\|\hat\U_\perp^\top \U_{r'}\right\| \leq C\left(\frac{\sqrt{p_1}}{\sigma_{r'}(\X)} + \frac{\sqrt{p_1p_2}}{\sigma_{r'}^2(\X)}\right).
    \end{equation*}
    Here, $\|\Z_{ij}\|_{\psi_2} $ denotes the sub-Gaussian norm of $\Z_{ij}$.
\end{Lemma}

Applying Lemma \ref{lm:new-perturbation-subGaussian} to $\tilde\U_k$ and $\U_k'$, we can obtain a neat bound for $\|\tilde\U_{k\perp}^\top \U_k'\|$, which can be further used to analyze the estimation error of $\cX'$ via the bound of $\|\hat\U_{k\perp}\cM_k(\cX')\|_\tF$. The complete proof is postponed to Section \ref{sec:proof-HOOI-singular-free}.

\begin{Remark}
We highlight the difference between the analyses of clustering in the matrix case (order $d=2$) and the high-order tensor case (order $d\geq 3$) and illustrate the necessity of developing these techniques here. Note that in Gaussian mixture or bi-clustering models, the estimator $\tilde\Y_1$ is essentially the rank-$r_1$ truncated-SVD estimator of $\cY \in \bbR^{p_1 \times p_2}$ and it be can directly obtained that
\begin{equation*}
        \begin{split}
            \left\|\tilde \Y_1 - \cX\right\|_\tF^2 & \leq 2r\left\|\tilde\Y_1 - \cX\right\|^2 \leq 4r\left(\left\|\tilde\Y_1 - \cY\right\|^2 + \left\|\cX - \cY\right\|^2\right) \\
            & \overset{(*)}{\leq} 8 r_1\left\|\cX - \cY\right\|^2 = 8 r_1\left\|\cZ\right\|^2 \overset{(**)}{\leq} 8(p_1+p_2)r_1,\qquad w.h.p.
        \end{split}
\end{equation*}
Here, \emph{($*$)} comes from the optimality of SVD while \emph{($**$)} comes from the spectral norm of Gaussian random matrix. Note that the above argument is free of any singular value gap as it does not involve any analysis on singular subspace estimators. 
In contrast, there is no such an analytic property for the estimator $\tilde \Y_1$ when $d\geq 3$ and we have to analyze the accuracy of the singular subspace estimators $\{\tilde\U_k\}_{k=1}^d$ in a more complicated way. This makes the theoretical analysis fundamentally more difficult without the singular value gap condition in the tensor setting.
\end{Remark}

\subsection{Proof Sketch of Theorem \ref{thm:HLloyd}}\label{sec:sketch-HLloyd}

To prove Theorem \ref{thm:HLloyd}, we establish a local contraction property on $l_k^{(t)}$. Inspired by the convergence analysis of iterative algorithm for single discrete structure ~\citep{gao2019iterative}, we introduce an oracle clustering procedure $\tilde\cS, \tilde z_k$ as follows:
\begin{equation}\label{eq:oracle-z}
    \begin{split}
        \tilde\cS_{i_1,\ldots,i_d} &= \text{Average}\left(\left\{\cY_{j_1,\ldots,j_d}\colon (z_{k})_{j_k} = i_k, \forall k \in [d]\right\}\right), \qquad i_k \in [r_k],~~\\
        (\tilde z_k)_j &= \argmin_{a\in[r_k]} \left\|\left(\cM_k(\cY)\right)_{j:}\V_k - \left(\cM_k(\tilde\cS)\right)_{a:}\right\|_2^2,\qquad j \in [p_k],
    \end{split}
\end{equation}
where $\V_k := \W_1 \otimes  \cdots \W_{k-1} \otimes \W_k \otimes \cdots \otimes \W_d$ and $\W_k := \M_k\left(\diag\left(1_{p_k}^\top \M_k\right)\right)^{-1}$ is the weighted membership matrix.

Consider the mode-$1$ clustering, for the oracle procedure \eqref{eq:oracle-z}, by definition $j \in [p_1]$, $(\tilde z_1)_j \neq (z_1)_j$ if and only if there exists $b \in [r_1]\backslash (z_1)_j$, such that
\begin{equation}\label{ineq:oracle-wrong}
    \left\|\left(\cM_1(\cY)\right)_{j:}\V_1 - \left(\cM_1(\tilde\cS)\right)_{(z_1)_j:}\right\|_2^2 > \left\|\left(\cM_1(\cY)\right)_{j:}\V_1 - \left(\cM_1(\tilde\cS)\right)_{b:}\right\|_2^2.
\end{equation}
 Note that \eqref{ineq:oracle-wrong} is equivalent to 
\begin{equation*}
    \begin{split}
        & \left\langle \left(\cM_1(\cE)\right)_{j:}\V_1, \left(\cM_1(\tilde\cS)\right)_{{(z_1)}_j:} - \left(\cM_1(\tilde\cS)\right)_{b:} \right\rangle \\
        < &  \frac{1}{2}\left(-\left\|\left(\cM_1(\cS)\right)_{(z_1)_j:} - \left(\cM_1(\tilde\cS)\right)_{b:}\right\|^2 + \left\|\left(\cM_1(\cS)\right)_{(z_1)_j:} - \left(\cM_1(\tilde\cS)\right)_{(z_1)_j:}\right\|^2\right) \\
        \approx & -\frac{1}{2}\left\|\left(\cM_1(\cS)\right)_{(z_1)_j:} - \left(\cM_1(\cS)\right)_{b:}\right\|^2.
    \end{split}
\end{equation*}
Therefore, we define the quantity
\begin{equation*}
    A_{1j} := \frac{\left\langle \left(\cM_1(\cE)\right)_{j:}\V_1, \left(\cM_1(\tilde\cS)\right)_{{(z_1)}_j:} - \left(\cM_1(\tilde\cS)\right)_{b:} \right\rangle}{\left\|\left(\cM_1(\cS)\right)_{(z_1)_j:} - \left(\cM_1(\cS)\right)_{b:}\right\|^2}
\end{equation*}
and the following oracle loss with some small ``tolerace'' constant $\delta \in (0,1)$:
\begin{equation}\label{eq:oracle-loss}
    \xi_1 = \frac{1}{p_1} \sum_{j=1}^{p_1} \sum_{b \in [r_1]/(z_1)_j} \left\|\left(\cM_1(\cS)\right)_{(z_1)_j:} - \left(\cM_1(\cS)\right)_{b:}\right\|_2^2 \cdot \bbI\left\{A_{1j} \leq -\frac{1-\delta}{2}\right\}.
\end{equation}
We can similarly define $\xi_k$ for each $k\in [d]$. Intuitively speaking, $\xi_k$ can be taken as a surrogate for the misclassification loss of the oracle clustering estimator $\tilde z_k$: we expect $\xi_k$ is closely related to $l_k^{(t)}$ in \eqref{eq:loss} and may serve as the iteration limiting point. Different from the regular clustering, HLloyd algorithm involves iterations of all $d$ memberships $z_k^{(t)}$; and $l_k^{(t+1)}$, the misclassification loss at step $(t+1)$, depends on $l_{k'}^{(t)}$ for all $k'\in [d]$. This makes the error contraction analysis for HLloyd much more involved. With a dedicated loss decomposition and error estimation, we can prove the following contraction inequality: 
\begin{equation}\label{ineq:lt-contraction}
    l_k^{(t+1)} \leq \frac{3}{2}\xi_k + \frac{1}{2}\max_{k\in [d]}l_k^{(t)}.
\end{equation}
By sub-Gaussian concentration, one can further prove that with high probability
\begin{equation}\label{ineq:xi-bound}
    \xi_k \lesssim \sigma^2\exp\left(-\frac{c_2p_{-k}}{r_{-k}}\cdot\frac{\Delta_k^2}{\sigma^2}\right).
\end{equation}
Combining \eqref{ineq:lt-contraction}, \eqref{ineq:xi-bound}, and Lemma \ref{lm:loss-relation}, we establish the loss convergence \eqref{ineq:step-t-lk-bound} in Theorem \ref{thm:HLloyd}.

\bibliography{reference}
\bibliographystyle{apalike}

\appendix
\newpage
\setcounter{page}{1}

\begin{center}
{\bf \large Supplement to ``Exact Clustering in Tensor Block Model:}
\end{center}
\begin{center}
{\bf \large Statistical Optimality and Computational Limit"}
\end{center}

\bigskip

\begin{center}
{Rungang Han\footnotemark[1]\footnotemark[2],~ Yuetian Luo\footnotemark[1],~ Miaoyan Wang\footnotemark[1], ~and~ Anru R. Zhang\footnotemark[3]\footnotemark[1]}
\end{center}

\footnotetext[1]{Department of Statistics, University of Wisconsin-Madison}
\footnotetext[2]{Department of Statistical Science, Duke University}
\footnotetext[3]{Department of Biostatistics \& Bioinformatics, Duke University}

\section{Additional Numeric Results}\label{sec:additional-numerics}
The following Table \ref{tab: airport clustering-HOSVD} provides the clustering results of airports by HOSVD based clustering algorithm. Compared to the clustering results in Table \ref{tab: airport clustering}, we can see the HOSVD based clustering method clusters Chinese airports into two clusters and mix some Europe and southeast Asia airports.

In the following Tables \ref{tab: airline clustering-CP} and \ref{tab: airport clustering-CP}, we provide the clustering results of airlines and airports by the CP decomposition based clustering algorithm. Again, there are more mixture clusters for CP decomposition based method compared to the clustering results in Tables \ref{tab: airline clustering} and \ref{tab: airport clustering}.

\newpage
\begin{table}
	\caption{\label{tab: airport clustering-HOSVD}Clustering results of airports by the HOSVD based clustering algorithm. } 
	\centering
	\begin{tabular}{c | c  }
		\hline
		& Airports  \\
		\hline
		\multirow{1}{4em}{Cluster 1}  & PHX, SFO, LAX, EWR, IAH, MIA, DFW, etc (North America)  \\
		\multirow{1}{4em}{Cluster 2} & CAN, XIY, KMG, HGH, CKG, CTU, TPE (China 2)  \\
		Cluster 3 & LHR, AMS, MAD, CDG, FCO, MEX, ATL  (Europe) \\
		Cluster 4 & PEK, PVG(China 1)\\
		Cluster 5 & BRU, FRA, DUS (Europe), NRT, HKG, DEL(Southeast Asia), etc \\\hline
	\end{tabular}
\end{table}
\begin{table}
	\caption{\label{tab: airline clustering-CP}Clustering results of airlines by the CPD based clustering algorithm. } 
	\centering
	\begin{tabular}{c | c  }
		\hline
		& Airlines  \\
		\hline
		Cluster 1  & AA, US (USA)  \\
		Cluster 2 &  UA (USA)  \\
		Cluster 3 &  SU, AB, AI, AM, NH, AC, AS, FL, DE, etc (Mixture)  \\
		Cluster 4 &  CA, MU, CZ, HU, 3U, ZH (China)  \\
		Cluster 5 & AF, AZ, KL (Europe), DL (USA) \\\hline
	\end{tabular}
\end{table}
\begin{table}
	\caption{\label{tab: airport clustering-CP}Clustering results of airports by the CP based clustering algorithm. } 
	\centering
	\begin{tabular}{c | c  }
		\hline
		& Airports  \\
		\hline
		\multirow{1}{4em}{Cluster 1}  & PHX, SFO, LAX, EWR, IAH, ATL, DEN, LAS,  \\
		\multirow{1}{4em}{}  & YYZ, MEX (North America)  \\
		\multirow{1}{4em}{Cluster 2} & LWG, AMS, VIE (Europe),  \\
		\multirow{1}{4em}{} & TPE, HKG, DEL, etc (Southeast Asian)  \\
		Cluster 3 & FRA, LHR, MAD, CDG, FCO (Europe),\\
		& ATL, JFK (USA) \\
		Cluster 4 & PEK, CAN, XIY, KMG, HGH, CKG, CTU, PVG(China)\\
		Cluster 5 & PHX, MIA, DFW, PHL (USA), \\
		& BRU, DUS, MUC, MAN, etc (Europe)\\\hline
	\end{tabular}
\end{table}
\clearpage

\section{Proofs of Main Results}\label{sec:proofs-main}

In this section, we present the proofs of Theorem \ref{thm:HO-SC}, Theorem \ref{thm:HLloyd} and Proposition \ref{prop:HOOI-singular-free}.

\subsection{Proof of Theorem \ref{thm:HO-SC}}
We assume $\sigma=1$ without loss of generality. By Proposition \ref{prop:HOOI-singular-free}, we know that with probability at least $1-\exp(-c\underline p)$,
\begin{equation}\label{ineq:tensor-error-decompose-bound}
    \left\|\cY \times_1 \hat \U_1\hat\U_1^\top \times \cdots \times_d \hat\U_d\hat\U_d^\top - \cX\right\|_\tF^2 \leq C\left(r_* + \bar p \bar r^2 + p_*^{1/2}\bar r\right).
\end{equation}
Now we bound the misclassification rate of $z_k^{(0)}$ based on \eqref{ineq:tensor-error-decompose-bound}. We focus on the first mode, since the proofs for other modes are essentially the same.

Recall $\hat\Y_1 = \hat \U_1 \hat \U_1^\top\cM_1(\cY \times_2 \hat\U_2^\top \times \cdots \times_d \hat\U_d^\top)$ and note that
\begin{equation}
	\begin{split}
		& \min_{\substack{x_1,\ldots,x_{r_1} \in \bbR^{r_{-1}} \\ \bar{z}_1 \in [r_1]^{p_1}}}\sum_{j=1}^{p_1}\left\|(\hat \Y_1)_{j:} - x_{(\bar{z}_1)_{j}}^\top\right\|^2 \\
		&= \min_{\substack{\theta_1,\ldots,\theta_{r_1} \in \bbR^{p_{-1}} \\ \bar{z}_1 \in [r_1]^{p_1}}}\sum_{j=1}^{p_1}\left\|(\hat \Y_1)_{j:}(\hat \U_2\otimes \cdots \otimes \hat \U_d)^\top - \theta_{(\bar{z}_1)_{j}}^\top\right\|^2.
	\end{split}
\end{equation}

Denote $\bar \Y_1 = \hat \Y_1(\hat\U_2\otimes \cdots \otimes \hat\U_d)^\top \in \bbR^{p_1 \times p_{-1}}$. For any $a\in [r_1]$, let $j_a \in [p_1]$ be any one of the indices such that $(z_1)_{j_a} = a$ and we denote $\theta_a^* := (\cM_1(\cX))_{j_a:}^\top \in \bbR^{p_{-1}}$. Then, the $k$-means++ programming in Algorithm \ref{alg:HO-SC} leads to
\begin{equation}\label{ineq:init-1}
	\begin{split}
		\sum_{j=1}^{p_1} \left\|(\bar \Y_1)_{j:}^\top - \hat\theta_{(z_1^{(0)})_j}\right\|^2 &\leq M\min_{\substack{\theta_1,\ldots,\theta_{r_1} \in \bbR^{p_{-1}} \\ \bar{z}_1 \in [r_1]^{p_1}}}\sum_{j=1}^{p_1}\left\|(\bar \Y_1)_{j:}^\top - \theta_{(\bar{z}_1)_{j}}^*\right\|^2 \\
	    & \leq M \sum_{j=1}^{p_1}\left\|(\bar \Y_1)_{j:}^\top - \theta^{*}_{(z_1)_j}\right\|^2 \\
	    & = M \left\|\hat{\U}_1\hat{\U}_1^\top \cM_1(\cY)(\hat \U_2\hat \U_2^\top \otimes \cdots \otimes \hat\U_d\hat\U_d^\top) - \cM_1(\cX)\right\|_\tF^2 \\
	    & = M\left\|\cY\times_1 \hat{\U}_1\hat{\U}_1^\top \times_2 \hat\U_2\hat\U_2^\top \times \cdots \times_d \hat\U_d\hat\U_d^\top - \cX\right\|_\tF^2 \\
	    & \overset{\eqref{ineq:tensor-error-decompose-bound}}{\leq} CM\left(r_* + \bar p \bar r^2 + p_*^{1/2}\bar r\right).
	\end{split}
\end{equation}
Therefore, we have
\begin{equation}\label{ineq:init-2}
	\begin{split}
		\sum_{j=1}^{p_1} \left\|\hat \theta_{(z_1^{(0)})_j} - \theta^*_{(z_1)_j}\right\|^2 &\leq 2\sum_{j=1}^{p_1}\left(\left\|(\bar \Y_1)_{j:}^\top - \hat\theta_{(z_1^{(0)})_j}\right\|^2 + \left\|(\bar \Y_1)_{j:}^\top - \theta^{*}_{(z_1)_j}\right\|^2\right) \\
		& \overset{\eqref{ineq:init-1}}{\leq} 4CM\left(r_* + \bar p \bar r^2 + p_*^{1/2}\bar r\right).
	\end{split}
\end{equation}
By Assumption \ref{asmp:balance-size}, we have 
\begin{equation}\label{ineq:init-cluster-size}
    \sqrt{\alpha p_k/r_k}  \leq \lambda_{r_k}(\M_k) \leq \sqrt{\beta p_k/r_k},\qquad \forall k \in [d]
\end{equation}
Therefore, for any $a \neq b \in [r_1]$,
\begin{equation}\label{ineq:init-3}
    \begin{split}
        \|\theta^*_a - \theta^*_b\| & = \|(\cM_1(\cX))_{j_a:} - (\cM_1(\cX))_{j_b:}\| \\
        & = \|\left((\cM_1(\cS))_{a:} - (\cM_1(\cS))_{b:}\right) (\M_2 \otimes \cdots \otimes \M_d)^\top\| \\
        & \geq \|(\cM_1(\cS))_{a:} - (\cM_1(\cS))_{b:}\| \cdot \prod_{k=2}^d \lambda_{r_k}(\M_k)\\
        & \overset{\eqref{ineq:init-cluster-size}}{\geq} c_0\sqrt{p_{-1}/r_{-1}}\Delta_1
    \end{split}
\end{equation}
for some small constant $c_0$. Now we define the index set 
\begin{equation*}
	S = \left\{j \in [p_1]: \left\|\hat\theta_{(z_1^{(0)})_j} - \theta^*_{(z_1)_j}\right\|_2 \geq \sqrt{\frac{p_{-1}}{r_{-1}}}\frac{c_0\Delta_{1}}{2}\right\}.
\end{equation*}
Then we can bound the size of $S$:
\begin{equation}\label{ineq:S-size}
	|S| \leq \frac{\sum_{j=1}^{p_1} \left\|\hat \theta_{(z_1^{(0)})_j} - \theta^*_{(z_1)_j}\right\|^2}{\frac{c_0^2p_{-1}}{4r_{-1}}\Delta_1^2} \overset{\eqref{ineq:init-1}}{\leq} \frac{C}{\Delta_1^2}\frac{Mr_{-1}\left(r_* + \bar p \bar r^2 + p_*^{1/2}\bar r\right)}{p_{-1}} \leq \frac{\alpha p_1}{2r_1}.
\end{equation}
Here the last inequality comes from the SNR condition that $\Delta_1^2 \geq CM\left(\frac{\bar p r_*^2\bar r}{p_*} + \frac{r_*\bar r}{p_*^{1/2}}\right)$ for a sufficient large constant $C$.
Define the set $\cC_a = \{j \in [p_1]: (z_1)_j = a, j \in S^c\}$. Then we firstly have
\begin{equation}\label{ineq:property-S}
    |\cC_a| \geq \sum_{j=1}^{p_1} \bbI\{(z_1)_j = a\} - |S|  \overset{\eqref{ineq:S-size}}{\geq} \frac{\alpha p_1}{2r_1}.
\end{equation}
In the meantime, we know that for any $a\neq b \in [r_1]$ and any $j_1 \in \cC_a$ and $j_2 \in \cC_b$, we must have $(z_1^{(0)})_{j_1}  \neq (z_1^{(0)})_{j_2}$. Otherwise
\begin{equation*}
    \left\|\theta^*_{a} - \theta^*_{b}\right\| \leq \left\|\theta^*_{a} - \hat \theta_{(z_1^{(0)})_{j_1}}\right\| + \left\|\theta^*_{b} - \hat \theta_{(z_1^{(0)})_{j_2}}\right\| \leq c_0\sqrt{\frac{p_{-1}}{r_{-1}}}\Delta,
\end{equation*}
which contradicts \eqref{ineq:init-3}. This implies that for all $j_1 \in \cC_a$, $(z_1^{(0)})_{j_1}$ shares the same labels and is different from all other $(z_1^{(0)})_{j_2}$ with $j_2 \in \cC_b$ for any $b \neq a$. As a result, we can find a permutation $\pi_1$ on $[r_1]$ such that $(z_1^{(0)})_j = \pi_1(z_1)$ for all $j \in S^c$. Thus, for any $a \in [r_1]$,
\begin{equation}\label{ineq:center-diff-bound}
    \begin{split}
        \left\|\theta_a^* - \hat\theta_{\pi_1(a)}\right\|_2^2 & = \frac{\sum_{j \in \cC_a}\left\|\theta^*_{(z_1)_j} - \hat \theta_{(z_1^{(0)})_j}\right\|_2^2}{|\cC_a|} \\
        & \overset{\eqref{ineq:property-S}}{\leq} C\frac{\sum_{j=1}^{p_1} \left\|\theta_{(z_1)_j}^* - \hat \theta_{(z_1^{(0)})_j}\right\|^2}{p_1/r_1} \\
        & \overset{\eqref{ineq:init-2}}{\leq} CM\frac{r_1}{p_1}\left(r_* + \bar p \bar r^2+ p_*^{1/2}\bar r\right).
    \end{split}
\end{equation}
Then,
\begin{equation*}
	\begin{split}
		l_1^{(0)} & = \frac{1}{p_1}\sum_{j\in[p_1]}\left\|(\cM_1(\cS))_{(z_1)_j:} - (\cM_1(\cS))_{\pi_1^{-1}\left((z_1^{(0)})_j\right):}\right\|_2^2 \\
		& = \frac{1}{p_1}\sum_{j\in[p_1]}\left\|(\cM_1(\cS))_{(z_1)_j:} - (\cM_1(\cS))_{\pi_1^{-1}\left((z_1^{(0)})_j\right):}\right\|_2^2 \cdot\bbI\{(z_1^{(0)})_j \neq \pi_1 (z_1)_j\} \\
		& \leq C\frac{1}{p_1} \cdot \prod_{k=2}^d\lambda_{r_k}^{-2}(\M_k) \sum_{j\in[p_1]}\left\|\theta^*_{(z_1)_j} - \theta^*_{\pi_1^{-1}\left((z_1^{(0)})_j\right)}\right\|_2^2  \cdot \bbI\{(z_1^{(0)})_j \neq \pi_1 (z_1)_j\}\left(\cM_1(\cS)\right)_{(z_1)_j:} \\
		& \overset{\eqref{ineq:init-cluster-size}}{\leq} C\frac{r_{-1}}{p_*} \sum_{j\in[p_1]}\left\|\theta^*_{(z_1)_j} - \theta^*_{\pi_1^{-1}\left((z_1^{(0)})_j\right)}\right\|_2^2 \cdot \bbI\{(z_1^{(0)})_j \neq \pi_1 (z_1)_j\} \\
		& \leq 2C\frac{r_{-1}}{p_*}\sum_{j\in [p_1]}\left(\left\|\theta^*_{(z_1)_j} - \hat \theta_{(z_1^{(0)})_{j}}\right\|_2^2 + \left\|\hat \theta_{(z_1^{(0)})_j} -  \theta^*_{\pi_1^{-1}\left((z_1^{(0)})_j\right)}\right\|_2^2 \right) \bbI\{(z_1^{(0)})_j \neq \pi_1 (z_1)_j\}. \\
		& \leq 2C\frac{r_{-1}}{p_*}\left(\sum_{j\in [p_1]}\left\|\theta^*_{(z_1)_j} - \hat \theta_{(z_1^{(0)})_{j}}\right\|_2^2 + \max_{a \in [r_1]} \left\|\hat \theta_{a} - \theta^*_{\pi_1^{-1}(a)}\right\|_2^2 \sum_{j\in[p_1]} \bbI\{(z_1^{(0)})_j \neq \pi_1 (z_1)_j\}\right)\\
		& \leq 2C\frac{r_{-1}}{p_*}\left(\sum_{j\in [p_1]}\left\|\theta^*_{(z_1)_j} - \hat \theta_{(z_1^{(0)})_{j}}\right\|_2^2 + |S|\max_{a \in [r_1]} \left\|\hat \theta_{a} - \theta^*_{\pi_1^{-1}(a)}\right\|_2^2 \right) \\
		& \overset{\eqref{ineq:init-2}\eqref{ineq:center-diff-bound}}{\lesssim} M\frac{r_{-1}}{p_*}\left(r_* + \bar p \bar r^2 + p_*^{1/2}\bar r\right).
	\end{split}
\end{equation*}
Now Theorem \ref{thm:HO-SC} follows by applying Lemma \ref{lm:loss-relation}. \qquad\qquad $\square$

\subsection{Proof of Proposition \ref{prop:HOOI-singular-free}}\label{sec:proof-HOOI-singular-free}
Without loss of generality, we assume $\sigma=1$. We start by introducing several notations and assumptions. For each $k=1,\ldots,d$, denote 
    \begin{equation*}
        \X_k = \cM_k(\cX),~\Z_k = \cM_k(\cZ),~\Y_k = \cM_k(\cY).
    \end{equation*}
    Recall that $\rank(\X_k) \leq r_k$. We further denote $\U_k = \SVD_{r_k}(\X_k)$ and $\tilde \U_k = \SVD_{r_k}(\Y_k)$. For some constant $C_0$ which will be specified later, define
    \begin{equation*}
        r_k' = \max\left\{r' \in \{0,\ldots,r_k\}: \sigma_{r'}(\X_k) \geq C_0  (p_*^{1/4} \vee \bar p^{1/2})\right\}.
    \end{equation*}
    We set $r_k'=0$ if $\sigma_{1}(\X_k) < C_0  (p_*^{1/4} \vee \bar p^{1/2})$. We use $\U_{k}'$ to denote the leading $r_k'$ singular vectors of $\U_k$ and use $\V_k'$ to denote the rest $r_k-r_k'$ singular vectors and thus $\U_k$ can be written as $[\U_k'~\V_k']$. We next define 
    \begin{equation*}
        \X_k' = \X_k \left(\bbP_{\U_{k+1}'} \otimes \cdots \otimes \bbP_{\U_d'} \otimes \bbP_{\U_1'} \otimes \cdots \otimes \bbP_{\U_{k-1}'}\right)
    \end{equation*}
    We also denote 
    \begin{equation*}
        \begin{split}
            \bar\Y_k & = \Y_k (\tilde\U_{k+1} \otimes \cdots \otimes \tilde\U_d \otimes \tilde\U_1 \otimes \cdots \otimes \tilde\U_{k-1}), \\
            \bar\X_k & = \X_k (\tilde\U_{k+1} \otimes \cdots \otimes \tilde\U_d \otimes \tilde\U_1 \otimes \cdots \otimes \tilde\U_{k-1}), \\
            \bar\Z_k & = \Z_k (\tilde\U_{k+1} \otimes \cdots \otimes \tilde\U_d \otimes \tilde\U_1 \otimes \cdots \otimes \tilde\U_{k-1}).
        \end{split}
    \end{equation*}
    Now we define the following events under which we conduct the subsequent analysis. 
    \begin{equation}\label{eq-HOOI-A1}
		\begin{split}
			A_1 = \left\{\left\|\tilde\U_{k\perp}^{\top}\U_k'\right\| \leq \frac{1}{\sqrt{2}},\quad k=1,\ldots,d.\right\}
		\end{split}
	\end{equation}
    \begin{equation}\label{eq-HOOI-A2}
		\begin{split}
			A_2 = \left\{\left\|\bar\Z_{k}\right\| \leq C(\sqrt{p_k} + \sqrt{r_{-k}} + \sum_{l \neq k} \sqrt{p_lr_l}),\quad k=1,\ldots,d.\right\}
		\end{split}
	\end{equation}
	\begin{equation}\label{eq-HOOI-A3}
		\begin{split}
			A_3 = \left\{\left\|\cZ \times_1 \hat\U_1 \times \cdots \times_d \hat\U_d\right\|_\tF \leq C(\sqrt{r_*} + \sum_{k=1}^d \sqrt{p_kr_k})\right\}
		\end{split}
	\end{equation}
	By Lemma \ref{lm:new-perturbation-subGaussian}, with probability at least $1-C\exp(-\underline p)$, for each $k \in [d]$,
	\begin{equation*}
	    \left\|\tilde\U_{k\perp}^{\top}\U_k'\right\| \leq \frac{C\sqrt{p_k}(\sigma_{r_k'}(\X_k)+\sqrt{p_{-k}})}{\sigma_{r_k'}^2(\X_k)} \leq \frac{C}{C_0}\left(\frac{\sqrt{p_k}}{\sqrt{\bar p}} + \frac{\sqrt{p_*}}{\sqrt{p_*}}\right) \leq \frac{1}{\sqrt{2}},
	\end{equation*}
	where the last inequality is obtained by specifying $C_0 = 2\sqrt{2}C$. Meanwhile,
	By Lemma \ref{lm:concentration-ineq}, $\bbP(A_2\cap A_3) \geq 1-\exp(-c\underline p)$. Therefore, $\bbP(A_1 \cap A_2 \cap A_3) \geq 1-\exp(-c\underline p)$. Now we prove the Theorem under $A_1 \cap A_2 \cap A_3$.
	
	We provide an upper bound for $\left\|\hat\U_{k\perp}^{\top}\X_k\right\|_\tF$. First of all,
	\begin{equation}\label{ineq:HOOI-1}   
        \begin{split}
            \left\|\hat\U_{k\perp}^\top\X_k\right\|_\tF & = \left\|\hat\U_{k\perp}^\top(\X_k' + \X_k - \X_k')\right\|_\tF \\
            & \leq \left\|\hat\U_{k\perp}^\top\X_k'\right\|_\tF + \left\|\hat\U_{k\perp}^\top(\X_k-\X_k')\right\|_\tF \\
            & \leq \left\|\hat\U_{k\perp}^\top\X_k'\right\|_\tF + \left\|\X_k-\X_k'\right\|_\tF.
        \end{split}
    \end{equation}
	To bound $\left\|\hat\U_{k\perp}^\top\X_k'\right\|_\tF$, we notice that 
	\begin{equation}\label{ineq:HOOI-2}
        \begin{split}
            & \left\|\hat\U_{k\perp}^\top\X_k'(\tilde\U_{k+1} \otimes \cdots \otimes \tilde\U_d \otimes \tilde\U_1 \otimes \cdots \otimes \tilde\U_{k-1})\right\|_\tF \\
            \leq & \left\|\hat\U_{k\perp}^\top\bar\X_k\right\|_\tF + \left\|\hat\U_{k\perp}^\top(\X_k-\X_k')(\tilde\U_{k+1} \otimes \cdots \otimes \tilde\U_d \otimes \tilde\U_1 \otimes \cdots \otimes \tilde\U_{k-1})\right\|_\tF \\
            \leq & \left\|\hat\U_{k\perp}^\top\bar\X_k\right\|_\tF + \left\|\X_k-\X_k'\right\|_\tF.
        \end{split}
    \end{equation}
    Also, since the right singular space of $\X_k'$ is $\U_{k+1}' \otimes \cdots \otimes \U_d' \otimes \U_1' \otimes \cdots \otimes \U_{k-1}'$, we have
    \begin{equation}\label{ineq:HOOI-3}
        \begin{split}
            & \left\|\hat\U_{k\perp}^\top\X_k'(\tilde\U_{k+1} \otimes \cdots \otimes \tilde\U_d \otimes \tilde\U_1 \otimes \cdots \otimes \tilde\U_{k-1})\right\|_\tF \\
           =  &  \left\|\hat\U_{k\perp}^\top\X_k'(\bbP_{\U_k'}\tilde\U_k \otimes \cdots \otimes \bbP_{\U_d'}\tilde\U_d \otimes \bbP_{\U_1'}\tilde\U_1 \otimes \cdots \otimes \bbP_{\U_{k-1}'}\tilde\U_{k-1})\right\|_\tF \\
           \geq  &  \left\|\hat\U_{k\perp}^\top\X_k'\right\|_\tF \cdot \prod_{l \neq k}\sigma_{r_k'}(\U_k^{'\top}\tilde\U_k) \\
           =  & \left\|\hat\U_{k\perp}^\top\X_k'\right\|_\tF \cdot \prod_{l \neq k}\sqrt{1-\left\|\tilde\U_{k\perp}^{\top}\U_k'\right\|^2} \overset{\eqref{eq-HOOI-A1}}{\geq} \frac{1}{\sqrt{2}^{d-1}}\left\|\hat\U_{k\perp}^\top\X_k'\right\|_\tF.
        \end{split}
    \end{equation}
    Combining \eqref{ineq:HOOI-1}, \eqref{ineq:HOOI-2} and \eqref{ineq:HOOI-3}, we obtain
    \begin{equation}\label{ineq:HOOI-4}
    \begin{split}
    &\left\|\hat\U_{k\perp}^\top\X_k'\right\|_\tF \leq 2^{(d-1)/2} \left( \left\|\hat\U_{k\perp}^\top\bar\X_k\right\|_\tF + \left\|\X_k-\X_k'\right\|_\tF \right) \\
       & \left\|\hat\U_{k\perp}^\top\X_k\right\|_\tF \leq 2^{(d-1)/2}\left\|\hat\U_{k\perp}^\top\bar\X_k\right\|_\tF + (2^{(d-1)/2}+1)\left\|\X_k-\X_k'\right\|_\tF.
    \end{split}
    \end{equation}
    By Lemma \ref{lm:low-rank-perturbation}, since $\bar \Y_k = \bar\X_k + \bar\Z_k$, $\SVD_{r_k \wedge r_{-k}}(\bar\Y_k) = \hat\U_k$, we have
	\begin{equation}\label{ineq:HOOI-5}
        \left\|\hat \U_{k\perp}^\top \bar\X_k\right\|_\tF \leq 2\sqrt{r_k \wedge r_{-k}}\left\|\bar \Z_k\right\| \overset{\eqref{eq-HOOI-A2}}{\lesssim} \sqrt{r_*} + \sum_{l=1}^d \sqrt{p_lr_l\bar r}.
    \end{equation}
    Now it suffices to bound $\|\X_k - \X_k'\|_\tF$. For notation simplicity, we focus on $k=1$, while the analysis for other modes can be similarly carried on.
    \begin{equation*}
        \begin{split}
            & \left\|\X_1 - \X_1'\right\|_\tF  = \left\|\X_1\left((\bbP_{\U_{2}'}+\bbP_{\V_2'})\otimes \cdots \otimes (\bbP_{\U_{d}'}+\bbP_{\V_d'}) - \bbP_{\U_{2}'} \otimes \cdots \otimes \bbP_{\U_{d}'} \right)\right\|_\tF \\
            = & \Big\|\X_1\big(\bbP_{\V_2'} \otimes \I_{p_3} \otimes \cdots \otimes \I_{p_d}  + \bbP_{\U_2'}\otimes \bbP_{\V_3'} \otimes \cdots \otimes \I_{p_d} + \cdots + \\
            & ~~ \bbP_{\U_2'} \otimes \cdots \otimes\bbP_{\U_{d-1}'} \otimes \bbP_{\V_{d}'} \big)\Big\|_\tF \\
            \leq & \sum_{k=2}^d\left\|\V_k^{'\top}\cM_k(\cX)\right\|_\tF 
            \leq \sum_{k=2}^d \sqrt{r_k-r_k'}\sigma_{r_k'+1}(\X_k) \leq \sum_{k=2}^d C_0( p_*^{1/4} + \bar p^{1/2})\sqrt{r_k}.
        \end{split}
    \end{equation*}
    Here, the last inequality comes from the definition of $r_k'$, i.e., the $r_k'+1$th singular value of $\X_k$ is smaller than $C_0 (p_*^{1/4} \vee \bar p^{1/2})$.
    In general, for any $k \in [d]$, we have
    \begin{equation}\label{ineq:HOOI-6}
        \left\|\X_k - \X_k'\right\|_\tF \leq C_0d(p_*^{1/4} \vee \bar p^{1/2})\bar r^{1/2}.
    \end{equation}
    Combining \eqref{ineq:HOOI-4}, \eqref{ineq:HOOI-5} and \eqref{ineq:HOOI-6}, it follows that
    \begin{equation}\label{ineq:HOOI-7}
        \left\|\hat\U_{k\perp}^\top \X_k\right\|_\tF \leq C_d\left(\sqrt{r_*} +  p_*^{1/4}\bar r^{1/2} + \bar p^{1/2}\bar r\right).
    \end{equation}
	Now we are ready to bound $\left\|\hat\cX - \cX\right\|$. Recall that $\hat\cX = \cY \times_1 \bbP_{\hat\U_1} \times \cdots \times_d \bbP_{\hat\U_d}$. Then,
	\begin{equation}\label{ineq:HOOI-8}
    \begin{split}
        & \left\|\cY \times_1 \hat \U_1\hat\U_1^\top \times \cdots \times_d \hat\U_d\hat\U_d^\top - \cX\right\|_\tF \\
        \leq & \left\|\cX \times_1 \hat \U_1\hat\U_1^\top \times \cdots \times_d \hat\U_d\hat\U_d^\top - \cX\right\|_\tF + \left\|\cZ \times_1 \hat \U_1^\top \times \cdots \times_d \hat\U_d^\top\right\|_\tF.
    \end{split}
\end{equation}
On the one hand, following the same argument on obtaining \eqref{ineq:HOOI-6}, we have
\begin{equation}\label{ineq:HOOI-9}
    \begin{split}
        \left\|\cX \times_1 \hat \U_1\hat\U_1^\top \times \cdots \times_d \hat\U_d\hat\U_d^\top - \cX\right\|_\tF
        \overset{\text{Lemma \ref{lm:orthogonal-perpendicular}}}{\leq}  \sum_{k=1}^d \left\|(\I - \hat\U_k\hat\U_k^\top)\X_k\right\|_\tF = \sum_{k=1}^d \left\|\hat\U_{k\perp}^\top\X_k\right\|_\tF.
    \end{split}
\end{equation}
On the other hand, under $A_3$,
\begin{equation}\label{ineq:HOOI-10}
    \left\|\cZ \times_1 \hat \U_1^\top \times \cdots \times_d \hat\U_d^\top\right\|_\tF \overset{\eqref{eq-HOOI-A3}}{\leq} C\left(\sqrt{r_*} + \sum_{k=1}^d \sqrt{p_kr_k}\right).
\end{equation}
Combining \eqref{ineq:HOOI-7}, \eqref{ineq:HOOI-8}, \eqref{ineq:HOOI-9} and \eqref{ineq:HOOI-10}, we finally obtain
\begin{equation*}
    \left\|\hat\cX - \cX\right\|_\tF \leq C_d\left(r_*^{1/2}+p_*^{1/4}\bar r^{1/2} + \bar p^{1/2}\bar r\right)
\end{equation*}
and the proof is finished.\qquad\qquad $\square$

\subsection{Proof of Theorem \ref{thm:HLloyd}}\label{sec:proof}
\begin{proof}
Assume $\sigma = 1$ and $\pi_k^{(0)}$ is the identity mapping on $[r_k]$ without loss of generality. Since the proof is fairly complicated, we divide the proof into several steps to facilitate the presentation. We start by introducing several notations, conditions and technical Lemmas, then we establish the proof based on these ingredients.
\begin{enumerate}[leftmargin=*]
	\item[Step 1] (Notations, Conditions and Lemmas)
Recall for any $k \in [d]$, $\M_k, \M_k^{(t)} \in \{0,1\}^{p_k \times r_k}$ denote the membership matrices for the $k$th mode, i.e., 
	\begin{equation*}
		\begin{split}
			(\M_{k})_{ja} &= 1 \text{ if and only if } (z_{k})_j = a;\\
			(\M_{k}^{(t)})_{ja} &= 1 \text{ if and only if } (z_{k}^{(t)})_j = a.
		\end{split}
	\end{equation*}
	We additionally introduce the following notations.
	\begin{enumerate}[label=(\subscript{N}{\arabic*})]
		\item Normalized membership matrices,  $\forall k =1,\ldots, d$,
			\begin{equation*}
				\W_k := \M_k\left(\diag\left(1_{p_k}^\top \M_k\right)\right)^{-1}, \quad \W_k^{(t)} := \M_k^{(t)}\left(\diag\left(1_{p_k}^\top \M_k^{(t)}\right)\right)^{-1}.
			\end{equation*}
		\item Estimator of block mean at step $t$ ($\cS^{(t)}$) and the oracle estimator given true clusters $z_k$ ($\tilde\cS$),
			\begin{equation*}
				\begin{split}
					\cS^{(t)} &:= \cY \times_1 \W_1^{(t)\top} \times \cdots \times_d \W_d^{(t)\top} \in \bbR^{r_1\times \cdots \times r_d}; \\
					\tilde\cS &:= \cY \times_1 \W_1^{\top} \times \cdots \times_d \W_d^{\top} \in \bbR^{r_1\times \cdots \times r_d}.
				\end{split}
			\end{equation*}
		\item Dual normalized membership matrices, $\forall k =1,\ldots, d$,
			\begin{equation*}
				\begin{split}
					\V_k &:= \W_{k+1} \otimes \cdots \otimes \W_d \otimes \W_1 \otimes \cdots \otimes \W_{k-1}; \\
					\V_k^{(t)} &:=  \W_{k+1}^{(t)} \otimes \cdots \otimes \W_d^{(t)} \otimes \W_1^{(t)} \otimes \cdots \otimes \W_{k-1}^{(t)}.
				\end{split}
			\end{equation*}
			
		\item Matricizations of tensor, $\forall k =1,\ldots, d$,
			\begin{equation*}
				\begin{split}
					 \S_k = \cM_k(\cS),\qquad \S_k^{(t)} =  \cM_k(\cS^{(t)}),\qquad \tilde\S_k = \cM_k(\tilde\cS),;\\
					\Y_k = \cM_k(\cY),\qquad \X_k = \cM_k(\cX), \qquad \E_k = \cM_k(\cE).
				\end{split}
			\end{equation*}
		\item Oracle error, $\forall k =1,\ldots, d$,
				\begin{equation*}
					\begin{split}
						\xi_k := \frac{1}{p_k}\sum_{j=1}^{p_k}\sum_{b \in [r_k]/(z_k)_j} &\bbI\Big\{\left\langle (\E_{k})_{j:}\V_k, (\tilde\S_k)_{{(z_k)}_j:} - (\tilde\S_k)_{b:} \right\rangle \\
						& \leq -\frac{1}{4}\left\|(\S_k)_{(z_k)_j:} - (\S_k)_{b:}\right\|^2\Big\} \cdot \left\|(\S_k)_{{(z_k)}_j:} - (\S_k)_{b:}\right\|^2
					\end{split}
				\end{equation*}
				Here, $\xi_k$ can be taken as the oracle statistical loss for the mode $k$ clustering when the true block structures of all modes are known, see more discussions in Section \ref{sec:sketch-HLloyd}.
	\end{enumerate}
	We next introduce the following conditions and then complete the proof given these conditions. 
	
	$\forall k \in [d], \forall a \in [r_k]$, 
	\begin{equation}\label{hpc-1-1}
		\begin{split}
			& \left\|\E_k \V_k\right\| \leq C\sqrt{\frac{r_{-k}}{p_{-k}}}\left(\sqrt{p_k}+\sqrt{r_{-k}}\right),\quad \|\E_k\V_k\|_\tF \lesssim \sqrt{\frac{p_1r_{-1}^2}{p_{-1}}},\\ 
			& \|(\W_k)_{:a}^\top \E_k\V_k\| \leq C\frac{r_*}{\sqrt{p_*}};
		\end{split}
	\end{equation}
	\begin{equation}\label{hpc-1-2}
	\begin{split}
	    & \sup_{\substack{\U_l \in \bbO_{p_k,r_k} \\ l=1,\ldots,d}}\left\|\E_k (\U_{k+1} \otimes \cdots \otimes \U_d \otimes \U_1 \otimes \cdots \otimes \U_{k-1})\right\| \\
	    \leq & C\left(\sqrt{r_{-k}}+\sum_{l \in [d]}\sqrt{p_lr_l}\right),
	\end{split}
	\end{equation}
    \begin{equation}\label{hpc-1-3}
		\begin{split}
			& \sup_{\substack{\U_l \in \bbO_{p_k,r_k} \\ l=1,\ldots,d}}\left\|\E_k (\U_{k+1} \otimes \cdots \otimes \U_d \otimes \U_1 \otimes \cdots \otimes \U_{k-1})\right\|_\tF \\
			\leq & C\left(\sqrt{p_kr_{-k}}+\sum_{l \in [d]}\sqrt{p_lr_l}\right).
		\end{split}
	\end{equation}
	\begin{equation}\label{hpc-2}
		\begin{split}
		& \xi_k \leq \exp\left(-c_1\frac{\Delta_k^2p_{-k}}{r_{-k}}\right).
		\end{split}
	\end{equation}
	\begin{equation}\label{hpc-3}
			l_k^{(t)} \leq c_2\frac{\Delta_{\min}^2}{r_k} \leq \frac{\Delta_k^2}{r_{k}}.
	\end{equation}
	Here $C,c_1,c_2$ are some universal constants. In Step 5 we will verify that above conditions \eqref{hpc-1-1}-\eqref{hpc-3} hold with high probability under our assumptions.
	
	Note that we only need to prove the error contraction inequality for the first mode and the others follow in the same way. For the simplicity of presentation, we omit the sub-subscripts of $\W_1^{(t)}, \V_1^{(t)}, \S_1, \Y_1, \E_1$, etc.
	
	Now we present several technical lemmas which will be used throughout the proof.

The first lemma quantifies the cluster sizes of $z_k^{(t)}$ at each iteration and the spectra of the (weighted) membership matrices.
\begin{Lemma}\label{lm:mebmership-spectra}
    Suppose \eqref{hpc-3} holds. Then, for any $k \in [d]$, $a \in [r_k]$, we have $cr_k/p_k \leq \left|j \in [p_k]: (z_k^{(t)})_j = a\right| \leq Cr_k/p_k$. Moreover, 
    \begin{equation*}
        \begin{split}
            & c\sqrt{p_k/r_k} \leq \lambda_{r_k}(\M_k) \leq  \|\M_k\| \leq C\sqrt{p_k/r_k}, \\
            & c\sqrt{r_k/p_k} \leq \lambda_{r_k}(\W_k) \leq  \|\W_k\| \leq C\sqrt{r_k/p_k}.
        \end{split}
    \end{equation*}
    The above two inequalities are also true by replacing $\M_k$, $\W_k$ to $\M_k^{(t)}$ and $\W_k^{(t)}$ respectively.
\end{Lemma}
\begin{proof}[Proof of Lemma \ref{lm:mebmership-spectra}]
See Section \ref{sec:lm:mebmership-spectra}.
\end{proof}
The next lemma provides the estimation error bounds for several intermediate quantities in our analysis.
	\begin{Lemma}\label{lm-1}
	Suppose conditions \eqref{hpc-1-1}-\eqref{hpc-3} hold. Then,
	\begin{equation}\label{ineq:lm-1-0}
		\|\V - \V^{(t)}\| \lesssim  \sqrt{\frac{r_{-1}}{p_{-1}}}\sum_{k=2}^d\frac{r_k}{\Delta_k^2}l_k^{(t)}, 
	\end{equation}
	\begin{equation}\label{ineq:lm-1-0.5}
		\|\E(\V - \V^{(t)})\|_\tF \lesssim  \sqrt{\frac{r_{-1}(p_1r_{-1}+\bar p \bar r)}{p_{-1}}} \sum_{k=2}^d \frac{r_kl_k^{(t)}}{\Delta_k^2}, 
	\end{equation}
	\begin{equation}\label{ineq:lm-1-1}
		\max_{b \in [r_1]}\left\|\left(\W_{:b} - \W_{:b}^{(t)}\right)^\top \Y\V\right\|_2 \lesssim r_1\frac{l_1^{(t)}}{\Delta_1} +  \sqrt{\frac{r_*^2 + p_1r_1r_*}{p_*}}\cdot\frac{\sqrt{l_1^{(t)}}}{\Delta_1},
	\end{equation}
	\begin{equation}\label{ineq:lm-1-2}
		\begin{split}
		 \max_{b \in [r_1]} \left\|\W_{:b}^{(t)\top} \Y\left(\V - \V^{(t)}\right)\right\|_2 &\lesssim \sqrt{\frac{\bar rr_*^2 + \bar p \bar r^2 r_*}{p_*}} \cdot \sum_{k = 2}^d \frac{\sqrt{l_k^{(t)}}}{\Delta_k}  +  \sum_{k=2}^d \frac{r_kl_k^{(t)}}{\Delta_k}.
		\end{split}
	\end{equation}
	\begin{equation}\label{ineq:lm-1-3}
		\begin{split}
			\max_{b\in [r_1]}\left\|(\W_{:b} - \W_{:b}^{(t)})^\top \Y\V^{(t)}\right\|_2 & \lesssim \sqrt{\frac{\bar rr_*^2 + \bar p \bar r^2 r_*}{p_*}} \cdot \sum_{k\geq 1} \frac{\sqrt{l_k^{(t)}}}{\Delta_k}  + \frac{r_1}{\Delta_1} l_1^{(t)}.
		\end{split}
	\end{equation}
\end{Lemma}
\begin{proof}[Proof of Lemma \ref{lm-1}]
See Section \ref{sec:lm-1}.
\end{proof}
	\item[Step 2] In this step, we decompose $l_1^{(t+1)}$ into several parts. Recall the membership of $j$th entry along mode-$1$ is updated via nearest neighbor search among the mode-$k$ slices of $\cY_k^{(t)}$, which is equivalent to:
	\begin{equation*}
	    (z_1^{(t+1)})_j = \argmin_{a \in [r_1]} \left\|\Y_{j:}\V^{(t)} - \S_{a:}^{(t)}\right\|_2^2,
	\end{equation*}
	here we use the fact $\cM_1(\cY_1^{(t)}) = \Y \V^{(t)}$. Thus, for each fixed $j\in [p_1]$, $b \in [r_1]$, we have
	\begin{equation*}
		\bbI\left\{(z_{1}^{(t+1)})_j = b\right\} = \bbI\left\{(z_{1}^{(t+1)})_j = b, \left\|\Y_{j:}\V^{(t)} - \S_{b:}^{(t)}\right\|_2^2  \leq \left\|\Y_{j:}\V^{(t)} - \S_{(z_{1})_j:}^{(t)}\right\|_2^2\right\}.
	\end{equation*}
	One can check that $\left\|\Y_{j:}\V^{(t)} - \S_{b:}^{(t)}\right\|_2^2  \leq \left\|\Y_{j:}\V^{(t)} - \S_{{(z_1)}_j:}^{(t)}\right\|_2^2$ is equivalent to
	\begin{equation}\label{ineq:equivalent-event}
	2\left\langle \E_{j:}\V, \tilde\S_{{(z_1)}_j:} - \tilde\S_{b:} \right\rangle \leq -\left\|\S_{{(z_1)}_j:} - \S_{b:}\right\|^2  + F_{jb}^{(t)} + G_{jb}^{(t)} + H_{jb}^{(t)},
	\end{equation}
	where $F_{jb}^{(t)}, G_{jb}^{(t)}, H_{jb}^{(t)}$ are defined as
\begin{equation}\label{eq:def-F}
	\begin{split}
	F_{jb}^{(t)}:= & 2\left\langle \E_{j:}\V^{(t)}, (\tilde\S_{{(z_1)}_j:} - \S_{{(z_1)}_j:}^{(t)}) - (\tilde\S_{b:} - \S_{b:}^{(t)})\right\rangle \\
	& + 2\left\langle \E_{j:}(\V - \V^{(t)}), \tilde\S_{{(z_1)}_j:}  - \tilde\S_{b:}\right\rangle
	\end{split}
\end{equation}
\begin{equation}\label{eq:def-G}
	\begin{split}
		G_{jb}^{(t)} &:= \left(\left\|\X_{j:}\V^{(t)} - \S_{{(z_1)}_j:}^{(t)}\right\|_2^2 - \left\|\X_{j:}\V^{(t)} - \W_{:{(z_1)}_j}^{\top}\Y\V^{(t)}\right\|_2^2\right) \\
		& \qquad - \left(\left\|\X_{j:}\V^{(t)} - \S_{b:}^{(t)}\right\|_2^2 - \left\|\X_{j:}\V^{(t)} - \W_{:b}^{\top}\Y\V^{(t)}\right\|_2^2\right)
	\end{split}
\end{equation}
\begin{equation}\label{eq:def-H}
	\begin{split}
		H_{jb}^{(t)} &:= \left\|\X_{j:}\V^{(t)} - \W_{:{(z_1)}_j}^{\top}\Y\V^{(t)}\right\|_2^2   - \left\|\X_{j:}\V^{(t)} - \W_{:b}^{\top}\Y\V^{(t)}\right\|_2^2   + \left\|\S_{{(z_1)}_j:} - \S_{b:}\right\|^2
	\end{split}
\end{equation}
Then, $\bbI\{(z_{1}^{(t+1)})_j = b\}$ can be upper bounded by 
\begin{equation}\label{ineq:step1-1}
	\begin{split}
		& \bbI\left\{(z_{1}^{(t+1)})_j = b, \left\langle \E_{j:}\V, \tilde\S_{{(z_1)}_j:} - \tilde\S_{b:} \right\rangle \leq -\frac{1}{4}\left\|\S_{(z_1)_j:} - \S_{b:}\right\|^2\right\} \\
		& \qquad + \bbI\left\{(z_{1}^{(t+1)})_j = b, \frac{1}{2}\left\|\S_{(z_1)_j:} - \S_{b:}\right\|^2 \leq F_{jb}^{(t)} + G_{jb}^{(t)} + H_{j,b}^{(t)}\right\}.
	\end{split}
\end{equation}
Recall the definition of $l_1^{(t+1)}$:
\begin{equation*}
    \begin{split}
        l_1^{(t+1)} & = \frac{1}{p_1}\left\|\cS \times_1 (\M_1^{(t+1)} - \M_1 )\right\|_\tF^2  = \frac{1}{p_1} \sum_{j=1}^{p_1}\sum_{b = 1}^{r_1}\bbI\left\{(z_{1}^{(t+1)})_j = b\right\}\|\S_{(z_1)_j:} - \S_{b:}\|^2.
    \end{split}
\end{equation*}
Then, one can take the weighted summation of \eqref{ineq:step1-1} over $j\in [p_1]$ and obtain:
	\begin{equation}\label{ineq:decompose-l(t+1)}
		\begin{split}
			l_1^{(t+1)}  & \leq \xi_1 + \frac{1}{p_1}\sum_{j=1}^{p_1}\sum_{b \in [r_1]/(z_{1})_j}\zeta_{jb}^{(t)}, \\			
			\zeta_{jb}^{(t)} := \left\|\S_{(z_1)_j:} - \S_{b:}\right\|^2 & \bbI\left\{(z_{1}^{(t+1)})_j = b,\frac{1}{2}\left\|\S_{(z_1)_j:} - \S_{b:}\right\|^2 \leq F_{jb}^{(t)} + G_{jb}^{(t)} + H_{jb}^{(t)}\right\},
		\end{split}
	\end{equation}

	\item[Step 3] In this step, we establish the following upper bounds of $F^{(t)}_{jb}$, $G^{(t)}_{jb}$ and $H^{(t)}_{jb}$:
	\begin{equation}\label{ineq-bound-F}
		\begin{split}
			\max_{j \in [p_1]}\max_{b \neq (z_1)_j}\frac{\left(F_{jb}^{(t)}\right)^2}{\left\|\S_{(z_1)_j:} - \S_{b:}\right\|^2} & \lesssim \frac{\sum_{k=1}^d r_kl_k^{(t)}}{\Delta_1^2}\left\|\E_{j:}\V\right\|^2 \\
				&  \qquad + \left\|\E_{j:}(\V - \V^{(t)})\right\|^2\left(1 + \frac{\sum_{k=1}^d r_kl_k^{(t)}}{\Delta_1^2}\right);
		\end{split}
	\end{equation}
	\begin{equation}\label{ineq-bound-G}
		\max_{j \in [p_1]}\max_{b \neq (z_1)_j}\frac{\left(G_{jb}^{(t)}\right)^2}{\left\|\S_{(z_1)_j:} - \S_{b:}\right\|^2} \leq \frac{1}{512d}\left(\Delta_1^2 + \sum_{k=1}^d l_k^{(t)} \right);
	\end{equation}
	\begin{equation}\label{ineq-bound-H}
		\max_{j \in [p_1]}\max_{b \neq (z_1)_j}\frac{\left|H_{jb}^{(t)}\right|}{\left\|\S_{(z_1)_j:} - \S_{b:}\right\|^2} \leq \frac{1}{4};
	\end{equation}
    We prove \eqref{ineq-bound-F}, \eqref{ineq-bound-G} and \eqref{ineq-bound-H} separately.
	\begin{itemize}[leftmargin=*]
		\item[(a)]  Let $b \in [r_1]$ and $b\neq (z_1)_j$. First of all, by \eqref{eq:def-F},
		\begin{equation}\label{ineq:F-decompose}
			\begin{split}
				\left(F_{jb}^{(t)}\right)^2 \leq & 8\left|\left\langle \E_{j:}\V^{(t)}, (\tilde\S_{{(z_1)}_j:} - \S_{{(z_1)}_j:}^{(t)}) - (\tilde\S_{b:} - \S_{b:}^{(t)})\right\rangle\right|^2 \\
				& + 8\left|\left\langle \E_{j:}(\V - \V^{(t)}), \tilde\S_{{(z_1)}_j:}  - \tilde\S_{b:}\right\rangle\right|^2 \\
				\leq & 32 \left\|\E_{j:}\V^{(t)}\right\|^2\cdot \max_{b \in [r_1]} \left\|\tilde\S_{b:}-\S_{b:}^{(t)}\right\|^2 \\
				& + 8\left\|\E_{j:}(\V - \V^{(t)})\right\|^2 \cdot \left\|\tilde\S_{{(z_1)}_j:}  - \tilde\S_{b:}\right\|^2 \\
				\leq & 64\left(\left\|\E_{j:}\V\right\|^2 + \left\|\E_{j:}(\V-\V^{(t)})\right\|^2\right)\cdot \max_{a \in [r_1]} \left\|\tilde\S_{a:}-\S_{a:}^{(t)}\right\|^2 \\
				& + 8\left\|\E_{j:}(\V - \V^{(t)})\right\|^2 \cdot \left\|\tilde\S_{{(z_1)}_j:}  - \tilde\S_{b:}\right\|^2.
			\end{split}
		\end{equation}
		
		By Lemma \ref{lm-1}, for any $a \in [r_1]$, 
		\begin{equation}\label{ineq:F-bound-1}
			\begin{split}
			\left\|\tilde\S_{a:} - \S_{a:}^{(t)}\right\|^2 & = \left\|\left(\W_{:a} - \W_{:a}^{(t)}\right)^\top \Y\V + \W_{:a}^{(t)\top} \Y\left(\V - \V^{(t)}\right)\right\|^2 \\
			 \leq  & 2\left\|\left(\W_{:a} - \W_{:a}^{(t)}\right)^\top \Y\V\right\|^2 + 2\left\|\W_{:a}^{(t)\top} \Y\left(\V - \V^{(t)}\right)\right\|^2 \\
			 \overset{\eqref{ineq:lm-1-1}\eqref{ineq:lm-1-2}}{\lesssim} & \frac{r_1^2}{\Delta_1^2}\left(l_1^{(t)}\right)^2 + \sum_{k=2}^d\frac{r_{k}^2}{ \Delta_k^2} \left(l_k^{(t)}\right)^2 + \frac{r_*(p_1r_1+r_*)}{p_*\Delta_1^2}l_1^{(t)} \\
			 & + \sum_{k=2}^d \frac{(\bar rr_*^2+\bar p \bar r r_*)}{p_*\Delta_k^2}l_k^{(t)} \\
			\overset{\eqref{hpc-3}}{\leq} & \left(r_1 + \frac{r_*(p_1r_1+r_*)}{p_*\Delta_1^2} \right)l_1^{(t)} + \sum_{k=2}^d\left(r_{k}+\frac{\bar rr_*^2+\bar p \bar r r_*}{p_*\Delta_k^2}\right)l_k^{(t)}  \\
			\lesssim & \sum_{k=1}^d r_kl_k^{(t)},
			\end{split}
		\end{equation}
		where the last inequality comes from the assumption that $\Delta_{\min}^2 \geq C\max\{\bar r r_*^2/p_*, \bar p \bar r r_*/p_*\}$. 
		
		In the meantime, we have
		\begin{equation}\label{ineq:F-bound-2} 
		    \begin{split}
		        \left\|\tilde{\S}_{(z_1)_j:} - \tilde \S_{b:}\right\|^2 & = \left\|\tilde\S_{(z_1)_j:} - \S_{(z_1)_j:} + \S_{(z_1)_j:} - \S_{b:} + \S_{b:} - \tilde\S_{b:}\right\|_2^2\\
		        &\leq  3\left\|\S_{(z_1)_j:} - \S_{b:}\right\|_2^2 + 6\max_{a \in [r_1]}\left\|\tilde\S_{a:} - \S_{a:}\right\|_2^2 \\
		        & = 3\left\|\S_{(z_1)_j:} - \S_{b:}\right\|_2^2 + 6\max_{a \in [r_1]}\left\|\W_{:a}^\top \E \V\right\|_2^2 \\
		        & \overset{\eqref{hpc-1-1}}{\lesssim} \left\|\S_{(z_1)_j:} - \S_{b:}\right\|_2^2 + \frac{r_*^2}{p_*} \lesssim \left\|\S_{(z_1)_j:} - \S_{b:}\right\|_2^2.
		    \end{split}
		\end{equation}  
		Here again the last inequality comes from the assumption on $\Delta_1^2$.
		Combining \eqref{ineq:F-decompose}, \eqref{ineq:F-bound-1} and \eqref{ineq:F-bound-2}, we obtain 
		\begin{equation*}
			\begin{split}
				\frac{\left(F_{jb}^{(t)}\right)^2}{\left\|\S_{(z_1)_j:} - \S_{b:}\right\|^2} & \lesssim \frac{\sum_{k=1}^d r_kl_k^{(t)}}{\Delta_1^2}\left\|\E_{j:}\V\right\|^2  + \left\|\E_{j:}(\V - \V^{(t)})\right\|^2\left(1 + \frac{\sum_{k=1}^d r_kl_k^{(t)}}{\Delta_1^2}\right),
			\end{split}
		\end{equation*}
		which proves \eqref{ineq-bound-F}.
		
		\item[(b)] By \eqref{eq:def-G},
			\begin{equation*}
				\begin{split}
					G_{jb}^{(t)} &= \left(\left\|\X_{j:}\V^{(t)} - \S_{(z_1)_j:}^{(t)}\right\|^2 - \left\|\X_{j:}\V^{(t)} - \W_{:(z_1)_j}^{\top}\Y\V^{(t)}\right\|^2\right) \\
					& \qquad - \left(\left\|\X_{j:}\V^{(t)} - \S_{b:}^{(t)}\right\|^2 - \left\|\X_{j:}\V^{(t)} - \W_{:b}^{\top}\Y\V^{(t)}\right\|^2\right) \\
					& = \left(\left\|\W_{:(z_1)_j}^{\top}\Y\V^{(t)} - \S_{(z_1)_j:}^{(t)}\right\|^2 - \left\|\W_{:b}^{\top}\Y\V^{(t)} - \S_{b:}^{(t)}\right\|^2\right) \\
					& \qquad + 2\left\langle \X_{j:}\V^{(t)} - \W_{:(z_1)_j}^{\top}\Y\V^{(t)}, \W_{:(z_1)_j}^{\top}\Y\V^{(t)} - \S_{(z_1)_j:}^{(t)} \right\rangle \\
					& \qquad - 2\left\langle \X_{j:}\V^{(t)} - \W_{:b}^{\top}\Y\V^{(t)}, \W_{:b}^{\top}\Y\V^{(t)} - \S_{b:}^{(t)} \right\rangle.
 				\end{split}
			\end{equation*}
			Let $\N := \M_2 \otimes \cdots \otimes \M_d$. Noticing $\X_{j:} = \W_{:(z_1)_j}^{\top}\X = \S_{(z_1)_j:}\N^{\top}$, we further have
			\begin{equation}\label{ineq:G-decompose}
				\begin{split}
				\left|G_{jb}^{(t)}\right| \leq & \left|\left\|\W_{:(z_1)_j}^{\top}\Y\V^{(t)} - \S_{(z_1)_j:}^{(t)}\right\|^2 - \left\|\W_{:b}^{\top}\Y\V^{(t)} - \S_{b:}^{(t)}\right\|^2\right| \\
				& \qquad + 4 \max_{a \in [r_1]} \left|\left\langle \W_{:a}^{\top}\E\V^{(t)}, (\W_{:a} - \W_{:a}^{(t)})^\top \Y\V^{(t)} \right\rangle\right| \\
				& \qquad + 2\left|\left\langle (\S_{(z_1)_j:} - \S_{b:})\N^{\top}\V^{(t)}, (\W_{:b}-\W_{:b}^{(t)})^\top \Y\V^{(t)} \right\rangle\right|.
				\end{split}
			\end{equation}
			We analyze the three terms in \eqref{ineq:G-decompose} separately. 
			\begin{itemize}
			    \item Firstly, by Lemma \ref{lm-1}, 
			\begin{equation}\label{ineq:G-1}
				\begin{split}
					& \left|\left\|\W_{:(z_1)_j}^{\top}\Y\V^{(t)} - \S_{(z_1)_j:}^{(t)}\right\|^2 - \left\|\W_{:b}^{\top}\Y\V^{(t)} - \S_{b:}^{(t)}\right\|^2\right|^2 \\
					\leq & \max_{a \in [r_1]} \left\|\W_{:a}^{\top}\Y\V^{(t)} - \S_{a:}^{(t)}\right\|^4 \\
					= & \max_{a \in [r_1]} \left\|\left(\W_{:a} - \W_{:a}^{(t)}\right)^\top \Y\V^{(t)}\right\|^4 \\
					\overset{\eqref{ineq:lm-1-3}}{\leq} & C\left(\frac{r_1^4}{\Delta_1^4}\left(l_1^{(t)}\right)^4 + \sum_{k=1}^d \frac{\bar r^2r_*^4 + r_*^2\bar p^2 \bar r^4}{p_*^2}\frac{(l_k^{(t)})^2}{\Delta_k^4 }\right) \\
					= & C\left(\frac{r_1^4}{\Delta_1^4}\left(l_1^{(t)}\right)^4 + \sum_{k=1}^d \frac{\bar r^2r_*^4 + r_*^2\bar p^2 \bar r^4}{p_*^2}\frac{(l_k^{(t)})^2}{\Delta_k^4 \Delta_1^2}\cdot\Delta_1^2\right)\\
					\leq & c\left(\Delta_1^4 + \sum_{k=1}^d \Delta_1^2 l_k^{(t)} \right).
				\end{split}
			\end{equation}
			Here, we use \eqref{hpc-3} and the assumption $\Delta_{\min}^2 \geq C\max\{\bar r r_*^2 /p_*,  \bar p\bar r^2 r_*/p_*\}$  to obtain the last inequality.
			    \item Next,
			\begin{equation}\label{ineq:G-2}
				\begin{split}
					& \max_a \left|\left\langle \W_{:a}^{\top}\E\V^{(t)}, (\W_{:a} - \W_{:a}^{(t)})^\top \Y\V^{(t)} \right\rangle\right|^2 \\
					\leq & \max_a\left\|\W_{:a}^{\top}\E\V^{(t)}\right\|^2 \cdot \max_a\left\|(\W_{:a} - \W_{:a}^{(t)})^\top \Y\V^{(t)}\right\|^2 \\
					\overset{\eqref{ineq:lm-1-3}}{\leq} & C \max_a \|\W_{:a}\|^2 \cdot \|\E\V^{(t)}\| \cdot \left(\frac{r_1^2}{\Delta_1^2}\left(l_1^{(t)}\right)^2 + \sum_{k=1}^d \frac{\bar r r_*^2 + \bar p \bar r^2 r_* }{p_*}\frac{l_k^{(t)}}{\Delta_k^2}\right) \\
					\overset{\text{Lemma \ref{lm:mebmership-spectra}},~\eqref{hpc-1-2}}{\leq} & C\frac{(r_{-1}+\sum_{k}p_kr_k)r_*}{p_*}\left(\frac{r_1^2}{\Delta_1^2}\left(l_1^{(t)}\right)^2 + \sum_{k=1}^d \frac{\bar r r_*^2 + \bar p \bar r^2 r_*}{p_*}\frac{l_k^{(t)}}{\Delta_k^2}\right) \\
					\leq &  c\sum_{k=1}^d \Delta_1^2l_k^{(t)}.
				\end{split}
			\end{equation}
			\item For the last term, we have:
				\begin{equation}\label{ineq:G-3}
					\begin{split}
						&\left|\left\langle (\S_{(z_1)_j:} - \S_{b:})\N^{\top}\V^{(t)}, (\W_{:b}-\W_{:b}^{(t)})^\top \Y\V^{(t)} \right\rangle\right|^2 \\
						\leq  &\left\|(\S_{(z_1)_j:} - \S_{b:})\N^{\top}\V^{(t)}\right\|_2^2 \cdot \left\|(\W_{:b}-\W_{:b}^{(t)})^\top \Y\V^{(t)}\right\|_2^2 \\
						\leq & \left\|\S_{(z_1)_j:} - \S_{b:}\right\|_2^2 \cdot \left\|\N\right\|^2 \cdot\left\|\V^{(t)}\right\|^2 \cdot \left\|(\W_{:b}-\W_{:b}^{(t)})^\top \Y\V^{(t)}\right\|_2^2 \\
						\leq & C \left\|\S_{(z_1)_j:} - \S_{b:}\right\|_2^2 \cdot \left\|(\W_{:b}-\W_{:b}^{(t)})^\top \Y\V^{(t)}\right\|_2^2 \\
						\overset{\eqref{ineq:lm-1-3}}{\leq} & C \left\|\S_{(z_1)_j:} - \S_{b:}\right\|_2^2 \cdot \left(\frac{r_1^2}{\Delta_1^2}\left(l_1^{(t)}\right)^2 + \sum_{k=1}^d \frac{\bar r r_*^2 + \bar p \bar r^2 r_*}{p_*}\frac{l_k^{(t)}}{\Delta_k^2}\right) \\
						\overset{\eqref{hpc-3}}{\leq} & \left\|\S_{(z_1)_j:} - \S_{b:}\right\|_2^2 \cdot c\left( \Delta_1^2 + \sum_{k=1}^d l_k^{(t)}\right).
					\end{split}
				\end{equation}
			\end{itemize}
			Combining \eqref{ineq:G-decompose}, \eqref{ineq:G-1}, \eqref{ineq:G-2} and \eqref{ineq:G-3} and applying Inequality of arithmetic and geometric means, we obtain
				\begin{equation*}
					\begin{split}
						\frac{\left(G^{(t)}_{jb}\right)^2}{\left\|\S_{(z_1)_j:} - \S_{b:}\right\|^2} \leq c\left(\Delta_1^2 + \sum_{k=1}^d l_k^{(t)}\right).
					\end{split}
				\end{equation*}
				and finish the proof of \eqref{ineq-bound-G}.
				
		\item[(c)] Recall the definition of $H_{jb}^{(t)}$:
			\begin{equation}\label{eq:H-decompose}
					\begin{split}
						H_{jb}^{(t)} &:= \left\|\X_{j:}\V^{(t)} - \W_{:{(z_1)}_j}^{\top}\Y\V^{(t)}\right\|_2^2   - \left\|\X_{j:}\V^{(t)} - \W_{:b}^{\top}\Y\V^{(t)}\right\|_2^2  + \left\|\S_{{(z_1)}_j:} - \S_{b:}\right\|^2 \\
						& = \left\|\W_{:{(z_1)}_j}^{\top}\E\V^{(t)}\right\|^2 + \left(\left\|\S_{{(z_1)}_j:} - \S_{b:}\right\|^2 - \left\|\X_{j:}\V^{(t)} - \W_{:b}^{\top}\X\V^{(t)}\right\|^2\right) \\
						& \qquad - \left(\left\|\X_{j:}\V^{(t)} - \W_{:b}^{\top}\Y\V^{(t)}\right\|^2 - \left\|\X_{j:}\V^{(t)} - \W_{:b}^{\top}\X\V^{(t)}\right\|^2\right) \\
						& = \left(\left\|\S_{{(z_1)}_j:} - \S_{b:}\right\|^2 - \left\|\X_{j:}\V^{(t)} - \W_{:b}^{\top}\X\V^{(t)}\right\|^2\right) \\
						& \qquad + \left(\left\|\W_{:{(z_1)}_j}^{\top}\E\V^{(t)}\right\|^2 - \left\|\W_{:b}^{\top}\E\V^{(t)}\right\|^2\right)  \\
						& \qquad + 2\left\langle (\S_{(z_1)_j:}-\S_{b:})\N^{\top}\V^{(t)}, \W_{:b}^{\top}\E\V^{(t)}\right\rangle.
					\end{split}				
			\end{equation}
		Again, we analyze the three terms in \eqref{eq:H-decompose} separately. 
		\begin{itemize}
		    \item Firstly,
		\begin{equation}\label{ineq:H-1-1}
			\begin{split}
				& \left|\left\|\S_{{(z_1)}_j:} - \S_{b:}\right\|^2 - \left\|\X_{j:}\V^{(t)} - \W_{:b}^{\top}\X\V^{(t)}\right\|^2\right| \\
				= & \left|\left\|\S_{{(z_1)}_j:} - \S_{b:}\right\|^2 - \left\|\left(\S_{{(z_1)}_j:} - \S_{b:}\right)\N^{\top}\V^{(t)}\right\|^2\right|.
			\end{split}
		\end{equation}
		By the same argument as \eqref{ineq:lm-1-0-2} in the proof of Lemma \ref{lm-1}, we have 
		\begin{equation*}
		    \left\|\N^\top \V^{(t)} - \I\right\| \leq C\sum_{k=2}^d \left\|\M_k^\top\W_k^{(t)} - \I\right\| \leq C\sum_{k=2}^d \frac{r_k l_k^{(t)}}{\Delta_k^2} \overset{\eqref{hpc-3}}{\leq} c
		\end{equation*}
		for some small constant $c$. Then,
		\begin{equation}\label{ineq:H-1-2}
			\begin{split}
			\left\|\N^{\top}\V^{(t)}\right\| \leq 1 + \left\|\I - \N^{\top}\V^{(t)}\right\| \leq 1 + \left\|\I - \N^{\top}\V^{(t)}\right\|_\tF \leq 1+c \\
			\sigma_{r_{-1}}\left(\N^{\top}\V^{(t)}\right) \geq 1 - \left\|\I - \N^{\top}\V^{(t)}\right\| \geq 1 - \left\|\I - \N^{\top}\V^{(t)}\right\|_\tF \geq 1-c.
			\end{split}
		\end{equation}	
		Combining \eqref{ineq:H-1-1} and \eqref{ineq:H-1-2}, we obtain
		\begin{equation}\label{ineq:H-1}
			\begin{split}
				\left|\left\|\S_{{(z_1)}_j:} - \S_{b:}\right\|^2 - \left\|\left(\S_{{(z_1)}_j:} - \S_{b:}\right)\N^{\top}\V^{(t)}\right\|^2\right| & \leq (2c+c^2)  \left\|\S_{{(z_1)}_j:} - \S_{b:}\right\|^2 \\ 
				& \leq \frac{1}{12}\left\|\S_{{(z_1)}_j:} - \S_{b:}\right\|^2.
			\end{split}
		\end{equation}
		\item Next,
		\begin{equation}\label{ineq:H-2}
			\begin{split}
				\left|\left\|\W_{:{(z_1)}_j}^{\top}\E\V^{(t)}\right\|^2 - \left\|\W_{:b}^{\top}\E\V^{(t)}\right\|^2\right| & \leq \max_{a\in [r_1]}\left\|\W_{:a}^{\top}\E\V^{(t)}\right\|^2 \\
				& \overset{\eqref{hpc-1-2}}{\leq} C\frac{(r_{-1}+\sum_{k}p_kr_k)r_*}{p_*} \\
				& \leq C\frac{(r_{-1}+\sum_{k}p_kr_k)r_*}{p_*\Delta_1^2} \left\|\S_{(z_1)_j:} - \S_{b:}\right\|^2 \\
				& \leq \frac{1}{12}\left\|\S_{(z_1)_j:} - \S_{b:}\right\|^2.
			\end{split}
		\end{equation}		
		\item Finally, 
		\begin{equation}\label{ineq:H-3}
			\begin{split}
				\left|\left\langle (\S_{{(z_1)}_j:} - \S_{b:})\N^{\top}\V^{(t)}, \W_{:b}^{\top} \E \V^{(t)} \right\rangle\right| & \leq \left\|(\S_{{(z_1)}_j:} - \S_{b:})\N^{\top}\V^{(t)}\right\| \cdot \left\|\W_{:b}^{\top}\E\V^{(t)}\right\| \\
				& \overset{\eqref{hpc-1-2}}{\leq}  C\frac{\sqrt{r}(\sqrt{r_{-1}}+\sum_k\sqrt{p_kr_k})}{\sqrt{p_{-1}}}\left\|\S_{{(z_1)}_j:} - \S_{b:}\right\| \\
				& \leq c \Delta_1 \left\|\S_{{(z_1)}_j:} - \S_{b:}\right\| \\
				& \leq \frac{1}{24}\left\|\S_{{(z_1)}_j:} - \S_{b:}\right\|^2.
			\end{split}
		\end{equation}
		\end{itemize}
		Now \eqref{ineq-bound-H} follows by \eqref{eq:H-decompose}, \eqref{ineq:H-1}, \eqref{ineq:H-2} and \eqref{ineq:H-3}.
	\end{itemize}

	\item[Step 4] In this step, we combine \eqref{ineq-bound-F}, \eqref{ineq-bound-G} and \eqref{ineq-bound-H} to obtain the error contraction from $l_1^{(t)}$ to $l_1^{(t+1)}$. By the definition of $\zeta_{jb}^{(t)}$, we have
	\begin{equation}\label{ineq:l(t+1)-1}
		\begin{split}
			\zeta_{jb}^{(t)} & = \left\|\S_{(z_1)_j:} - \S_{b:}\right\|^2\bbI\left\{(z_{1}^{(t+1)})_j = b,\frac{1}{2}\left\|\S_{(z_1)_j:} - \S_{b:}\right\|^2 \leq F_{jb}^{(t)} + G_{jb}^{(t)} + H_{jb}^{(t)}\right\} \\
			& \overset{\eqref{ineq-bound-H}}{\leq}   \left\|\S_{(z_1)_j:} - \S_{b:}\right\|^2\bbI\left\{(z_{1}^{(t+1)})_j = b,\frac{1}{4}\left\|\S_{(z_1)_j:} - \S_{b:}\right\|^2 \leq F_{jb}^{(t)} + G_{jb}^{(t)} \right\} \\
			& \leq \bbI\left\{(z_{1}^{(t+1)})_j = b\right\}\cdot 64\left( \frac{(F_{jb}^{(t)})^2}{\left\|\S_{(z_1)_j:} - \S_{b:}\right\|^2} + \frac{(G_{jb}^{(t)})^2}{\left\|\S_{(z_1)_j:} - \S_{b:}\right\|^2}\right).
		\end{split}
	\end{equation}
	We first calculate the summation over $(F_{jb}^{(t)})^2$,
	\begin{equation}\label{ineq:sum-F}
		\begin{split}
			& \frac{1}{p_1}\sum_{j=1}^{p_1}\sum_{b\in [r_1]/(z_1)_j} \bbI\left\{(z_{1}^{(t+1)})_j = b\right\}\frac{\left(F_{jb}^{(t)}\right)^2}{\left\|\S_{z_{1j}:} - \S_{b:}\right\|^2} \\
			\leq & \frac{1}{p_1} \sum_{j=1}^{p_1}\max_{b\in [r_1]/(z_1)_j}\frac{\left(F_{jb}^{(t)}\right)^2}{\left\|\S_{z_{1j}:} - \S_{b:}\right\|^2} \\
			\overset{\eqref{ineq-bound-F}}{\lesssim} & \frac{1}{p_1} \sum_{j=1}^{p_1} \frac{\sum_{k=1}^d r_kl_k^{(t)}}{\Delta_1^2}\left\|\E_{j:}\V\right\|^2  + \frac{1}{p_1} \sum_{j=1}^{p_1}\left\|\E_{j:}(\V - \V^{(t)})\right\|^2\left(1 + \frac{\sum_{k=1}^d r_kl_k^{(t)}}{\Delta_1^2}\right) \\
			= &  \frac{\sum_{k=1}^d r_kl_k^{(t)}}{p_1\Delta_1^2}\left\|\E\V\right\|_\tF^2  + \frac{1}{p_1} \left\|\E(\V - \V^{(t)})\right\|_\tF^2\left(1 + \frac{\sum_{k=1}^d r_kl_k^{(t)}}{\Delta_1^2}\right) .
		\end{split}
	\end{equation}
	We bound the two terms in \eqref{ineq:sum-F} separately. By the condition $\Delta_1^2 \geq C\bar p \bar r r_*^2 /p_*$,
	\begin{equation*}
		\begin{split}
			\frac{\sum_{k=1}^d r_kl_k^{(t)}}{p_1\Delta_1^2}\left\|\E\V\right\|_\tF^2 &  \overset{\eqref{hpc-1-1}}{\leq} C\frac{\sum_{k=1}^d r_kl_k^{(t)}}{p_1\Delta_1^2}\frac{p_1r_{-1}^2}{p_{-1}} \\
			& = C \frac{\sum_{k=1}^d r_{-1}^2 r_k l_k^{(t)}}{p_{-1}\Delta_1^2}  \leq\frac{1}{1024d}\sum_{k=1}^d l_k^{(t)},
		\end{split}
	\end{equation*}
	\begin{equation*}
		\begin{split}
			& \frac{1}{p_1} \sum_{j=1}^{p_1}\left\|\E_{j:}(\V - \V^{(t)})\right\|^2\left(1 + \frac{\sum_{k=1}^d r_kl_k^{(t)}}{\Delta_1^2}\right) \\
			\leq & \frac{1}{p_1} \left\|\E(\V - \V^{(t)})\right\|_\tF^2\left(1 + \frac{\sum_{k=1}^d r_kl_k^{(t)}}{\Delta_1^2}\right) \\
			\overset{\eqref{ineq:lm-1-0.5}}{\leq} & C \frac{r_{-1}(p_1r_{-1}+\bar p \bar r)}{p_{*}} \sum_{k=2}^d \frac{r_k^2(l_k^{(t)})^2}{\Delta_k^4}\cdot\left(1 + \frac{\sum_{k=1}^d r_kl_k^{(t)}}{\Delta_1^2}\right) \\
			\overset{\eqref{hpc-3}}{\leq} & C \frac{r_{-1}(p_1r_{-1}+\bar p \bar r)}{p_{*}} \sum_{k=2}^d \frac{r_kl_k^{(t)}}{\Delta_k^2}\cdot\left(1 + \frac{\sum_{k=1}^d r_kl_k^{(t)}}{\Delta_1^2}\right) \\
			\overset{\eqref{hpc-3}}{\leq} & C \left(\sum_{k=2}^d\frac{p_1\bar r r_*^2 + \bar p \bar r^2 r_*}{p_*\Delta_k^2}l_k^{(t)} +  \frac{(p_1r_{-1}^2+r_{-1}\bar p \bar r)\sum_{k=1}^d r_kl_k^{(t)}}{p_*\Delta_1^2}\right)\\
			 \leq& \frac{1}{1024d}\sum_{k=1}^d l_k^{(t)}; 
		\end{split}
	\end{equation*}
	Thus,
	\begin{equation}\label{ineq:sum-F-bound}
		\begin{split}
			\frac{1}{p_1}\sum_{j=1}^{p_1}\sum_{b\in [r_1]/(z_1)_j} \bbI\left\{(z_{1}^{(t+1)})_j = b\right\}\frac{\left(F_{jb}^{(t)}\right)^2}{\left\|\S_{z_{1j}:} - \S_{b:}\right\|^2} \leq \frac{1}{512d}\sum_{k=1}^d l_k^{(t)}.
		\end{split}
	\end{equation}
	Next, we analyze the summation over $(G_{jb}^{(t)})^2$.
	\begin{equation}\label{ineq:sum-G}
		\begin{split}
			& \frac{1}{p_1}\sum_{j=1}^{p_1}\sum_{b \in [r_1]/(z_{1})_j} \bbI\left\{(z_{1}^{(t+1)})_j = b\right\}\frac{\left(G_{jb}^{(t)}\right)^2}{\left\|\S_{(z_{1})_j:} - \S_{b:}\right\|^2} \\
			\leq & \frac{1}{p_1}\sum_{j=1}^{p_1}\bbI\left\{(z_{1}^{(t+1)})_j \neq (z_{1})_j\right\} \max_{b \in [r_1]/(z_{1})_j} \frac{\left(G_{jb}^{(t)}\right)^2}{\left\|\S_{z_{1j}:} - \S_{b:}\right\|^2} \\
			\overset{\eqref{ineq-bound-G}}{\leq} & \frac{1}{p_1}\sum_{j=1}^{p_1}\bbI\left\{(z_{1}^{(t+1)})_j \neq (z_{1})_j\right\} \frac{1}{512d}\left(\Delta_1^2 + \sum_{k=1}^dl_k^{(t)}\right) \\
			\overset{\text{Lemma \ref{lm:loss-relation}}}{\leq} & \frac{1}{512d}\left(l_1^{(t+1)} +\sum_{k=1}^d l_k^{(t)}\right).
		\end{split}
	\end{equation}
	Now by combining \eqref{ineq:decompose-l(t+1)}, \eqref{ineq:l(t+1)-1}, \eqref{ineq:sum-F-bound} and \eqref{ineq:sum-G}, we obtain
	\begin{equation*}
	    \begin{split}
	        l^{(t+1)}_1 & \leq \xi_1 + \frac{1}{8}l_1^{(t+1)} + \frac{1}{4d}\sum_{k=1}^d l_k^{(t)}. \\
	        & \leq \xi_1 + \frac{1}{8}l_1^{(t+1)} + \frac{1}{4}\max_{k\in [d]} l_k^{(t)}.
	    \end{split}
	\end{equation*}
	which implies 
	\begin{equation}\label{ineq:local-contraction}
		l^{(t+1)}_1 \leq \frac{3}{2}\xi_1 + \frac{1}{2}\max_{k\in [d]}l_k^{(t)}.
	\end{equation}
	
	\item[Step 5] In this step, we prove that conditions \eqref{hpc-1-1}-\eqref{hpc-3} hold with high probability and establish the upper bound for $l_1^{(t)}$ for arbitrary $t \geq 0$. By sub-Gaussian concentration (Lemma \ref{lm:concentration-ineq}), \eqref{hpc-1-1}, \eqref{hpc-1-2} and \eqref{hpc-1-3} holds with probability at least $1-\exp(-c\underline p)$. Now we focus on proving \eqref{hpc-2} (and we only need to prove for $k=1$). Recall
	\begin{equation*}
		\begin{split}
			\xi_1 := \frac{1}{p_1}\sum_{j=1}^{p_1}\sum_{b \in [r_1]/(z_1)_j} &\bbI\left\{\left\langle \E_{j:}\V, \tilde\S_{{(z_1)}_j:} - \tilde\S_{b:} \right\rangle \leq -\frac{1}{4}\left\|\S_{(z_1)_j:} - \S_{b:}\right\|^2\right\} \\
			& \cdot \left\|\S_{{(z_1)}_j:} - \S_{b:}\right\|^2.
		\end{split}
	\end{equation*}
	We denote $e_j = \E_{j:} \V$, then one can see that $e_j$ are independent mean-zero sub-Gaussian random vectors in $\bbR^{r_{-1}}$. Moreover, each coordinate of $e_j$ are independent with sub-Gaussian norm bounded by $C_0\sqrt{\frac{r_{-1}}{p_{-1}}}$ for some constant $C_0$ which only depends on $\alpha,\beta$. 

	Now notice that
	\begin{equation}\label{ineq:delta-event-decompose}
		\begin{split}
			& \bbP\left(\left\langle e_j, \tilde\S_{{(z_1)}_j:} - \tilde\S_{b:} \right\rangle \leq -\frac{1}{4}\left\|\S_{(z_1)_j:} - \S_{b:}\right\|^2\right) \\
			\leq & \bbP\left(\left\langle e_j, \S_{{(z_1)}_j:} - \S_{b:} \right\rangle \leq -\frac{1}{8}\left\|\S_{(z_1)_j:} - \S_{b:}\right\|^2\right)\\
			& \quad + \bbP\left(\left\langle e_j, \tilde\S_{{(z_1)}_j:} - \S_{(z_1)_j:} \right\rangle \leq -\frac{1}{16}\left\|\S_{(z_1)_j:} - \S_{b:}\right\|^2\right)\\
			& \quad + \bbP\left(\left\langle e_j, \S_{b:} - \tilde\S_{b:} \right\rangle \leq -\frac{1}{16}\left\|\S_{(z_1)_j:} - \S_{b:}\right\|^2\right).
		\end{split}
	\end{equation}
	The following Lemma can be used to bound the three terms in \eqref{ineq:delta-event-decompose} separately. 
    \begin{Lemma}\label{lm-2}
	Under the same notations and conditions of Theorem \ref{thm:HLloyd}, there exist universal constants $c$ such that
			\begin{equation}\label{lm-2-1}
				\begin{split}
					& \bbP\left(\left\langle e_j, \S_{{(z_1)}_j:} - \S_{b:} \right\rangle \leq -\frac{1}{8}\left\|\S_{(z_1)_j:} - \S_{b:}\right\|^2\right)
					\leq \exp\left(-\frac{cp_{-1}}{r_1}\left\|\S_{(z_1)_j:} - \S_{b:}\right\|^2\right),
				\end{split}
			\end{equation} 
			\begin{equation}\label{lm-2-2}
				\begin{split}
					& \bbP\left(\left\langle e_j, \tilde\S_{{(z_1)}_j:} - \S_{(z_1)_j:} \right\rangle \leq -\frac{1}{16}\left\|\S_{(z_1)_j:} - \S_{b:}\right\|^2\right) \leq 2\exp\left(-\frac{c p_{-1}}{r_{-1}}\left\|\S_{(z_1)_j:} - \S_{b:}\right\|^2\right),
				\end{split}
			\end{equation} 
			and
			\begin{equation}\label{lm-2-3}
				\begin{split}
					\bbP\left(\left\langle e_j, \S_{b:} - \tilde\S_{b:} \right\rangle \leq -\frac{1}{16}\left\|\S_{(z_1)_j:} - \tilde \S_{b:}\right\|^2\right) \leq 2\exp\left(-\frac{c p_{-1}}{r_{-1}}\left\|\S_{(z_1)_j:} - \S_{b:}\right\|^2\right).
				\end{split}
			\end{equation} 
	\end{Lemma}
	\begin{proof}[Proof of Lemma \ref{lm-2}]
	    See Section \ref{sec:lm-2}.
	\end{proof}
	With Lemma \ref{lm-2}, we can bound
	\begin{equation*}
		\begin{split}
			\bbE \xi_1 & = \frac{1}{p_1}\sum_{j=1}^{p_1}\sum_{b\in [r_1]/(z_1)_j}\left\|\S_{b:}-\S_{(z_1)_j:}\right\|^2\bbP\left(\left\langle e_j, \tilde\S_{{(z_1)}_j:} - \tilde\S_{b:} \right\rangle \leq -\frac{1}{4}\left\|\S_{(z_1)_j:} - \S_{b:}\right\|^2\right) \\
			& \leq \frac{5}{p_1}\sum_{j=1}^{p_1}\sum_{b\in [r_1]/(z_1)_j}\left\|\S_{b:}-\S_{(z_1)_j:}\right\|^2\exp\left(-\frac{c p_{-1}}{r_{-1}}\left\|\S_{b:}-\S_{(z_1)_j:}\right\|^2\right) \\
			& \leq \exp\left(-\frac{c p_{-1}}{2r_{-1}}\left\|\S_{b:}-\S_{(z_1)_j:}\right\|^2\right) \leq \exp\left(-\frac{c p_{-1}}{2r_{-1}}\Delta_1^2\right).
		\end{split}
	\end{equation*}
	Then, by Markov inequality,
	\begin{equation*}
	    \begin{split}
	        &\bbP\left(\xi_1 \leq \bbE\xi_1\exp\left(\frac{c p_{-1}}{4r_{-1}}\Delta_1^2\right)\right) \geq 1 - \exp\left(-\frac{c p_{-1}}{4r_{-1}}\Delta_1^2\right) \geq 1-  \exp\left(-\frac{c p_*}{4r_{*}\bar p}\Delta_{\min}^2\right).
	    \end{split}
	\end{equation*}
	Since the same argument holds for each mode, we know that with probability at least $1-  \exp\left(-\frac{c p_*}{4r_{*}\bar p}\Delta_{\min}^2\right)$ that \eqref{hpc-2} holds.
	
	Finally we use induction to prove \eqref{hpc-3} holds under given \eqref{hpc-1-1}-\eqref{hpc-2}. By the initialization condition, we know that
	\begin{equation*}
	    l_k^{(0)} \leq c_2\frac{\Delta_{\min}^2}{r_k}
	\end{equation*}
	and \eqref{hpc-3} holds at $t = 0$. Now suppose it also holds for all $t \leq t_0$. Then by \eqref{ineq:local-contraction},
	\begin{equation*}
	    \begin{split}
	        l^{(t_0+1)}_k & \leq \frac{3}{2}\xi_k + \frac{1}{2}\max_{k\in [d]}l_k^{(t_0)} \\
	        & \overset{\eqref{hpc-2}}{\leq} \exp\left(-\frac{c_1p_{-k}}{r_{-k}}\Delta_k^2\right) + \frac{c_2\Delta_{\min}^2}{2r_k}.
	    \end{split}
	\end{equation*}
	Since $\Delta_k^2 \geq \Delta_{\min}^2 \geq C\frac{\bar p r_*^2\bar r\log \bar p }{p_{*}}$, we can bound
	\begin{equation*}
	    \exp\left(-\frac{c_2p_{-k}}{r_{-k}}\Delta_k^2\right) \leq \frac{c_2\Delta_{\min}^2}{2r_k}.
	\end{equation*}
	Thus by induction \eqref{hpc-3} also holds at step $t_0+1$ and \eqref{hpc-3} is proved.
	
	Finally, by taking induction on \eqref{ineq:local-contraction}, we have 
	\begin{equation*}
	    \begin{split}
	        l_k^{(t)} \overset{\text{Lemma \ref{lm:loss-relation}}}{\leq} 3\xi_k + \frac{1}{2^t}\max_{k' \in [d]}l_k^{(0)} & \overset{\eqref{hpc-2},\eqref{hpc-3}}{\leq} \exp\left(-\frac{c_1 p_*}{r_*\bar p}\Delta_{\min}^2\right) + \frac{c_2\Delta_{\min}^2}{2^t} 
	    \end{split}
	\end{equation*}
	and the proof is completed.
\end{enumerate} 
\end{proof}

\section{Additional Proofs}\label{sec:proofs-other}
We collect the proofs of Theorem \ref{thm:tensor-est-hp} (Estimation error bound), Theorem \ref{thm:mcr-lower-bound} (Statistical lower bound), Theorem \ref{thm:computational-lower-bound} (Computational lower bound) and Lemma \ref{lm:new-perturbation-subGaussian} (Perturbation bound on subspaces of different dimensions) in this section.

\subsection{Proof of Theorem \ref{thm:tensor-est-hp}}
Recall $\W_k := \M_k\left(\diag\left(1_{p_k}^\top \M_k\right)\right)^{-1}$ is the weighted membership matrix. Let $\hat \cS = \cY \times_1 \W_1^{\top} \times \cdots \times_d \W_d^{\top}$.  We define the following two events:
\begin{equation*}
    \begin{split}
        A_1 = \left\{z_k^{(T)} = z_k, k=1,\ldots,d\right\}, \\
        A_2 = \left\{\|\hat\cS - \cS\|_\tF^2 \leq C\frac{r^2_*\sigma^2}{p_*}\right\}.
    \end{split}
\end{equation*}
First, by Theorem \ref{coro:exact-recovery}, $A_1$ holds up to some permutations $\{\pi_k\}_{k=1}^d$ with probability at least $1-\exp(-c\underline p)-\exp\left(-\frac{c p_*}{4r_{*}\bar p}\frac{\Delta_{\min}^2}{\sigma^2}\right)$. We assume the permutations are identity without loss of generality. Then conditional on $A_1$,
\begin{equation*}
    \begin{split}
        \|\hat\cS - \cS\|_\tF^2 & = \|\cE \times_1 \W_1^{\top} \times \cdots \times_d \W_d^{\top}\|_\tF^2.
    \end{split}
\end{equation*}
Note that $\cE \times_1 \U_1^{\top} \times \cdots \times \W_d^{\top}$ is a $r_1\times \cdots \times r_d$ random tensors with independent mean-zero entries whose sub-Gaussian norm is bounded by $C\sqrt{r_*/p_*}$, By Bernstein inequality, we know that with probability at least $1-\exp(-cr_*)$,  
\begin{equation*}
    \|\hat\cS - \cS\|_\tF^2 \leq  C\frac{r_*^2\sigma^2}{p_*}.
\end{equation*}
Here the last inequality comes from the fact that $\|\W_k\|^2 \lesssim \frac{r_k}{p_k}$.
In other words, $\bbP\left(A_2\right) \geq 1- \exp(-cr)$. By union bound, $\bbP(A_1 \cap A_2) \geq 1-\exp(-cr_*)-\exp(-c\underline p) - \exp\left(-\frac{c p_*}{4r_{*}\bar p}\Delta_{\min}^2\right)$.
Under $A_1 \cap A_2$, we have
\begin{equation*}
    \begin{split}
        \left\|\hat \cX - \cX\right\|_\tF^2 & = \left\|(\hat \cS - \cS)\times_1 \M_1 \times \cdots \times_d \M_d\right\|_\tF^2 \\
        & \leq \left\|\hat{\cS} - \cS\right\|_\tF^2 \cdot\prod_{k\in [d]}\|\M_k\|^2 \leq Cr_*\sigma^2.
    \end{split}
\end{equation*}
Thus we finish the proof. \qquad\qquad $\square$

\subsection{Proof of Theorem \ref{thm:mcr-lower-bound}}
We adopt the proof idea from \cite{gao2018community}. Without loss of generality, we only need to establish the estimation risk lower bound for mode $1$. For any specific $z_1 \in [r_1]^{p_1}$, We define $n_a(z_1) = \sum_{j=1}^{p_1}\bbI\{(z_1)_j = a\}$ for each $a \in [r_1]$. We first construct a $z_1^* \in [r_1]^{p_1}$ such that 
\begin{equation*}
    \lceil \frac{\alpha p_1}{r_1} \rceil = n_1(z_1^*) = n_2(z_1^*) \leq n_3(z_1^*) \leq \cdots \leq n_{r_1}(z_1^*).
\end{equation*}
In the meantime, we construct $z_2^* \in [r_2]^{p_2},\ldots,z_d^* \in [r_d]^{p_d}$ such that \eqref{ineq:asmp-balance} is satisfied. We also specify a $\cS^*$ such that
$\left\|\left(\cM_1(\cS^*)\right)_{1:} - \left(\cM_1(\cS^*)\right)_{2:}\right\|_2 = \Delta_1(\cS^*) = \Delta_1$ and $\Delta_k(\cS) \geq \Delta_k$ for all $k =2,\ldots,d$. Therefore, we have $(\cS^*, z_1^*, z_2^*, \ldots ,z_d^*) \in \Theta\left(\{\Delta_k\}_{k=1}^d, \alpha, \beta\right)$.

Now for each $a \in [r_1]$, let $T_a$ be a fixed subset of $\{j \in [p_1]: (z_1^*)_j = a\}$ with cardinality $|T_a| = \lceil n_a(z_1^*) - \frac{\alpha p_1}{4r_1^2} \rceil$. Define $T := \cup_{a \in [r_1]}T_a$ and 
\begin{equation*}
    \cZ_T := \left\{z_1: z_1 \text{ satisfies \eqref{ineq:asmp-balance} and } (z_1)_j = (z_1^*)_j \text{ for all } j \in T \right\}.
\end{equation*}
Consider any $z_1 \neq \tilde z_1 \in \cZ_T$. We firstly have  
\begin{equation*}\label{ineq:mcr-lb-permutation-1}
    \frac{1}{p_1} \sum_{j=1}^{p_1}\bbI\left\{(z_1)_j \neq (\tilde z_1)_j\right\} \leq \frac{1}{p_1}|T^c| \leq  \frac{r_1}{p_1}\frac{\alpha p_1}{4r_1^2} = \frac{\alpha}{4r_1}.
\end{equation*}
In addition, we also have
\begin{equation*}\label{ineq:mcr-lb-permutation-2}
        \frac{1}{p_1} \sum_{j=1}^{p_1}\bbI\left\{\pi\left((z_1)_j\right) \neq (\tilde z_1)_j\right\} \geq \frac{1}{p_1} \min_a |T_a| \geq \frac{1}{p_1}\left(\frac{\alpha p_1}{r_1} - \frac{\alpha p_1}{4r_1^2}\right) \geq \frac{3\alpha}{4r_1}
\end{equation*}
for any non-identical permutation $\pi$ on $[r_1]$. This implies that identical mapping is the best permutation in the definition of misclassification rate $h_1(z_1,\tilde z_1)$, i.e.,
\begin{equation*}
    h_1(z_1, \tilde z_1) = \frac{1}{p_1}\sum_{j\in [p_1]} \bbI\{(z_1)_j \neq (\tilde z_1)_j\}.
\end{equation*}
Now following the proof of \cite[Theorem 2]{gao2018community}, for fixed $\cS^*$, we have
\begin{equation}\label{ineq:mcr-lb-1}
    \inf_{\hat z_1} \sup_{z_1} \bbE h(\hat z_1, z_1) \geq \frac{\alpha }{6r_1|T^c|} \sum_{j \in T^c}\left[\frac{1}{2r_1^2} \inf_{(\hat z_1)_j} \left(\bbP_1((\hat z_1)_j = 2)\right) + \left(\bbP_2((\hat z_1)_j = 1)\right)\right],
\end{equation}
where the supreme is taken over all $z_1$ satisfying \eqref{ineq:asmp-balance} and $\bbP_t$ denotes the probability distribution given $(z_1^*)_j = t$. By Neyman-Peason Lemma, the infimum of the right hand side of \eqref{ineq:mcr-lb-1} is achieved by the likelihood ratio test
$(\hat z_1)_j = \argmin_{a \in \{1,2\}} \left\|\left(\cM_1(\cY)\right)_{j:} - \left(\cM_1(\cX)\right)_{j_a:}\right\|_2^2$, where 
\begin{equation*}
    \cY = \cX + \cE,\qquad \cX = \cS^* \times_1 \M_1^* \times \cdots \times_d \M_d^*,
\end{equation*}
$\M_k^*$ is the membership matrix of $z_k^*$ and $j_a$ is some index such that $(z_1^*)_{j_a} = a$.

Note that
\begin{equation}\label{ineq:mcr-lb-theta-delta}
    \begin{split}
        \left\|\left(\cM_1(\cX)\right)_{j_1:} - \left(\cM_1(\cX)\right)_{j_2:}\right\|_2 &= \left\|\left((\cM_1(\cS))_{1:} - (\cM_1(\cS))_{2:}\right) \times_1 \M_1^* \times \cdots \times_d \M_d^*\right\|_\tF \\
        & \leq \left\|(\cM_1(\cS))_{1:} - (\cM_1(\cS))_{2:}\right\|_2 \cdot \prod_{k=2}^d \left\|\M_k^*\right\| \\
        & \leq C\sqrt{\frac{p_{-1}}{r_{-1}}}\Delta_1 \leq \frac{1}{2}\sigma.
    \end{split}
\end{equation}

Then, we can calculate
\begin{equation*}
    \begin{split}
        & \inf_{(\hat z_1)_j}\left(\bbP_1((\hat z_1)_j = 2)\right) + \left(\bbP_2((\hat z_1)_j = 1)\right)  \\
        & \qquad = 2\bbP\left(\|\left(\cM_1(\cX)\right)_{j_1:} - \left(\cM_1(\cX)\right)_{j_2:} + (\cM_1(\cE))_{j:} \|^2 \leq \|(\cM_1(\cE))_{j:}\|^2\right) \\
        & \qquad = 2\bbP\left(\|\left(\cM_1(\cX)\right)_{j_1:} - \left(\cM_1(\cX)\right)_{j_2:}\|^2 \leq 2(\left(\cM_1(\cX)\right)_{j_1:} - \left(\cM_1(\cX)\right)_{j_2:})(\cM_1(\cE))_{j:}^\top \right).
    \end{split}
\end{equation*}
Since $(\left(\cM_1(\cX)\right)_{j_1:} - \left(\cM_1(\cX)\right)_{j_2:})(\cM_1(\cE))_{j:}^\top \sim N\left(0, \sigma^2 \left\|\left(\cM_1(\cX)\right)_{j_1:} - \left(\cM_1(\cX)\right)_{j_2:}\right\|_2^2\right)$, we have
\begin{equation}\label{ineq:mcr-lb-2}
    \begin{split}
        \inf_{(\hat z_1)_j}\left(\bbP_1((\hat z_1)_j = 2)\right) + \left(\bbP_2((\hat z_1)_j = 1)\right)  &=  \bbP\left(N(0,1) \geq \frac{\left\|\left(\cM_1(\cX)\right)_{j_1:} - \left(\cM_1(\cX)\right)_{j_2:}\right\|_2}{2\sigma}\right) \\
        & \overset{\eqref{ineq:mcr-lb-theta-delta}}{\geq} \frac{1}{\sqrt{2\pi}} \int_{1/2}^\infty e^{-t^2/2} dt \geq c.
    \end{split}
\end{equation}

Combining \eqref{ineq:mcr-lb-1} and \eqref{ineq:mcr-lb-2}, we finish the proof. \qquad\qquad $\square$

\subsection{Proof of Theorem \ref{thm:computational-lower-bound}}
The idea to show the computational limit is to show the computational lower bound for a special class of high-order clustering model via average-case reduction. See more about the idea of average-case reduction in \cite{brennan2018reducibility,luo2020tensor}. Without loss of generality, we assume $\sigma = 1$. The special class of high-order clustering model we consider is the following. Suppose the core tensor $\cS$ in \eqref{eq:model-tensor} is a rank-$1$ tensor and satisfies:
\begin{equation*}
	\cS \in \bbR^{3 \times 3 \times \cdots \times 3}, \quad \cS = \Delta_{\min} \cdot \begin{bmatrix}
		1\\
		-1\\
		0
	\end{bmatrix} \circ \begin{bmatrix}
		1\\
		-1\\
		0
	\end{bmatrix} \circ \cdots \circ \begin{bmatrix}
		1\\
		-1\\
		0
	\end{bmatrix}.
\end{equation*}

In this special case, we can rewrite model \eqref{eq:model-tensor} as a rank-1 high-order clustering model
\begin{equation} \label{eq:mode-rank1}
	\cY = \Delta_{\min} \cdot m_1 \circ \cdots \circ m_d + \cE = \Delta_{\min} \prod_{i=1}^d \|m_i\|_2 \cdot \frac{m_1}{\|m_1\|_2}\circ \cdots \circ \frac{m_d}{\|m_d\|_2} + \cE,
\end{equation} where $m_i = \M_i \times \begin{bmatrix}
		1\\
		-1\\
		0
	\end{bmatrix}$. 

The rank-1 high-order clustering model \eqref{eq:mode-rank1} has one planted cluster supported on $S(m_1) \times \cdots \times S(m_d)$ and its statistical and computational limits for recovering $\{S(m_i) \}_{i=1}^d$ have been considered in \cite{luo2020tensor}. Specifically, define
\begin{equation*}
	\beta := - \lim_{p \to \infty} \frac{\log (\Delta_{\min} \prod_{i=1}^d \|m_i\|_2 /p^{d/2})}{\log p} = - \gamma/2,\quad  \lim_{p \to \infty} \frac{\log (|S(m_1)|) }{\log p} = \cdots =  \lim_{p \to \infty} \frac{\log (|S(m_d)|) }{\log p} = 1 =: \alpha.
\end{equation*}
Following the proof of Theorem 17 of \cite{luo2020tensor}, when $\beta > (\alpha d/2- d/4) \vee 0$, i.e., $\gamma < - d/2$, then for any sequence of polynomial-time algorithm $\phi: \cY \to (\hat{S}(m_1),\ldots, \hat{S}(m_d))$, 
\begin{equation*}
	\liminf_{p\to \infty }\sup_{\Theta}\bbP(\phi(\cY) \neq (S(m_1),\ldots, S(m_d))) \geq 1/2,
\end{equation*} under the HPC detection conjecture. By further noticing that recovering $\M_i$ is harder than recovering $S(m_i)$, we finish the proof of this Theorem. \quad $\square$

\subsection{Proof of Lemma \ref{lm:new-perturbation-determin}}

First of all, $\Y$ admits the following singular value decomposition:
\begin{equation*}
    \Y = \begin{bmatrix} \hat\U & \hat\U_{\perp} \end{bmatrix} \begin{bmatrix} \SSigma_1 & \\ & \SSigma_2 \end{bmatrix} \begin{bmatrix} \hat\V^\top \\ \hat\V_{\perp}^\top \end{bmatrix},
\end{equation*}
where $\SSigma_1$ and $\SSigma_2$ are diagonal matrices.
We introduce several transformations of the matrix $\Y$. First, let $\widetilde \Y = [\W~\W_\perp]^\top \Y$. Since $\widetilde \Y$ shares the same singular values and right singular space with $\Y$, we can decompose it as
\begin{equation*}
    \widetilde\Y = \begin{bmatrix} \widetilde\U & \widetilde\U_{\perp} \end{bmatrix} \begin{bmatrix} \SSigma_1 & \\ & \SSigma_2 \end{bmatrix} \begin{bmatrix} \hat\V^\top \\ \hat\V_{\perp}^\top \end{bmatrix}.
\end{equation*}
Here $\widetilde\U \in \bbO_{p_1,r}$, $\widetilde \U_\perp \in \bbO_{p_1,p_1-r}$, and it holds that
\begin{equation}\label{ineq:1st-transform}
    \hat \U_\perp^\top\W = \widetilde\U_\perp^\top \cdot \begin{bmatrix}\I_{r'} \\ 0 \end{bmatrix}.
\end{equation}
Next, we introduce $\widetilde \Y_1 = \W^\top\Y \in \bbR^{r' \times p_2}$ and $\widetilde \Y_2 = \W_\perp^\top\Y \in \bbR^{(p_1-r') \times p_2}$ be the first $r'$ and the rest $p_1-r'$ rows of $\widetilde \Y$, and let $\widetilde\Y_1 = \bar\U\bar\SSigma\bar\V^\top$ be the singular value decomposition with $\bar\U \in \bbO_{r'}, \bar\V \in \bbO_{p_2,r'}$. We denote
\begin{equation}\label{eq:construct-bar-Y}
    \begin{split}
        \bar \Y & = \begin{bmatrix}
    \bar\U^\top & \\  & \I_{p_1-r'} 
    \end{bmatrix} \cdot \widetilde \Y \cdot 
    \begin{bmatrix} \bar\V & \bar\V_\perp \end{bmatrix}.
    \end{split}
\end{equation}
Let $\U\in \bbO_{p_1,r}$, $\U_\perp \in \bbO_{p_1,p_1-r}$ be the left singular vectors of $\bar \Y$, then we have
\begin{equation*}
    \begin{bmatrix}
        \U & \U_\perp 
    \end{bmatrix} = \begin{bmatrix}
    \bar\U^\top & \\  & \I_{p_1-r'} 
    \end{bmatrix} \cdot \begin{bmatrix} \widetilde\U & \widetilde\U_{\perp}\end{bmatrix},
\end{equation*}
and it follows that
\begin{equation}\label{ineq:2nd-transform}
    \left\|\U_\perp^\top \cdot \begin{bmatrix}\I_{r'} \\ 0 \end{bmatrix}\right\| = \left\|\widetilde\U_\perp^\top \cdot \begin{bmatrix}\bar\U\\ 0 \end{bmatrix}\right\| = \left\|\widetilde\U_\perp^\top \cdot \begin{bmatrix}\I_{r'}\\ 0 \end{bmatrix}\right\|.
\end{equation}
By \eqref{ineq:1st-transform} and \eqref{ineq:2nd-transform}, it suffices to bound
$\left\|\U_\perp^\top \cdot \begin{bmatrix}\I_{r'} \\ 0 \end{bmatrix}\right\|$. We further denote the left singular vectors of $\bar\Y$ as $[u^{(1)}, \cdots, u^{(p_1)}]$ (with corresponding singular values descending) and write
\begin{equation*}
    u^{(k)} = \begin{bmatrix}
    \alpha^{(k)} \\ \beta^{(k)}
    \end{bmatrix},\qquad \alpha^{(k)} \in \bbR^{r'}, \beta^{(k)} \in \bbR^{p_1-r'}.
\end{equation*}
Then we have
\begin{equation}\label{ineq:3rd-transform}
    \left\|\U_\perp^\top \cdot \begin{bmatrix}\I_{r'} \\ 0 \end{bmatrix}\right\| = \left\|\begin{bmatrix}
        \alpha^{(r+1)} & \cdots \alpha^{(p_1)}
    \end{bmatrix}\right\|.
\end{equation}
Note that by the construction of $\bar\Y$ (Eqn. \eqref{eq:construct-bar-Y}), we have
\begin{equation*}
    \bar \Y = \begin{bmatrix}
        \bar\SSigma & 0 \\
        \widetilde\Y_2\bar \V & \widetilde\Y_2\bar \V_\perp
        \end{bmatrix}
\end{equation*}
and
\begin{equation*}
    \begin{split}
        \bar \Y\bar\Y^\top = \begin{bmatrix}
        \bar\SSigma^2 & \bar\SSigma\bar\V^\top \widetilde\Y_2^\top \\
        \widetilde\Y_2\bar \V\bar\SSigma & \widetilde\Y_2\widetilde\Y_2^\top
        \end{bmatrix}.
    \end{split}
\end{equation*}
Denote $ \widetilde\Y_2\bar\V= [y^{(1)} \ldots y^{(r')}] \in \bbR^{(p_1-r') \times r'}$. Since $\bar \SSigma = \diag(\sigma_1(\widetilde\Y_1),\ldots \sigma_{r'}(\widetilde\Y_1))$, we can further write
\begin{equation*}
    \begin{split}
        \bar \Y\bar\Y^\top = \begin{bmatrix}
        \sigma_1^2(\widetilde\Y_1) & & & \sigma_1(\widetilde\Y_1)y^{(1)\top} \\
         & \ddots & & \vdots \\
         & & \sigma_{r'}^2(\widetilde\Y_1) & \sigma_{r'}(\widetilde\Y_1)y^{(r')\top} \\ 
        \sigma_1(\widetilde\Y_1)y^{(1)} & \cdots & \sigma_{r'}(\widetilde\Y_1)y^{(r')} & \widetilde\Y_2\widetilde\Y_2^\top
        \end{bmatrix}.
    \end{split}
\end{equation*}
Since $\sigma_k(\widetilde\Y)$ is the $k$th singular value of $\widetilde\Y$, it is also the $k$th singular value of $\bar\Y$ due to the orthogonal transformation \eqref{eq:construct-bar-Y}. By the relationship between SVD and eigenvalue decomposition, we have $\bar\Y \bar \Y^\top u^{(k)} = \sigma_k^2(\widetilde\Y)u^{(k)}$. This implies, for any $1\leq i \leq r'$, $r+1\leq k \leq p_1$:
\begin{equation*}
    \begin{split}
        &\sigma_{i}^2(\widetilde\Y_1)\alpha_i^{(k)} + \sigma_i(\widetilde\Y_1)y^{(i)\top}\beta^{(k)} = \sigma_k^2(\widetilde\Y)\alpha_i^{(k)} \\
        & \implies \alpha_i^{(k)} = \frac{-\sigma_i(\widetilde\Y_1)}{\sigma_i^2(\widetilde\Y_1) - \sigma_k^2(\widetilde\Y)}y^{(i)\top}\beta^{(k)}.
    \end{split}
\end{equation*}
Now we are ready to bound $\left\|[\alpha^{(r+1)}~\ldots~\alpha^{(p_1)}]\right\|$ in \eqref{ineq:3rd-transform}. Let $s = (s_{r+1},\ldots,s_{p_1})$ be any norm-1 vector in $\bbR^{p_1-r}$, one can calculate
\begin{equation*}
    \begin{split}
        \sum_{k=r+1}^{p_1} s_k\alpha_i^{(k)} & = \sum_{k=r+1}^{p_1} \frac{-s_k\sigma_i(\widetilde\Y_1)}{\sigma_i^2(\widetilde\Y_1) - \sigma_k^2(\widetilde\Y)}y^{(i)\top}\beta^{(k)} \\
        & = \sum_{k=r+1}^{p_1} -\frac{s_k}{\sigma_i(\widetilde\Y_1)}\frac{1}{1 - \sigma_k^2(\widetilde\Y)/\sigma_i^2(\widetilde\Y_1)}y^{(i)\top}\beta^{(k)} \\
        & \overset{(a)}{=} \sum_{k=r+1}^{p_1}\sum_{j=0}^\infty -\frac{s_k\sigma_k^{2j}(\widetilde\Y)}{\sigma_i^{2j+1}(\widetilde\Y_1)}y^{(i)\top}\beta^{(k)} \\
        & = \sum_{j=0}^\infty \frac{-y^{(i)\top}}{\sigma_i^{2j+1}(\widetilde\Y_1)}\left(\sum_{k=r+1}^{p_1}s_k\sigma_k^{2j}(\widetilde\Y)\beta^{(k)}\right).
    \end{split}
\end{equation*}
Here in (a), the expansion is valid because $\sigma_k^2(\widetilde\Y) \leq \sigma_{r+1}^2(\widetilde\Y) = \sigma_{r+1}^2(\Y) < \sigma_{r'}^2(\widetilde\Y_1) \leq \sigma_{i}^2(\Y_1)$ by the assumption.
Therefore,
\begin{equation}\label{ineq:bound-pertu-spect}
    \begin{split}
        \left\|\sum_{k=r+1}^{p_1} s_k\alpha^{(k)}\right\|_2 = & \left\|\sum_{j=0}^\infty \bar \SSigma^{-(2j+1)} \bar\V^\top \widetilde\Y_2^\top \cdot \left(\sum_{k=r+1}^{p_1}s_k\sigma_k^{2j}(\widetilde\Y)\beta^{(k)}\right)\right\|_2 \\
        \leq & \sum_{j=0}^\infty \left\|\bar\SSigma^{-(2j+1)} \bar\V^\top \widetilde\Y_2^\top\right\| \cdot \left\|\sum_{k=r+1}^{p_1}s_k\sigma_k^{2j}(\widetilde\Y)\beta^{(k)}\right\|_2 \\
        \leq & \sum_{j=0}^\infty \left\|\bar\SSigma^{-(2j+1)} \bar\V^\top \widetilde\Y_2^\top\right\| \cdot \left\|[\beta^{(r+1)} \cdots \beta^{(p_1)}]\right\| \\
        & \cdot \left\| (s_{r+1}\sigma_{r+1}^{2j}(\widetilde\Y),\cdots,s_{p_1}\sigma_{p_1}^{2j}(\widetilde\Y))\right\| \\
        \overset{(a)}{\leq} & \sum_{j=0}^\infty \frac{\sigma_{r+1}^{2j}(\widetilde\Y)\|\widetilde\Y_2\bar \V\|}{\sigma_{r'}^{2j+1}(\widetilde\Y_1)} = \frac{\sigma_{r'}(\widetilde\Y_1)\cdot\left\|\widetilde\Y_2\bar\V\right\|}{\sigma_{r'}^2(\widetilde\Y_1) - \sigma_{r+1}^2(\widetilde\Y)}.
    \end{split}
\end{equation}
Here in the last inequality, we use the facts that $\left\|[\beta^{(r+1)} \cdots \beta^{(p_1)}]\right\| \leq 1$ since it is a submatrix of an orthogonal matrix; and
\begin{equation*}
    \begin{split}
        \left\| (s_{r+1}\sigma_{r+1}^{2j}(\widetilde\Y),\cdots,s_{p_1}\sigma_{p_1}^{2j}(\widetilde\Y))\right\| & \leq \left\| (s_{r+1}\sigma_{r+1}^{2j}(\widetilde\Y),\cdots,s_{p_1}\sigma_{r+1}^{2j}(\widetilde\Y))\right\| \\ & = \sigma_{r+1}^{2j}(\widetilde\Y)\cdot \|s\|_2 =  \sigma_{r+1}^{2j}(\widetilde\Y).
    \end{split}
\end{equation*}
Combining \eqref{ineq:3rd-transform} and \eqref{ineq:bound-pertu-spect}, we conclude 
\begin{equation*}
    \left\|\hat\U_\perp^\top \W\right\| \leq \frac{\sigma_{r'}(\widetilde\Y_1)\cdot\left\|\widetilde\Y_2\bar\V\right\|}{\sigma_{r'}^2(\widetilde\Y_1) - \sigma_{r+1}^2(\widetilde\Y)}.
\end{equation*}
Then the Lemma is proved by the definitions $\widetilde\Y_1 = \W^\top \Y$, $\widetilde\Y_2 = \W_\perp^\top \Y$; and the facts that $\sigma_{r+1}(\widetilde\Y) = \sigma_{r+1}(\Y)$, $\bbP_{\bar\V} = \bbP_{\Y^\top\W}$.

\subsection{Proof of Lemma \ref{lm:new-perturbation-subGaussian}}
Applying Lemma \ref{lm:new-perturbation-determin} with $\W = \U_{r'}$, we have
\begin{equation*}
    \left\|\hat\U_\perp^\top \U_{r'}\right\| \leq \frac{\sigma_{r'}(\U_{r'}^\top\Y)\left\|\U_{r'\perp}^\top\Y\bbP_{\Y^\top \U_{r'}}\right\|}{\sigma_{r'}^2(\U_{r'}^\top\Y) - \sigma_{r+1}^2(\Y)}.
\end{equation*}
We adopt the concentration inequalities developed by \citep[Lemma 4]{cai2018rate} to obtain the probabilistic bounds for $\sigma_{r'}(\U_{r'}^\top\Y)$, $\sigma_{r+1}^2(\Y)$ and $\left\|\U_{r'\perp}^\top \Y \bbP_{\Y^\top\U_{r'}}\right\|$. To this end, we denote $\X_1 := \bbP_{\U_{r'}}\X$, $\X_2 := \X - \X_1$ and $\Y_1 = \X_1 + \Z$. Then \citep[Lemma 4]{cai2018rate} implies that for any $x>0$,
\begin{equation*}
    \bbP\left(\sigma_{r'}^2(\U_{r'}^\top\Y_1)\leq (\sigma_{r'}^2(\X_1)+p_2)(1-x)\right) \leq C\exp\left(Cr'-c(\sigma_{r'}^2(\X_1)+p_2)\cdot x^2\wedge x\right),
\end{equation*}
\begin{equation*}
    \bbP\left(\sigma_{r+1}^2(\Y)\geq p_2(1+x)\right) \leq C\exp\left(Cp_1-cp_2 \cdot x^2\wedge x\right),
\end{equation*}
and
\begin{equation*}
	\begin{split}
    \bbP\left(\left\|\U_{r'\perp}^\top\Y_1\bbP_{\Y_1^\top \U_{r'} }\right\| \geq x\right) \leq & C\exp\left(Cp_1-cx^2\wedge x\sqrt{\sigma_{r'}^2(\X)+p_2}\right) \\
    & - C\exp\left(-c(\sigma_{r'}^2(\X)+p_2)\right),
	\end{split}
\end{equation*}
Setting $x = \frac{\sigma_{r'}^2(\X)}{3(\sigma_{r'}^2(\X)+p_2)}, \frac{\sigma_{r'}^2(\X)}{3p_2}$ and $C\left(\sqrt{p_1}+\frac{p_1}{\sqrt{\sigma_{r'}^2(\X)+p_2}}\right)$ in the above three displays, respectively, we know that with probability at least 
\begin{equation*}
    \begin{split}
        & 1 - C\exp\left(Cr' - c\frac{\sigma_{r'}^4(\X)}{3(\sigma_{r'}^2(\X)+p_2)}\right) - C\exp\left(Cp_1 - c\frac{\sigma_{r'}^4(\X)}{p_2} \wedge \sigma_{r'}^2(\X)\right) - C\exp(-c\underline p) \\
        &\qquad \geq 1-C\exp(-c\underline p),
    \end{split}
\end{equation*}
the following holds:
\begin{equation*}
    \begin{split}
        \sigma_{r'}^2(\U_{r'}\Y) = \sigma_{r'}^2(\U_{r'}^\top\Y_1) \geq p_2 + \frac{2}{3}\sigma_{r'}^2(\X) ,\qquad \sigma_{r+1}(\Y) \leq p_2 + \frac{1}{3}\sigma_{r'}^2(\X),\\ \left\|\U_{r'\perp}^\top\Y_1\bbP_{\Y_1^\top \U_{r'}}\right\| \leq C\left(\sqrt{p_1}+\frac{p_1}{\sqrt{\sigma_{r'}^2(\X)+p_2}}\right) \leq C\sqrt{p_1}.
    \end{split}
\end{equation*}
Note that
\begin{equation*}
    \begin{split}
        \left\|\U_{r'\perp}^\top\Y\bbP_{\Y^\top \U_{r'}}\right\| & = \left\|\U_{r'\perp}^\top\Y\bbP_{\Y_1^\top \U_{r'}}\right\| \\
        & = \left\|\U_{r'\perp}^\top(\Y_1 + \X_2)\bbP_{\Y_1^\top \U_{r'}}\right\| \\
        & \leq \left\|\U_{r'\perp}^\top\Y_1\bbP_{\Y_1^\top \U_{r'}}\right\| + \left\|\U_{r'\perp}^\top\X_2\bbP_{\Y_1^\top \U_{r'}}\right\| \\
        & \leq C\sqrt{p_1} + \left\|\U_{r'\perp}^\top\X_2\bbP_{\Y_1^\top \U_{r'}}\right\|.
    \end{split}
\end{equation*}
Now we focus on bounding $\left\|\U_{r'\perp}^\top\X_2\bbP_{\Y_1^\top \U_{r'}}\right\|$. 
\begin{equation*}
    \left\|\U_{r'\perp}^\top\X_2\bbP_{\Y_1^\top \U_{r'}}\right\| \leq \frac{\left\|\U_{r'\perp}^\top\X_2(\Y_1^\top\U_{r'})\right\|}{\sigma_{r'}(\Y_1^\top\U_{r'})} = \frac{\left\|\U_{r'\perp}^\top\X_2\Z^\top\U_{r'}\right\|}{\sigma_{r'}(\Y_1^\top\U_{r'})} \leq \frac{3\left\|\U_{r'\perp}^\top\X_2\Z^\top\U_{r'}\right\|}{2\sigma_{r'}(\X)}
\end{equation*}
By Lemma \ref{lm:concentration-ineq}, we have with probability at least $1-\exp(-c\underline p)$,
\begin{equation*}
    \left\|\U_{r'\perp}^\top\X_2 \Z^\top \U_{r'}\right\| \leq C\|\X_2\|\cdot (\sqrt{r} + \sqrt{p_1}) \leq C\sigma_{r'}(\X)\sqrt{p_1}.
\end{equation*}
Therefore,
$$
\left\|\U_{r'\perp}^\top\X_2\bbP_{\Y_1^\top \U_{r'}}\right\| \leq C\sqrt{p_1}.
$$

Combining all the arguments above, we finally obtain
\begin{equation*}
    \left\|\hat\U_{\perp}^\top \U_{r'}\right\| \leq C\frac{(\sqrt{p_2} + \sigma_{r'}(\X))\sqrt{p_1}}{(p_2+\frac{2}{3}\sigma_{r'}^2(\X)) - (p_2+\frac{1}{3}\sigma_{r'}^2(\X))} \leq \frac{\sqrt{p_1}\sigma_{r'}(\X) + \sqrt{p_1p_2}}{\sigma_{r'}^2(\X)}.
\end{equation*}
Now the proof is completed. \qquad\qquad $\square$

\newpage

\section{Additional Technical Lemmas}

\begin{Lemma}\label{lm:orthogonal-perpendicular}
    Suppose $\U_1 \in \bbO_{p_1,r_1},\ldots, \U_d \in \bbO_{p_d,r_d}$ are orthogonal matrices and $\cX \in \bbR^{p_1 \times \cdots \times p_d}$ is a tensor. Then,
    \begin{equation*}
        \left\|\cX \times_1 \U_1\U_1^\top \times \cdots \times_d \U_d\U_d^\top - \cX\right\|_\tF \leq \sum_{k=1}^d \left\|(\I - \U_k\U_k^\top)\cM_k(\cX)\right\|_\tF.
    \end{equation*}
\end{Lemma}
\begin{proof}
See \cite[Lemma 1]{zhang2019optimal}.
\end{proof}

\begin{Lemma}\label{lm:low-rank-perturbation}
    Suppose $\X, \E \in \bbR^{m \times n}$, $\rank(\X) = r$. Let $\U \in \bbO_{m,r}$ be the leading $r$ singular vectors of $\Y = \X + \E$. Then,
    \begin{equation*}
        \left\|(\I - \U\U^\top)\X\right\| \leq 2\left\|\Z\right\|, \qquad \left\|(\I - \U\U^\top)\X\right\|_\tF \leq \min\{2\sqrt{r}\|\Z\|, 2\|\Z\|_\tF\}.
    \end{equation*}
\end{Lemma}
\begin{proof}
See \cite[Lemma 6]{zhang2018tensor}.
\end{proof}
\begin{Lemma}\label{lm:concentration-ineq}
    Let $\cE \in \bbR^{p_1 \times \cdots \times p_d}$ such that each entry has independent mean-zero sub-Gaussian distribution with bounded sub-Gaussian norm, i.e.,
    \begin{equation*}
        \bbE \exp\left(\lambda\cE_{j_1,\ldots,j_d}\right) \leq e^{\lambda^2/2},\qquad \forall \lambda \in \bbR.
    \end{equation*}
    Let $\E_k = \cM_k(\cE)$ for $k \in [d]$, fix $\U_k^* \in \bbO_{p_k,r_k}$. Then with probability at least $1-\exp(-c\underline p)$, the following holds:
    \begin{equation}\label{ineq:concentration-1}
            \left\|\E_k (\U_{k+1}^* \otimes \cdots \otimes \U_d^* \otimes \U_1^* \otimes \cdots \U_{k-1}^{*})\right\| \leq C\left(\sqrt{p_k} + \sqrt{r_{-k}}\right),
    \end{equation}
    \begin{equation}\label{ineq:concentration-1.5}
            \left\|\E_k (\U_{k+1}^* \otimes \cdots \otimes \U_d^* \otimes \U_1^* \otimes \cdots \U_{k-1}^{*})\right\|_\tF \leq C\left(\sqrt{p_kr_{-k}}\right),
    \end{equation}
    \begin{equation}\label{ineq:concentration-2}
            \sup_{\substack{\U_l \in \bbO_{p_k,r_k} \\ l=1,\ldots,d}}\left\|\E_k (\U_{k+1} \otimes \cdots \otimes \U_d \otimes \U_1 \otimes \cdots \otimes \U_{k-1})\right\| \leq C\left(\sqrt{p_k} + \sqrt{r_{-k}} + \sum_{l\neq k} \sqrt{p_lr_l}\right),
    \end{equation}
    \begin{equation}\label{ineq:concentration-3}
            \sup_{\substack{\U_l \in \bbO_{p_k,r_k} \\ l=1,\ldots,d}}\left\|\E_k (\U_{k+1} \otimes \cdots \otimes \U_d \otimes \U_{1}  \otimes\cdots \otimes \U_{k-1})\right\|_\tF \leq C\left(\sqrt{p_kr_{-k}} + \sum_{l\neq k} \sqrt{p_lr_l}\right).
    \end{equation}
    \begin{equation}\label{ineq:concentration-4}
            \sup_{\substack{\U_l \in \bbO_{p_k,r_k} \\ l=1,\ldots,d}}\left\|\cE \times_1 \U_1^{\top} \times \cdots \times \U_d^{\top} \right\|_\tF \leq C\left(\sqrt{r_*} + \sum_{l = 1}^d \sqrt{p_lr_l}\right).
    \end{equation}
    Here the constants only depend on $d$.
\end{Lemma}
\begin{proof}
See Section \ref{sec:lm-concentration}.    
\end{proof}

\section{Proof of Technical Lemmas}\label{sec:lemmas}

\subsection{Proof of Lemma \ref{lm:loss-relation}}\label{sec:lm-loss-relation}
\begin{proof}
    Without loss of generality, we may assume $\pi_k^{(0)}$ is identity mapping on $[r_k]$. Let $e^{i}_n$ be the $i$th canonical basis in $\bbR^n$. Then, 
\begin{equation*}
	\begin{split}
		 l_k^{(t)} &= \frac{1}{p_k}\sum_{j=1}^{p_k}\sum_{b \in [r_k]} \bbI\{(z_k^{(t)})_j = b\}\cdot\left\|\cS \times_k \left(e^{b}_{r_k} - e^{(z_k)_j}_{r_k}\right)\right\|_\tF^2 \\
		 & = \frac{1}{p_k}\sum_{j=1}^{p_k}\sum_{b \neq (z_k)_j} \bbI\{(z_k^{(t)})_j = b\}\cdot\left\|\cS \times_k \left(e^{b}_{r_k} - e^{(z_k)_j}_{r_k}\right)\right\|_\tF^2 \\
		& \geq \frac{1}{p_k}\sum_{j=1}^{p_k}\bbI\{(z_k^{(t)})_j \neq (z_k)_j\} \cdot \min_{b \in [r_k]/(z_k)_j}\left\|\cS \times_k \left(e^{b}_{r_k} - e^{(z_k)_j}_{r_k}\right)\right\|_\tF^2 \\
		& \geq \Delta_k^2 h_k^{(t)}.
	\end{split}
\end{equation*}
\end{proof}

\subsection{Proof of Lemma \ref{lm:mebmership-spectra}}\label{sec:lm:mebmership-spectra}
\begin{proof}
Since $\alpha p_k/r_k \leq \sum_{j=1}^{p_k}\bbI\{(z_{k})_j=a\} \leq \beta p_k/r_k$,
\begin{equation}\label{ineq:block-size-t}
    \begin{split}
        \sum_{j=1}^{p_k}\bbI\{(z_{k}^{(t)})_j=a\} &\geq \sum_{j=1}^{p_k}\bbI\{(z_{k})_j=a\} - \sum_{j=1}^{p_k}\bbI\{(z_{k}^{(t)})_j \neq (z_{k})_j\} \\
        & \overset{\text{Lemma \ref{lm:loss-relation}}}{\geq} \frac{\alpha p_k}{r_k} - \frac{p_k l_k^{(t)}}{\Delta_k^2} \overset{\eqref{hpc-3}}{\geq} \frac{\alpha p_k}{2 r_k}; \\
        \sum_{j=1}^{p_k}\bbI\{(z_{k}^{(t)})_j=a\} & \leq \sum_{j=1}^{p_k}\bbI\{(z_{k})_j=a\} + \sum_{j=1}^{p_k}\bbI\{(z_{k}^{(t)})_j \neq (z_{k})_j\} \\
        & \overset{\text{Lemma \ref{lm:loss-relation}}}{\leq} \frac{\beta p_k}{r_k} + \frac{p_k l_k^{(t)}}{\Delta_k^2} \overset{\eqref{hpc-3}}{\leq}\frac{2\beta p_k}{ r_k}.
    \end{split}
\end{equation}
This proves the argument on the bounds of $\left|j \in [p_k] : (z_k^{(t)})_j = a\right|$. On the other hand, note that $\M_k^\top\M_k = \diag\left(\left|j \in [p_k] : (z_k^{(t)})_j = 1\right|, \ldots, \left|j \in [p_k] : (z_k^{(t)})_j = r_k\right|\right)$, therefore the spectra of $\M_k$  can be upper and lower bounded by $\max_{a \in [r_k]}\left(\left|j \in [p_k] : (z_k)_j = a\right|\right)^{1/2}$ and $\min_{a \in [r_k]}\left(\left|j \in [p_k] : (z_k)_j = a\right|\right)^{1/2}$ respectively. Similar arguments can be applied to $\W_k,\M_k^{(t)}$ and $\W_k^{(t)}$.
\end{proof}

\subsection{Proof of Lemma \ref{lm-1}}\label{sec:lm-1}
\begin{proof}
Let $\N := \M_2 \otimes \cdots \otimes \M_d$. Note that 
\begin{equation*}
	\N^\top \V = (\M_2^{\top}\W_2) \otimes \cdots \otimes (\M_d^{\top}\W_d) = \I_{r_2} \otimes \cdots \otimes \I_{r_d} = \I_{r_{-1}}
\end{equation*}
and 
\begin{equation*}
	\lambda_{r_{-1}}(\N) = \prod_{k=2}^d \lambda_{r_k}(\M_k) \overset{\text{Lemma \ref{lm:mebmership-spectra}}}{\geq} c\sqrt{\frac{p_{-1}}{r_{-1}}}.
\end{equation*}
Thus we have
\begin{equation}\label{ineq:lm-1-0-1}
	\left\|\V - \V^{(t)}\right\| \leq \left(\lambda_{r_{-1}}(\N)\right)^{-1} \cdot \left\|\N^{\top}\V - \N^{\top}\V^{(t)}\right\| \leq C\sqrt{\frac{r_{-1}}{p_{-1}}} \left\|\I - \N^{\top}\V^{(t)}\right\|.
\end{equation}
In the meantime,
\begin{equation}\label{ineq:lm-1-0-2}
    \begin{split}
        & \left\|\N^\top\V^{(t)} - \I\right\| = \left\|(\M_2^\top\W_2^{(t)}) \otimes \cdots \otimes(\M_d^\top\W_d^{(t)})-\I\right\| \\
		= & \left\|(\M_2^\top\W_2^{(t)} - \I) \otimes \cdots \otimes (\M_d^\top\W_d^{(t)}) + \I\otimes (\M_3^\top\W_3^{(t)} ) \otimes \cdots \otimes(\M_d^\top\W_d^{(t)})- \I\right\| \\
		\leq & \left\|(\M_2^\top\W_2^{(t)} - \I) \otimes \cdots \otimes (\M_d^\top\W_d^{(t)})\right\| +\left\|(\M_3^\top\W_3^{(t)} ) \otimes \cdots \otimes (\M_d^\top\W_d^{(t)}) - \I\right\| \\
		\overset{(a)}{\leq} & C\left\|\M_2^\top\W_2^{(t)} - \I\right\| + \left\|(\M_3^\top\W_3^{(t)} ) \otimes \cdots \otimes (\M_d^\top\W_d^{(t)}) - \I\right\| \\
		 \leq & \cdots \leq C\sum_{k=2}^d \left\|\M_k^\top\W_k^{(t)} - \I\right\| \leq C\sum_{k=2}^d \left\|\M_k^\top \W_k^{(t)}-\I\right\|_\tF.
	\end{split}
\end{equation}

Here (a) comes from Lemma \ref{lm:mebmership-spectra} that $\|\M_k\| \lesssim \sqrt{p_k/r_k}$ and $\|\W_k^{(t)}\| \lesssim \sqrt{r_k/p_k}$ and thus $\prod_{k=2}^d\|\M_k^\top \W_k^{(t)}\| \leq C$ for some constant $C$. 

Now we fix $k\in [d]$. For any $b \in [r_k]$, denote $\delta_b:=1 - \left(\M_k^{\top}\W_k^{(t)}\right)_{bb}$ and $n^{(t)}_b = \sum_{j=1}^{p_k}\bbI\{(z_k^{(t)})_{j} = b\}$. Since for any $ b, b' \in [r_k]$, 
\begin{equation*}
	\left(\M_k^{\top}\W_k^{(t)}\right)_{b'b} = \frac{\sum_{j=1}^{p_k}\bbI\left\{(z_{k})_{j} = b', (z_{k}^{(t)})_{j} = b\right\}}{n_{b}^{(t)}},
\end{equation*}
one can observe that 
\begin{equation*}
	0 \leq \delta_b \leq 1, \qquad \sum_{b' \in [r_k]/b}\left(\M_k^{\top}\W_k^{(t)}\right)_{b'b} = \delta_b.
\end{equation*}
Therefore,
\begin{equation}\label{ineq:lm-1-0-3}
	\begin{split}
		\left\|\M_k^\top\W_k^{(t)} - \I\right\|_\tF & = \sqrt{\sum_{b\in [r_k]}\left(\delta_b^2 + \sum_{b'\in [r_k]/b}\left(\M_k^\top\W_k^{(t)}\right)_{b'b}^2\right)} \\
		& \leq \sqrt{\sum_{b \in [r_k]}\left(\delta_b^2 + \left(\sum_{b'\in [r_k]/b}\left(\M_k^\top\W_k^{(t)}\right)_{b'b}\right)^2\right)}  \\
		& = \sqrt{2\sum_{b\in [r_k]}\delta_b^2} \leq \sqrt{2}\sum_{b\in [r_k]}\delta_b \\
		& = \sqrt{2} \sum_{b\in [r_k]} \frac{\sum_{j=1}^{p_k}\bbI\left\{(z_{k})_{j} \neq b, (z_{k}^{(t)})_{j} = b\right\}}{n_{b}^{(t)}} \\
		& \leq \sqrt{2}\max_{b \in [r_k]}\left(n_{b}^{(t)}\right)^{-1} \cdot \sum_{j=1}^{p_k}\bbI\left\{(z_k)_j \neq (z_k^{(t)})_j\right\} \\
		& \overset{\text{Lemma \ref{lm:mebmership-spectra}}}{\leq} C\frac{r_{k}}{p_{k}}\cdot p_{k}h_k^{(t)} \overset{\text{Lemma \ref{lm:loss-relation}}}{\leq} C \frac{r_{k}l_k^{(t)}}{\Delta_k^2}.
	\end{split}
\end{equation}
Now \eqref{ineq:lm-1-0} is proved by combining \eqref{ineq:lm-1-0-1}, \eqref{ineq:lm-1-0-2}, and \eqref{ineq:lm-1-0-3}.

To prove \eqref{ineq:lm-1-0.5}, we define $\V^k := \W_{2} \otimes \cdots \otimes \W_{k} \otimes \W_{k+1}^{(t)} \otimes \cdots \otimes \W_{d}^{(t)}$ for $2 \leq k \leq d-1$ and denote $\V^1 = \V^{(t)}$, $\V^d = \V$. Then, it follows that
\begin{equation}\label{ineq:lm-1-0.5-1}
    \begin{split}
        & \left\|\E(\V-\V^{(t)})\right\| \leq \sum_{k=2}^d \left\|\E(\V^k - \V^{k-1})\right\| \\
        \leq &  \sum_{k=2}^d \left\|\E\left(\W_2\otimes \cdots \otimes \W_{k-1} \otimes (\W_{k}-\W_k^{(t)}) \otimes  \W_{k+1}^{(t)} \otimes \cdots \otimes \W_{d}^{(t)} \right)\right\| \\
        \overset{\eqref{hpc-1-2}}{\leq} & C\left(\sqrt{r_{-1}}+\sqrt{\bar p\bar r}\right)\sum_{k=2}^d \left\|\W_k - \W_k^{(t)}\right\| \cdot \prod_{1<l<k} \|\W_{l}\| \cdot \prod_{k<l<d}\|\W_l^{(t)}\|.  \\
        \lesssim & \sqrt{\frac{r_{-1}(r_{-1}+\bar p \bar r)}{p_{-1}}} \sum_{k=2}^d \frac{r_kl_k^{(t)}}{\Delta_k^2}.
    \end{split}
\end{equation}
Here the last inequality comes from the fact that $\|\W_{l}\|, \|\W_{l}^{(t)}\| \lesssim \sqrt{r_k/p_k}$ (Lemma \ref{lm:mebmership-spectra}) and 
$$\|\W_k - \W_k^{(t)}\| \lesssim \sqrt{r_k/p_k}\|\I - \M_k^\top \W_k^{(t)}\| \overset{\eqref{ineq:lm-1-0-3}}{\lesssim} \sqrt{\frac{r_k}{p_k}}\cdot \frac{r_k l_k^{(t)}}{\Delta_k^2}.$$
Similarly, one can prove
\begin{equation*}
    \begin{split}
        \left\|\E(\V-\V^{(t)})\right\|_\tF & \overset{\eqref{hpc-1-3}}{\leq}  C\left(\sqrt{p_1r_{-1}}+\sqrt{\bar p\bar r}\right)\sum_{k=2}^d \left\|\W_k - \W_k^{(t)}\right\| \cdot \prod_{1<l<k} \|\W_{l}\| \cdot \prod_{k<l<d}\|\W_l^{(t)}\| \\
        & \lesssim \sqrt{\frac{r_{-1}(p_1r_{-1}+\bar p \bar r)}{p_{-1}}} \sum_{k=2}^d \frac{r_kl_k^{(t)}}{\Delta_k^2}.
    \end{split}
\end{equation*}

Next we prove \eqref{ineq:lm-1-1}. On one hand, by noticing $\X\V = \M\S$, we can bound $\left\|\left(\W_{:b} - \W_{:b}^{(t)}\right)^\top \X\V\right\|_2$ as
\begin{equation}\label{ineq:lm-1-1-1}
	\begin{split}
		& \left\|\left(\W_{:b} - \W_{:b}^{(t)}\right)^\top \X\V\right\|_2 = \left\|\left(\W_{:b} - \W_{:b}^{(t)}\right)^\top \M_1\S\right\|_2 \\
		& \qquad = \left\|\S_{b:} - \frac{\sum_{j=1}^{p_1}\S_{(z_{1})_j:}\bbI\{(z_{1}^{(t)})_j=b\}}{\sum_{j=1}^{p_1}\bbI\{(z_{1}^{(t)})_j=b\}}\right\|_2 \\
		& \qquad = \left\|\frac{1}{\sum_{j=1}^{p_1}\bbI\{(z_{1}^{(t)})_j=b\}}\left(\sum_{j=1}^{p_1}\sum_{b' \in [r_1]/b} \bbI\{(z_{1})_j=b, (z_{1}^{(t)})_j=b'\}(\S_{b:} - \S_{b':})\right)\right\|_2 \\
		& \qquad \overset{\text{Lemma \ref{lm:mebmership-spectra}}}{\leq} \frac{2 r_1}{\alpha p_1} \sum_{j=1}^{p_1}\sum_{b' \in [r_1]/b} \bbI\{(z_{1})_j=b, (z_{1}^{(t)})_j=b'\}\left\|\S_{b:} - \S_{b':}\right\|)_2 \\
		& \qquad \leq \frac{2r_1}{\alpha p_1 \Delta_1}\sum_{j=1}^{p_1}\sum_{b' \in [r_1]/b} \bbI\{(z_{1})_j=b, (z_{1}^{(t)})_j=b'\}\left\|\S_{b:} - \S_{b':}\right\|_2^2 \leq \frac{2r_1}{\alpha \Delta_1}l_1^{(t)}.
	\end{split}
\end{equation}
On the other hand, by the similar argument for the proof of \eqref{ineq:lm-1-0}, one can show \begin{equation}\label{ineq:lm-1-W-Wt}
    \left\|\W - \W^{(t)}\right\| \lesssim \sqrt{\frac{r_1}{p_1}} \cdot \left\|\I - \M^{(t)\top}\W\right\| \lesssim \frac{r_1^{3/2}}{\sqrt{p_1}\Delta_1^2}l_1^{(t)}   
\end{equation}
and it follows that
\begin{equation}\label{ineq:lm-1-1-2}
	\begin{split}
		\left\|(\W_{:b} - \W_{:b}^{(t)})^\top \E\V\right\|_2 & \leq \left\|\W_{:b} - \W_{:b}^{(t)}\right\|_2 \cdot \left\|\E\V\right\| \\
		& \overset{\eqref{hpc-1-1}}{\leq} C\frac{r_1^{3/2}}{\sqrt{p_1}\Delta_1^2}l_1^{(t)}\cdot \sqrt{\frac{r_{-1}}{p_{-1}}}\left(\sqrt{p_1} + \sqrt{r_{-1}}\right) \\
		& \overset{\eqref{hpc-3}}{\leq} C\sqrt{\frac{r_*}{p_*}}\left(\sqrt{p_1r_1} + \sqrt{r_*}\right)\frac{\sqrt{l_1^{(t)}}}{\Delta_1},
	\end{split}
\end{equation}
Then \eqref{ineq:lm-1-1} follows by \eqref{ineq:lm-1-1-1}, \eqref{ineq:lm-1-1-2} and triangle inequality.

Now we prove \eqref{ineq:lm-1-2}. First of all, we develop an upper bound for $\|\X (\V - \V^{(t)})\|_\tF$:
\begin{equation*}
    \begin{split}
        & \left\|\X(\V-\V^{(t)})\right\|_\tF \leq \sum_{k=2}^d \left\|\X(\V^k - \V^{k-1})\right\|_\tF \\
        \leq &  \sum_{k=2}^d \left\|\X\left(\W_2\otimes \cdots \otimes \W_{k-1} \otimes (\W_{k}-\W_k^{(t)}) \otimes  \W_{k+1}^{(t)} \otimes \cdots \otimes \W_{d}^{(t)} \right)\right\|_\tF \\
        = & \sum_{k=2}^d \left\|\cX \times_2 \W_{2}^{\top} \times \cdots \times_{k-1} \W_{k-1}^{\top} \times_k (\W_k - \W_k^{(t)})^\top \times_{k+1} \W_{k+1}^{(t)\top} \times \cdots \times_d \W_d^{(t)\top}\right\|_\tF \\
        \leq & \sum_{k=2}^d \left\|(\W_k - \W_k^{(t)})^\top\X_k\right\|_\tF \cdot \prod_{1<l<k} \|\W_{l}\| \cdot \prod_{k<l<d}\|\W_l^{(t)}\|. 
    \end{split}
\end{equation*}
By the similar argument as  \eqref{ineq:lm-1-1-1}, we can show that
\begin{equation*}
    \begin{split}
        &\left\|(\W_k^{(t)}-\W_k)^\top\X_k\right\|_\tF \leq \sum_{b \in [r_k]} \left\|\left((\W_k^{(t)})_{:b}-(\W_k)_{:b}\right)^\top\M_k\S_k\right\| \cdot \sqrt{\frac{p_{-k}}{r_{-k}}} \\
        & \lesssim \sqrt{\frac{p_{-k}}{r_{-k}}}\frac{r_k}{p_k}\sum_{j=1}^{p_k}\sum_{b\neq b' \in [r_k]} \bbI\{(z_{k})_j=b, (z_{k}^{(t)})_j=b'\}\left\|(\S_k)_{b:} - (\S_k)_{b':}\right\|_2 \\
        & = \sqrt{\frac{p_{-k}}{r_{-k}}}\frac{r_k}{\Delta_k}l_k^{(t)}.
    \end{split}
\end{equation*}
As a result,
\begin{equation}\label{ineq:lm-1-2-1}
    \begin{split}
        & \left\|\W_{:b}^{(t)\top} \X\left(\V - \V^{(t)}\right)\right\|_2 \\
        \leq & \left\|\W_{:b}^{(t)}\right\| \cdot \sum_{k=2}^d  \left\|(\W_k^{(t)}-\W_k)^\top\X_k\right\|_\tF \cdot \prod_{1<l<k} \|\W_{l}\| \cdot \prod_{k<l<d}\|\W_l^{(t)}\| \\
        \lesssim & \sum_{k=2}^d\frac{r_k}{\Delta_k} l_k^{(t)}.
    \end{split}
\end{equation}

Next we develop an upper bound for $\left\|\W_{:b}^{(t)\top}\E(\V - \V^{(t)})\right\|_2$. By triangle inequality,
\begin{equation}\label{ineq:lm-1-2-2}
	\left\|\W_{:b}^{(t)\top}\E(\V - \V^{(t)})\right\|_2 \leq \left\|\W_{:b}^{\top}\E(\V - \V^{(t)})\right\|_2 + \left\|(\W_{:b} - \W_{:b}^{(t)})^\top \E(\V - \V^{(t)})\right\|_2.
\end{equation}
We firstly have
\begin{equation}\label{ineq:lm-1-2-3}
	\begin{split}
	\left\|\W_{:b}^{\top}\E(\V - \V^{(t)})\right\|_2 & \leq \left\|\W_{:b}\right\|_2 \cdot \left\|\E(\V - \V^{(t)})\right\| \\
	& \overset{\eqref{ineq:lm-1-0.5-1}}{\lesssim} \|\W_{:b}\|_2\cdot \sqrt{\frac{r_{-1}(r_{-1}+\bar p \bar r)}{p_{-1}}} \sum_{k=2}^d \frac{r_kl_k^{(t)}}{\Delta_k^2}\\
	& \overset{\text{Lemma \ref{lm:mebmership-spectra}}}{\lesssim} \sqrt{\frac{r_*(r_{-1}+\bar p \bar r)}{p_*}}\sum_{k=2}^d\frac{r_k}{\Delta_k^2}l_k^{(t)} \\
	& \overset{\eqref{hpc-3}}{\lesssim} \sqrt{\frac{\bar r r_*^2 + \bar p\bar r^2 r_* }{p_*}} \cdot \sum_{k\geq 2} \frac{\sqrt{l_k^{(t)}}}{\Delta_k}.
	\end{split}
\end{equation}
On the other hand, we also have
\begin{equation}\label{ineq:lm-1-2-4}
	\begin{split}
	\left\|(\W_{:b} - \W_{:b}^{(t)})^\top \E(\V - \V^{(t)})\right\|_2 & \leq \left\|\W_{:b} - \W_{:b}^{(t)}\right\|_2 \cdot \left\|\E(\V - \V^{(t)})\right\| \\
	& \overset{\eqref{ineq:lm-1-0.5-1}\eqref{ineq:lm-1-W-Wt}}{\lesssim} \frac{r_1^{3/2}}{\sqrt{p_1}\Delta_1^2}l_1^{(t)}\left(\sqrt{\frac{r_{-1}(r_{-1}+\bar p \bar r)}{p_{-1}}} \sum_{k=2}^d \frac{r_kl_k^{(t)}}{\Delta_k^2}\right)\\
	& \overset{\eqref{hpc-3}}{\lesssim} \sqrt{\frac{\bar rr_*^2 + \bar p \bar r^2 r_*}{p_*}} \cdot \sum_{k\geq 2} \frac{\sqrt{l_k^{(t)}}}{\Delta_k}.
	\end{split}
\end{equation}
Then \eqref{ineq:lm-1-2} follows by combining \eqref{ineq:lm-1-2-1}, \eqref{ineq:lm-1-2-2}, \eqref{ineq:lm-1-2-3} and \eqref{ineq:lm-1-2-4}. 

For \eqref{ineq:lm-1-3}, first note that
\begin{equation}\label{ineq:lm-1-3-1}
	\begin{split}
		\left\|(\W_{:b}-\W_{:b})^\top \X \V^{(t)}\right\|_2 & \leq \left\|(\W_{:b}-\W_{:b})^\top \M_1\S\right\|_2 \cdot \left\|\N^{\top}\V^{(t)}\right\| \\
		& \overset{\eqref{ineq:lm-1-1-1}}{\lesssim} \frac{r_1}{\Delta_1}l_1^{(t)};
	\end{split}
\end{equation}
In the meantime, 
\begin{equation}\label{ineq:lm-1-3-2}
	\begin{split}
		\left\|(\W_{:b} - \W_{:b}^{(t)})^\top \E \V^{(t)}\right\|_2 &\leq \left\|(\W_{:b} - \W_{:b})^\top \E (\V - \V^{(t)})\right\| \\
		& \qquad +  \left\|(\W_{:b} - \W_{:b}^{(t)})^\top \E \V \right\| \\
		& \overset{\eqref{ineq:lm-1-1-2}\eqref{ineq:lm-1-2-4}}{\lesssim}  \sqrt{\frac{\bar rr_*^2 + \bar p \bar r^2 r_*}{p_*}} \cdot \sum_{k\geq 1} \frac{\sqrt{l_k^{(t)}}}{\Delta_k}.
	\end{split}
\end{equation} 
Thus by triangle inequality we proved \eqref{ineq:lm-1-3} and additionally finish the proof of this Lemma.\qquad $\square$

\subsection{Proof of Lemma \ref{lm-2}}\label{sec:lm-2}
First of all, noticing that $\left\langle e_j, \S_{{(z_1)}_j:} - \S_{b:} \right\rangle$ are mean-zero sub-Gaussian with sub-Gaussian norm $\tilde\sigma \leq C_0\sqrt{\frac{r_{-1}}{p_{-1}}}\left\|\S_{{(z_1)}_j:} - \S_{b:} \right\|$, \eqref{lm-2-1} can be obtained by standard sub-Gaussian concentration inequality.

To show \eqref{lm-2-2}, note that
\begin{equation*}
	\begin{split}
		& \left\langle e_j, \tilde\S_{(z_1)_j:} - \S_{(z_1)_j:}\right\rangle = \left\langle e_j, \W^{\top}_{:(z_k)_j} \E \V\right\rangle \\
		= & \frac{\left\|e_j\right\|^2 + \sum_{l \in [p_1]/j}\bbI\{(z_1)_l = (z_1)_j\}e_j^\top e_l}{\sum_{l=1}^{p_1}\bbI\{(z_1)_l = (z_1)_j\}} \\
		\geq & \frac{ e_j^\top\sum_{l \in [p_1]/j}\bbI\{(z_1)_l = (z_1)_j\} e_l}{\sum_{l=1}^{p_1}\bbI\{(z_1)_l = (z_1)_j\}}.
	\end{split}
\end{equation*} 
Denote
\begin{equation*}
	\eta_j = \frac{\sum_{l \in [p_1]/j}\bbI\{(z_1)_l = (z_1)_j\} e_l}{\sum_{l=1}^{p_1}\bbI\{(z_1)_l = (z_1)_j\}}.
\end{equation*}
One can observe that $\eta_j$ is a random vector with independent entries whose sub-Gaussian norms are bounded by $\sqrt{\frac{C_1r_*}{p_*}}$ for all $i \in [r_{-1}]$. Thus, by Bernstein inequality,
\begin{equation*}
	\bbP\left(\left\|\eta_j\right\|^2 \geq \frac{C_1r}{p}\left(r_{-1}+2\sqrt{xr_{-1}}+2x\right)\right) \leq e^{-x}.
\end{equation*}
Then,
\begin{equation*}
	\begin{split}
		& \bbP\left(\left\langle e_j, \tilde\S_{{(z_1)}_j:} - \S_{(z_1)_j:} \right\rangle \leq -\frac{1}{16}\left\|\S_{(z_1)_j:} - \S_{b:}\right\|^2\right) \\
		\leq & \bbP\left(e_j^\top\eta_j \leq -\frac{1}{16}\left\|\S_{(z_1)_j:} - \S_{b:}\right\|^2\right) \\
		\leq & \bbP\left(\left.e_j^\top\eta_j \leq -\frac{1}{16}\left\|\S_{(z_1)_j:} - \S_{b:}\right\|^2\right|\left\|\eta_j^2\right\| \leq \frac{C_1r_*}{p_*}\left(r_{-1}+2\sqrt{xr_{-1}}+2x\right)\right) \\
		& \quad + \bbP\left(\left\|\eta_j^2\right\| \geq \frac{C_1r_*}{p_*}\left(r_{-1}+2\sqrt{xr_{-1}}+2x\right)\right)  \\
		\leq & \exp\left(-\frac{p_*p_{-1} \left\|\S_{(z_1)_j:} - \S_{b:}\right\|^4}{C r_*r_{-1}(r_{-1}+2\sqrt{xr_{-1}}+2x)}\right) + e^{-x}.
	\end{split}
\end{equation*}
Setting $x = \frac{p_{-1}}{Cr_{-1}}\left\|\S_{(z_1)_j:} - \S_{b:}\right\|^2 \geq r_*$, we obtain
 \begin{equation*}
	\begin{split}
		& \bbP\left(\left\langle e_j, \tilde\S_{{(z_1)}_j:} - \S_{(z_1)_j:} \right\rangle \leq -\frac{1}{16}\left\|\S_{(z_1)_j:} - \S_{b:}\right\|^2\right) \\
		\leq & 2\exp\left(-\frac{ p_{-1}}{Cr_{-1}}\left\|\S_{(z_1)_j:} - \S_{b:}\right\|^2\right)
	\end{split}
\end{equation*}
and \eqref{lm-2-2} is proved.

To show \eqref{lm-2-3}, we only need to notice that since $b \neq (z_1)_j$, one simply have
\begin{equation*}
	\left \langle e_j, \S_{b:} - \tilde\S_{b} \right\rangle = -\frac{ e_j^\top\sum_{l \in [p_1]/j}\bbI\{(z_1)_l = b\} e_l}{\sum_{l=1}^{p_1}\bbI\{(z_1)_l = b\}}.
\end{equation*}
Then \eqref{lm-2-3} can be similarly proved as \eqref{lm-2-2}.\qquad $\square$
\end{proof}

\subsection{Proof of Lemma \ref{lm:concentration-ineq}}\label{sec:lm-concentration}
By \cite[Lemma 3]{zhang2019optimal}, we know that \eqref{ineq:concentration-1}, \eqref{ineq:concentration-2} hold with probability at least $1-e^{-c\underline p}$; and \eqref{ineq:concentration-1.5} comes from the basic inequality:
\begin{equation*}
    \begin{split}
        & \left\|\E_k(\U_{k+1}^* \otimes \cdots \otimes \U_d^* \otimes \U_1^* \otimes \cdots \U_{k-1}^*)\right\|_\tF \\
        \leq & \left(\sqrt{p_k} \wedge \sqrt{r_k}\right) \left\|\E_k(\U_{k+1}^* \otimes \cdots \otimes \U_d^* \otimes \U_1^* \otimes \cdots \U_{k-1}^*)\right\|  \overset{\eqref{ineq:concentration-1}}{\leq} C\sqrt{p_kr_{-k}}.
    \end{split}
\end{equation*}

We focus on establishing the concentration of \eqref{ineq:concentration-3}. Without loss of generality, let $k=1$. Note that
\begin{equation*}
    \begin{split}
        & \sup_{\substack{\U_l \in \bbO_{p_l,r_l} \\ l=2,\ldots,d}}\left\|\E_1(\U_2 \otimes \cdots \otimes \U_d)\right\|_\tF = \sup_{\substack{\U_l \in \bbO_{p_l,r_l} l=2,\ldots,d \\ \cS \in \bbR^{p_1 \times r_{-1}}}} \left\langle \cE \times_2 \U_2^\top \times \ldots \times \times_d \U_d^\top, \cS \right\rangle \\
        = & \sup_{\substack{\U_l \in \bbO_{p_l,r_l} l=2,\ldots,d \\ \cS \in \bbR^{p_1 \times r_2 \times \cdots  \times r_d}, \|\cS\|_\tF \leq 1}} \left\langle \cE , \cS \times_2 \U_2 \times \ldots \times_d \U_d \right\rangle.
    \end{split}
\end{equation*}
By \cite[Lemma 7]{zhang2018tensor}, for any $k=2,\ldots,d$, we can construct $\varepsilon$-net $\left\{\U_k^{(1)}, \ldots, \U_k^{(N_k)}\right\}$ for $\left\{\V_k \in \bbR^{p_k \times r_k}: \left\|\V_k \right\|\leq 1\right\}$ such that
\begin{equation*}
	\sup_{\V_k: \left\|\V_k\right\|\leq 1}\min_{i \leq N_k} \left\|\U_k - \U_k^{(i)}\right\| \leq \varepsilon
\end{equation*}
with $N_k \leq \left(\frac{4+\varepsilon}{\varepsilon}\right)^{p_kr_k}$. Also, we can construct $\varepsilon$-net $\{\cS^{(1)}, \ldots, \cS^{(N_\cS)}\}$ for $\{\cS \in \bbR^{p_1\times r_2 \times \cdots \times r_d}: \left\|\cS\right\|_\tF \leq 1\}$ such that
\begin{equation*}
	\sup_{\cS: \left\|\cS\right\|_\tF \leq 1}\min_{i \leq N_\cS} \left\|\cS - \cS^{(i)}\right\|_\tF \leq \varepsilon
\end{equation*} 
with $N_\cS \leq ((2+\varepsilon)/\varepsilon)^{p_1r_{-1}}$. We use $\cI$ to denote the index set $[N_\cS] \times [N_2] \times \cdots \times [N_d]$. Now suppose we have 
\begin{equation}
	\begin{split}
		\left(\cS^*, \U_2^*, \cdots, \U_d^*\right) = \argmax_{\substack{\U_l \in \bbO_{p_l,r_l} l=2,\ldots,d \\ \cS \in \bbR^{p_1 \times r_2 \times \cdots  \times r_d}, \|\cS\|_\tF \leq 1}} \left\langle \cE , \cS \times_2 \U_2 \times \ldots \times_d \U_d \right\rangle. \\
	\end{split}
\end{equation}
and denote
\begin{equation*}
    T := \left\langle \cE , \cS^* \times_2 \U_2^* \times \ldots \times_d \U_d^* \right\rangle. 
\end{equation*}
Then we can find some index $i = (i_s, i_2, \ldots,i_d) \in \cI$, such that
\begin{equation*}
	\begin{split}
	    \left\|\cS^* - \cS^{(i_0)}\right\|_\tF &\leq \varepsilon,\\
		\left\|\U_k^* - \U_k^{(i_k)}\right\| &\leq \varepsilon, ~k=2,\ldots,d
	\end{split}
\end{equation*} 
Thus,
\begin{equation*}
	\begin{split}
		& T = \left\langle \cE , \cS^{(i_s)} \times_2 \U_2^{(i_2)} \times \ldots \times_d \U_d^{(i_d)} \right\rangle + T - \left\langle \cE , \cS^{(i_s)} \times_2 \U_2^{(i_2)} \times \ldots \times_d \U_d^{(i_d)} \right\rangle  \\
		& = \left\langle \cE , \cS^* \times_2 \U_2^* \times \ldots \times_d \U_d^* \right\rangle + \left|\left\langle \cE , \cS^* \times_2 \U_2^* \times \ldots \times_d \U_d^* - cS^{(i_s)} \times_2 \U_2^{(i_2)} \times \ldots \times_d \U_d^{(i_d)} \right\rangle\right| \\
		& \leq \left\langle \cE , \cS^* \times_2 \U_2^* \times \ldots \times_d \U_d^* \right\rangle + (2^{d-1}-1)\varepsilon T.
	\end{split}
\end{equation*}
Take $\varepsilon = 2^{d-2}$, we obtain $T \leq 2 \left\langle \cE , \cS^* \times_2 \U_2^* \times \ldots \times_d \U_d^* \right\rangle$.

Notice that for any fixed $\cS$, $\left\{\U_k\right\}_{k=2}^d$ with $\|\cS\|_\tF \leq 1$ and $\left\|\U_k\right\| \leq 1$, we have $\left\|\llbracket \cS; \V_1,\V_2,\V_3 \rrbracket\right\|_\tF \leq 1$. Since each entry of $\cE$ are independent sub-Gaussian random variable with sub-Gaussian norm bounded by $\sigma$, by Hoeffding-type inequality for sub-Gaussian random variables' summation \citep{vershynin2010introduction}, we have 
\begin{equation*}
	\begin{split}
		\bbP\left(\left\langle \cE , \cS^* \times_2 \U_2^* \times \ldots \times_d \U_d^* \right\rangle \geq t\right) \leq \exp\left(1 - \frac{c t^2 }{\sigma^2}\right).
	\end{split}
\end{equation*}
Then it follows by union bound that
\begin{equation*}
	\begin{split}
		\bbP\left(T \geq t \right) & \leq \bbP\left(\max_{j_0,j_1,j_2,j_3} \left|\left\langle \cZ, \llbracket \cS^{(j_0)}; \V_1^{(j_1)}, \V_2^{(j_2)}, \V_3^{(j_3)}\rrbracket \right\rangle\right| \geq \frac{t}{2}\right) \\
		& \leq N_\cS N_2\cdots N_d\exp\left(1-\frac{ct^2}{4\sigma^2}\right) \\
		& \leq \exp\left(1 - \frac{ct^2}{4\sigma^2} + C(\varepsilon)\left(p_1r_{-1} + \sum_{k=2}^d p_kr_k\right)\right).
	\end{split}
\end{equation*}
Now by taking $t = C\sigma \sqrt{p_1r_{-1} + \sum_{k=1}^d p_kr_k}$, we have with probability at least $1 - \exp\left(-c(p_1r_{-1} + \sum_{k=1}^d p_kr_k)\right)$,
\begin{equation*}
	T \leq C\sigma \left(p_1r_{-1} + \sum_{k=1}^d p_kr_k\right)^{1/2},
\end{equation*}
This proves \eqref{ineq:concentration-3}. \eqref{ineq:concentration-4} can be similarly proved as \eqref{ineq:concentration-3} and we omit them here. \qquad\qquad $\square$

\end{sloppypar}
\end{document}